\documentclass[manuscript,screen]{acmart}
\AtBeginDocument{%
  }

\setcopyright{acmlicensed}
\copyrightyear{2026}
\acmYear{2026}
\acmDOI{XXXXXXX.XXXXXXX}
\usepackage{tcolorbox}

\begin{document}

\title{On AI Safety and Security Technical Debt in Engineering AI-Enabled Systems}


\author{Muhammad Tukur}
\authornote{Corresponding author}
\email{mmt310@student.bham.ac.uk}
\orcid{0000-0003-1103-9659}
\affiliation{%
  \institution{Computer Science, University of Birmingham, Edgbaston, UK; CSE, HBKU}
  \country{Qatar}
}

\author{Hayatullahi B. Adeyemo}
\affiliation{%
  \institution{Computing and Informatics, Bournemouth University}
  \city{Bournemouth}
  \country{UK}}
\email{hadeyemo@bournemouth.ac.uk}

\author{Tao Chen}
\affiliation{%
  \institution{Computer Science, University of Birmingham, Edgbaston}
  \city{Birmingham}
  \state{}
  \country{UK}
}
\email{t.chen@bham.ac.uk}

\author{Nour Ali}
\affiliation{%
 \institution{Brunel University London, Uxbridge}
 \city{London}
 \state{}
 \country{UK}}
 \email{Nour.Ali@brunel.ac.uk}

\author{Anis Zarrad}
\affiliation{%
  \institution{Computer Science, University of Birmingham, Edgbaston}
  \city{Birmingham}
  \state{}
  \country{UK}
}
\email{a.zarrad@bham.ac.uk}

\author{Rick Kazman}
\affiliation{%
  \institution{University of Hawaii}
  \city{Honolulu}
  \state{}
  \country{USA}
}
\email{kazman@hawaii.edu}

\author{Marco Agus}
\affiliation{%
  \institution{College of Science and Engineering, HBKU}
  \city{}
  \state{}
  \country{Qatar}}
\email{magus@hbku.edu.qa}

\author{Rami Bahsoon}
\affiliation{%
\authornotemark[1]
  \institution{Computer Science, University of Birmingham, Edgbaston}
  \city{Birmingham}
  \state{}
  \country{UK}
}
\email{r.bahsoon@bham.ac.uk}

\renewcommand{\shortauthors}{Tukur et al.}

\begin{abstract}
  Artificial intelligence (AI) systems are increasingly deployed in high-stakes domains such as healthcare, autonomous driving, finance, and education. While these systems offer powerful data-driven and adaptive capabilities, their complexity, rapid evolution, and dependence on dynamic data pipelines introduce new forms of engineering liability collectively referred to as AI Technical Debts (AITDs). AITDs arise from root causes spanning data governance, model implementation, algorithm design, architectural decisions, operational processes, documentation practices, and testing adequacy. Unlike conventional technical debt, many AITDs are latent and propagate across tightly coupled AI pipelines, leading to maintenance challenges, reliability degradation, and heightened safety or security risks.
Guided by the principles of AI TRiSM (AI Trust, Risk, and Security Management), this study reinterprets technical debt through the interconnected dimensions of trustworthiness, focusing on AI safety and security technical debts. We conduct a systematic review of 60 primary studies and identify 31 distinct types of AITD, which are organized into a root-cause–oriented taxonomy comprising seven classes. The analysis examines how these debts map to 18 trust-related concerns, including 6 safety hazards and 12 security vulnerabilities. To support mitigation, the review synthesizes 34 actionable guidelines (8 safety and 26 security) targeting the prevention, detection, and reduction of AITDs across the AI lifecycle.
Building on these findings, we introduce AITD-MAP, an integrated framework that connects AITD taxonomy, quality and risk impacts, and mitigation strategies into a unified structure for risk-aware AI engineering. The framework aims at assisting AI Software Engineers in making AI Safety and Security technical debts visible, along their root causes and mitigating their presence.
\end{abstract}



\received{25 July 2026}
\received[revised]{}
\received[accepted]{}

\maketitle

\section{Introduction}
\label{sec:introduction}
Artificial Intelligence (AI)-based systems have become deeply embedded in critical domains such as healthcare~\cite{dave2023artificial}, autonomous vehicles~\cite{bendiab2023autonomous}, education~\cite{Srinivasa2022}, and finance~\cite{giudici2023safe}. An AI-based system is defined as a software-enabled environment that integrates AI techniques—such as machine learning (ML), deep learning (DL), or natural language processing (NLP)—to perform complex tasks with a degree of autonomy or intelligence beyond what is achievable through conventional rule-based automation~\cite{recupito2024technical}. These systems leverage such techniques to automate decision-making, adapt to dynamic environments, and solve problems that require reasoning and learning, thereby exceeding the capabilities of traditional rule-based systems~\cite{recupito2024technical}.

Despite their growing capabilities and adoption, AI-enabled systems introduce substantial engineering and operational challenges. These include issues of transparency, maintainability, robustness, and long-term reliability, especially in dynamic, data-intensive, and safety-critical environments~\cite{salhab2024systematic}. 
A significant contributor to these challenges is the accumulation of AI Technical Debt (AITD)—the hidden costs and risks associated with suboptimal architectural, design, or operational decisions made throughout the AI system lifecycle~\cite{cunningham1992wycash}. In this study, AI Technical Debt (AITD) is defined as “a metaphor to indicate the typical quality concerns of suboptimal solutions integrated into the building process of AI-enabled systems”~\cite{bogner2021characterizing}. This definition extends the classical notion of technical debt to the AI context, reflecting quality concerns unique to data-driven and learning-based development processes. As Sculley et al.~\cite{sculley2015hidden} observed, “ML systems have a special capacity for incurring technical debt, because they have all the maintenance problems of traditional code plus an additional set of ML-specific issues.” These issues include data instability, model brittleness, algorithmic bias, and the opacity of learned behaviors. 
Similar to conventional technical debt, AI Technical Debts (AITDs) may remain latent and only become visible when they manifest as performance degradation, ethical violations, or system failures. However, in AI-enabled systems, this latency is often intensified by uncertainty in the system behaviour, continuous learning, feedback loops, and complex data pipelines and drifts feeding into the learning, making tracing and identifying the root causes of the debts difficult - particularly in high-stakes domains~\cite{Gyevnár2025531, salhab2024systematic}.

In response to the growing need for trustworthy AI, the concept of AI TRiSM (AI Trust, Risk, and Security Management) has emerged as a guiding governance framework. Highlighted by Gartner as a top strategic technology trend for 2023~\cite{groombridge2022gartner} and formally developed by Habbal et al.~\cite{habbal2024artificial}, IBM~\cite{ibm2025aitrism}, NIST~\cite{nist2024aitrism}, and Avivah~\cite{avivah2024aitrism}, AI TRiSM frames trust as a multidimensional construct encompassing safety, security, fairness, reliability, and transparency. Within this framework, safety and security are treated as complementary and interdependent pillars, not siloed concerns.

AI safety refers to the assurance that AI systems operate reliably without causing unintended harm~\cite{Gyevnár2025531,10.1007/978-3-030-83906-2_17}, while AI security focuses on protecting these systems against adversarial threats, unauthorized access, and malicious manipulation~\cite{schneider2024designing, gnitko2024systematic}. Although their technical origins differ—accidental failures versus intentional attacks—their consequences frequently overlap. A security breach may trigger unsafe system behavior, while weaknesses in safety mechanisms can expose systems to exploitation. In adaptive and continuously learning AI environments, AI Safety and Security vulnerabilities make the system more "porous" (e.g., numerous gaps/holes weakening the protective safety/security boundary of the system) due to factors such as data drift, continuous model updates, feedback loops, and growing system autonomy, which allow failures and attacks to propagate across traditionally separate concerns. This can be observed through coupled safety–security incidents, shared failure modes, and the accumulation of interrelated technical debts. Despite this convergence, existing literature largely treats safety and security debts in isolation, limiting our ability to reason about their co-evolution and cumulative impact on system risk and trustworthiness over time. The AI TRiSM perspective therefore motivates integrated approaches that jointly identify, model, and mitigate these intertwined forms of debt.

While our study does not extend AI TRiSM in a technical implementation sense, it adopts AI TRiSM as a conceptual lens to reinterpret technical debt in AI-enabled systems. Our core innovation is to view AITDs not solely through architectural layers or lifecycle phases, but also through the risk-oriented dimensions of trust, risk, safety, and security, as advocated by AI TRiSM. This trust-oriented framing enables a structured re-evaluation of technical debt and helps consolidate fragmented discussions around AI safety and security. For example, debts such as ethical debt, undeclared consumers, feedback loops, and algorithmic bias are reinterpreted as systemic risks that degrade AI trustworthiness if left unmanaged. To the best of our knowledge, this is the first systematic review to explicitly examine AITDs through the lens of AI TRiSM. By doing so, we offer a risk-informed, actionable roadmap that aligns technical debt management with the broader goals of trustworthy and sustainable AI governance.

\subsection{Motivation}

AI-based systems are increasingly deployed in dynamic, high-stakes environments where long-term reliability and perceived trustworthiness are critical~\cite{aleksandra2025evaluating}. These systems often incur AI Technical Debt (AITD)—hidden costs arising from suboptimal decisions during design, development, or deployment~\cite{sculley2015hidden}. Unlike traditional technical debt, AITDs are frequently driven by AI-specific factors such as data drift, opaque models, feedback loops, and algorithmic bias, making them harder to detect and manage.
In the context of responsible AI, safety and security represent core system quality attributes that can be explicitly designed, implemented, and verified, while trust and risk reflect higher-level assessments of system behavior, assurance, and stakeholder confidence~\cite{li2023trustworthy, nist2024aitrism}. Recent governance frameworks, including AI TRiSM (AI Trust, Risk, and Security Management), emphasize the need to manage these dimensions in a coordinated manner~\cite{groombridge2022gartner, ibm2025aitrism, avivah2024aitrism, habbal2024artificial}. However, existing AITD research rarely examines how accumulated technical debt degrades safety and security properties and, in turn, shapes system-level risk exposure and trust outcomes over time.


This study is motivated by the need to bridge this gap. We adopt AI TRiSM as a conceptual lens—not to extend the framework but to reframe AITD through its dimensions. Our goal is to support a more holistic, risk-informed understanding of technical debt in AI systems, enabling more trustworthy and sustainable AI development.

\subsection{Research Questions and Contributions}
To address the gaps identified in the current literature and advance a root-cause–oriented understanding of AI Technical Debt (AITD), this study conducts a systematic review investigating the forms, impacts, and mitigation strategies associated with AITDs across the AI development lifecycle. The review is guided by the following research questions:

\begin{itemize}
    \item \textbf{RQ1:} What are the different types of technical debt found in AI-enabled systems? And which types are reported most frequently in the literature?

    \textit{This question seeks to identify, characterize, and classify AITDs across the full AI lifecycle, including debts related to data, models, algorithms, architecture, operational processes, documentation, and testing.}



    \item \textbf{RQ2:} How are the identified AITDs related to safety and security debt?
    
    \textit{This question examines the extent to which AITDs contribute to vulnerabilities and operational hazards, reinforcing the AI TRiSM view of trustworthiness.}
    
    \item \textbf{RQ3:} What mitigation strategies and management activities are suggested for addressing the identified AITDs?
    
    \textit{This question synthesizes actionable guidelines to support responsible, risk-aware AI development.}
    
\end{itemize}

\begin{figure}[t]
    \centering
    \includegraphics[width=0.8\columnwidth]{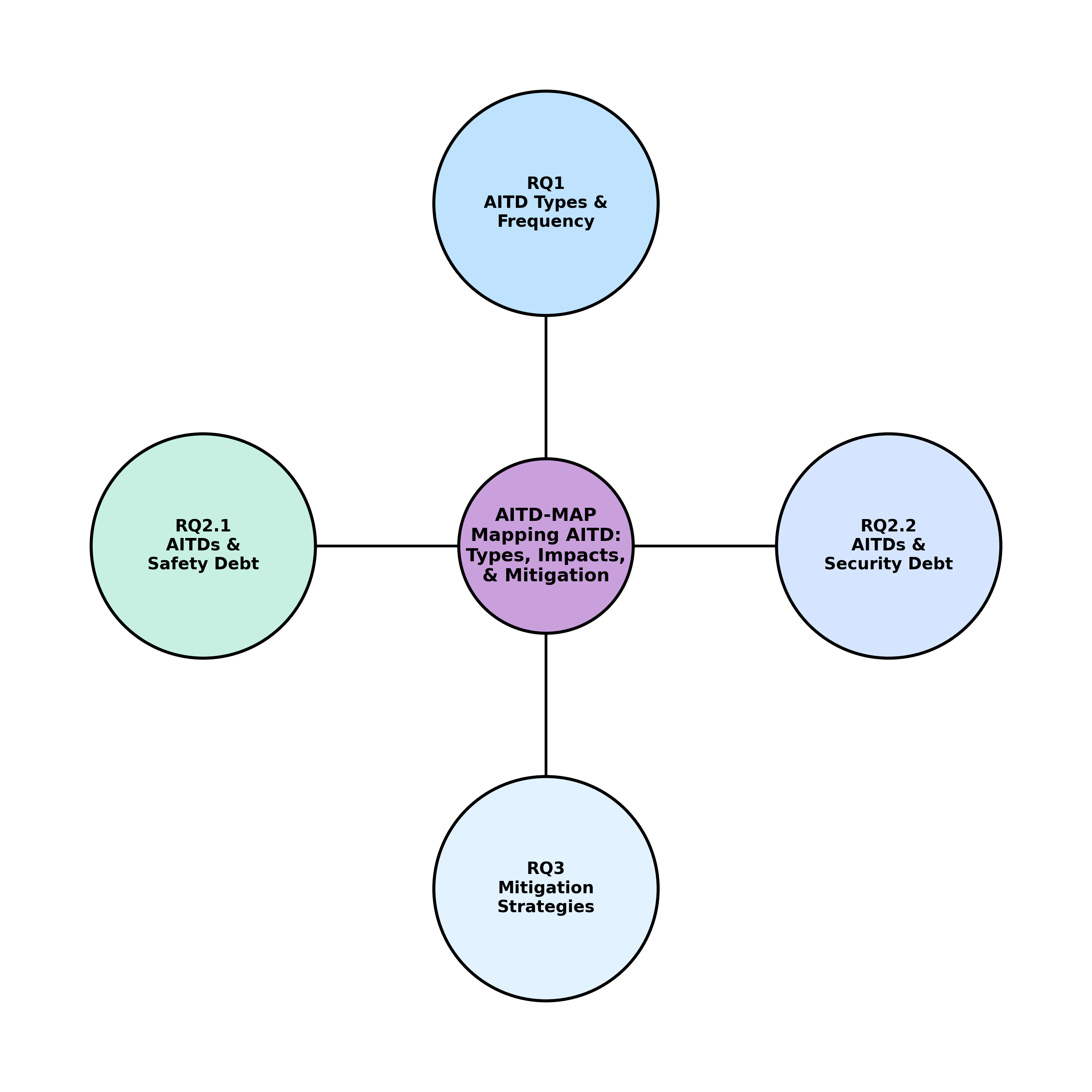}
    \caption{\textbf{Circular representation of AITD-MAP (Mapping AI Technical Debt: Types, Impacts, \& Guidelines)}, illustrating how each research question (RQ1–RQ3) contributes to the construction of the integrative framework. The diagram highlights the flow from taxonomy development and frequency analysis to the assessment of security and safety concerns and corresponding mitigation strategies.}\label{fig:AIM_map}

\end{figure}

Building on these research questions, the study offers the following key contributions:

\begin{itemize}
    \item A comprehensive, root-cause–oriented taxonomy of \textit{31 AITDs}, organized into seven major classes: Data \& Library–Related Debts, Model \& Code–Related Debts, Algorithm–Related Debts, Design \& Architecture Debts, Operational \& Lifecycle Debts, Documentation \& Communication Debts, and Testing \& Quality Assurance Debts.

    

    \item A \textit{novel mapping} between AITDs and trust-related risks, identifying \textit{6 safety} and \textit{12 security} concerns directly linked to specific forms of debt, thereby demonstrating how AITDs shape vulnerability paths and operational hazards in AI-enabled systems.

    \item A synthesis of \textit{34 actionable mitigation guidelines} (\textit{8 safety} and \textit{26 security}), providing concrete strategies for preventing, detecting, and reducing AITDs. These include practices such as Human-AI Control Mode Switching, Ethical Black Boxes, Continuous Behavioral and Drift Monitoring, Adversarial and Defensive Training, Out-of-Distribution Detection, and Formal Verification for high-risk applications.

    \item The introduction of AITD-MAP, a unified analytical framework that integrates AITD taxonomy, impact analysis, and mitigation strategies (See Figure \ref{fig:AIM_map}), designed to support risk-informed decision-making in AI system engineering and align with principles of trustworthy AI.

\end{itemize}

By consolidating fragmented insights and providing a structured, risk-aware roadmap for managing AI Technical Debt, this study advances responsible, resilient, and sustainable AI engineering. The findings reinforce the importance of integrating quality, safety and security considerations throughout the AI lifecycle, offering actionable guidance for practitioners and researchers working toward trustworthy and dependable AI systems.

\subsection{Paper Organization}
The rest of the paper is structured as follows: Section~\ref{sec:rw} presents the related works. Section~\ref{sec:methodology} describes the research methodology. Section~\ref{sec:AITDs} presents the taxonomy of AITDs. It further explores their impact on software quality attributes. 
Section~\ref{sec:secsafeconcerns} examines their connection to safety and security. Section~\ref{sec:secsafeguidelines} proposes mitigation guidelines. Section~\ref{sec:principalfindings} discusses principal findings and outlines strengths and limitations. Section~\ref{sec:conclusion} concludes with a summary and future directions.

\section{Related work}
\label{sec:rw}

Technical debt (TD) has been extensively studied in the context of traditional software  \cite{avgeriou2016managing} and systems  \cite{kleinwaks2023technical, kleinwaks2023ontology} engineering. Foundational reviews have explored its identification  \cite{alves2016identification}, management  \cite{avgeriou2016managing, li2015systematic}, prioritization  \cite{alfayez2020systematic}, financial implications  \cite{ampatzoglou2015financial}, and intelligent management techniques across diverse domains  \cite{albuquerque2022comprehending}. However, these studies generally overlook the unique characteristics and quality concerns associated with AI-based systems.

In response to the growing importance of Artificial Intelligence (AI), recent research has begun to explore technical debt in AI-enabled systems. These efforts, however, remain fragmented. Some works focus on specific debt types such as ethical debt  \cite{petrozzino2021pays}, bias and fairness  \cite{roselli2019managing}, self-admitted technical debt (SATD)  \cite{obrien202223}, algorithmic debt  \cite{simon2023algorithm}, architectural debt \cite{sas2023architectural}, and requirements engineering debt \cite{belani2019requirements}. Others investigate technical debt within particular application domains, such as AI-based competition platforms \cite{sklavenitis2024measuring}, recommender systems \cite{moreschini2024towards}, large language models (LLMs) \cite{menshawy2024navigating}, and complex systems \cite{belani2019requirements}.

In terms of comprehensive analyses, Bogner et al. \cite{bogner2021characterizing} presented a systematic mapping study on TD and anti-patterns in AI systems, offering a partial taxonomy with limited impact analysis and scattered mitigation approaches. Recupito et al. \cite{recupito2024technical} conducted a survey-based study that captures practitioner insights into architecture-level AITDs, but lacks a structured taxonomy and generalized mitigation framework. Washizaki et al. \cite{washizaki2019studying} attempted to classify SE design patterns for ML systems but focused more on conceptual classification than technical debt characterization.

Sculley et al. \cite{sculley2015hidden} provided the foundational industry perspective on ML-specific technical debt, such as glue code and entanglement, but did not offer empirical classification or mitigation strategies. Bhatia et al. \cite{bhatia2023empirical} conducted a large-scale empirical study on SATD in ML projects, contributing a refined SATD taxonomy and insights into long-term debt evolution, although security and safety aspects were not addressed. Menshawy et al. \cite{menshawy2024navigating} offered practical insights into TD associated with LLM deployment, highlighting engineering trade-offs and mitigation practices from real-world systems.


A comparative summary of these works is presented in Table~\ref{tab:related_reviews}. As shown, existing literature generally lacks a holistic view of AITDs across types, impacts, and mitigation strategies. In addition, few studies incorporate a systematic approach to evaluate the security and safety implications of AITDs or provide actionable guidelines for debt reduction.

In contrast, this study provides the first comprehensive review that systematically identifies and categorizes 31 distinct types of AITDs from 60
primary studies. It goes further to analyze their effects on system quality, with an emphasis on safety and security concerns, and synthesizes 16 actionable guidelines for mitigating these issues. Our findings address critical gaps in the previous literature by integrating taxonomic classification, impact assessment, and practical mitigation strategies within a single unified framework.

\begin{table}[ht]
\centering
\caption{Summary of Related Works on Technical Debt in AI-Based Systems}
\label{tab:related_reviews}
\footnotesize
\begin{tabular}{p{1.2cm} p{1.2cm} p{0.8cm} p{0.7cm} p{2.0cm} p{2.0cm} p{1.3cm} p{1.5cm} p{1.2cm} p{1.3cm}}
\toprule
\textbf{Study} & \textbf{Type} & \textbf{\# of Studies} & \textbf{Pub. Period} & \textbf{Objective} & \textbf{Methodological Framework} & \textbf{Taxonomy Provided} & \textbf{Impact Analysis} & \textbf{Security/ Safety Consideration} & \textbf{Mitigation Strategies} \\
\hline

Bogner et al. (2021) \cite{bogner2021characterizing} & Systematic Mapping Study (SMS) & 21 & Up to 2020 & Map the landscape of TD and antipatterns in AI systems & Wohlin \cite{wohlin2014guidelines} and Petersen \cite{petersen2008systematic} guidelines & Partial – categorized TDs \& antipatterns loosely & General discussion of impact & Mentioned generally & 46 scattered mitigation approaches \\
\hline

Recupito et al. (2024) \cite{recupito2024technical} & Survey Study & 53 practitioner responses & Up to 2024 & Investigate code- and architecture-level AITDs & Survey + qualitative content analysis & Focused on 9 AITD types & Moderate, practitioner-perceived impact & Mentioned generally & Mostly ad hoc, manual strategies \\
\hline

Sklavenitis et al. (2024) \cite{sklavenitis2024measuring} & Scoping Review & 100 & 2012–2023 & Measure AITDs in competitions; introduce Accessibility Debt & Scoping review + domain-specific questionnaire & Yes – 18 AITDs incl. Accessibility Debt & Structured analysis with practical examples & Not a primary emphasis & Not a primary emphasis \\
\hline

Washizaki et al. (2019) \cite{washizaki2019studying} & Preliminary SLR & 38 & Up to 2019 & Identify SE design patterns in ML systems & Literature review + ML lifecycle classification & Yes – 33 SE patterns & Partially – conceptual & Not explicitly analyzed & Not explicitly analyzed \\
\hline

Sculley et al. (2015) \cite{sculley2015hidden} & Conceptual Essay & N/A & Pre-2015 & Highlight ML-specific TD from industry view & Conceptual analysis (Google experience) & Yes – anti-patterns & Qualitative discussion & Limited (e.g., feedback loops) & Conceptual suggestions (e.g., monitoring) \\
\hline

Bhatia et al. (2023) \cite{bhatia2023empirical} & Empirical Study & 318 ML + 318 Non-ML & Up to 2023 & Analyze SATD in ML software & Mixed methods – SHAP, survival, manual coding & Yes – Extends SATD taxonomy & Evolution + predictors of SATD & Not directly addressed & Implicit suggestions \\
\hline

Menshawy et al. (2024) \cite{menshawy2024navigating} & Perspective Study & N/A & Up to 2024 & Discuss TD in LLM deployment & Experience-based synthesis + examples & Yes – LLM-specific TD types & Covers performance, memory, latency & Explicit (bias, hallucinations, feedback) & Prompt tuning, quantization, caching \\
\hline

\textbf{Ours (2025)} & Scoping Review & 60 & Up to 2025 & Identify, categorize, and analyze AITDs, their impacts, and mitigation strategies & PRISMA-ScR \cite{tricco2018prisma} & Comprehensive taxonomy of AITDs & Detailed impact on AI-system trustworthiness & Explicit analysis of security and safety implications & 34 synthesized mitigation guidelines \\
\bottomrule
\end{tabular}
\end{table}

\section{Methodology}
\label{sec:methodology}
To address the research questions outlined in the introduction, we adopted the guidelines set by the PRISMA Extension for Scoping Reviews (PRISMA-ScR) \cite{tricco2018prisma}. As illustrated in~Figure~\ref{fig:prismachart}.A, this framework provides a structured and systematic approach for conducting comprehensive scoping reviews, ensuring rigor and thoroughness. The literature search was conducted through the following stages:

\subsection{Search Strategy}
The search strategy encompassed the selection of bibliographic databases, formulation of search terms and strings, establishment of inclusion and exclusion criteria, and the process for selecting relevant studies for inclusion.

\subsection{Data Source Selection}
To ensure comprehensive coverage of relevant literature, we conducted searches in the following electronic databases: ACM Digital Library, IEEE Xplore, Scopus, and Springer. These databases were selected for their extensive coverage of research in software engineering and computer science in general.

\subsection{Venue Selection Criteria}
In addition to database and content filtering, venue selection was guided by stringent quality and relevance criteria. Priority was given to studies published in high-impact, peer-reviewed journals, top-tier conferences, workshops, and symposium with a clear focus on AI system engineering and/or software development practices. The final selection spans 34 distinct venues, with CAIN, Empirical Software Engineering, SEAA,  ICSE, and TechDebt, among the most represented (c.f. Table~\ref{tab:venue-distribution}). Emphasis was placed on venues indexed by ACM, IEEE, Springer, and Scopus, given their established reputation and consistent record of publishing research on software engineering. 



\begin{table*}[ht]
\small
\centering
\caption{Distribution of Selected Studies by Venue}
\begin{tabular}{|c|p{12cm}|c|}
\hline
\textbf{S/N} & \textbf{Venue} & \textbf{Count} \\
\hline
1 & International Conference on AI Engineering: Software Engineering for AI (CAIN) & 6 \\
2 & Empirical Software Engineering & 5 \\
3 & Euromicro Conference on Software Engineering and Advanced Applications (SEAA) & 4 \\
4 & International Conference on Software Engineering (ICSE) & 4 \\
5 & ACM/IEEE International Conference on Technical Debt (TechDebt) & 4 \\
6 & ACM Transactions on Software Engineering and Methodology & 3 \\
7 & IEEE International Conference on Software Maintenance and Evolution (ICSME) & 3 \\
8 & IEEE Transactions on Software Engineering & 2 \\
9 & Software Quality: Future Perspectives on Software Engineering Quality (SWQD) & 2 \\
10 & AI and Ethics & 2 \\
11 & International Conference on Mining Software Repositories (MSR) & 2 \\
12 & Journal of Systems and Software & 1 \\
13 & ACM Conference on Equity and Access in Algorithms, Mechanisms, and Optimization & 1 \\
14 & ACM Conference on Fairness, Accountability, and Transparency & 1 \\
15 & ACM Joint European Software Engineering Conference and Symposium on the Foundations of Software Engineering & 1 \\
16 & ACM SIGMOD Record & 1 \\
17 & ACM/IEEE Workshop on AI Engineering-Software Engineering for AI (WAIN) & 1 \\
18 & Advances in Neural Information Processing Systems & 1 \\
19 & IEEE International Conference on Big Data (Big Data) & 1 \\
20 & Information and Software Technology & 1 \\
21 & International Conference on Automated Software Engineering (ASE) & 1 \\
22 & International Conference on Green Computing and Internet of Things (ICGCIoT) & 1 \\
23 & International Conference on Industrial Informatics (INDIN) & 1 \\
24 & International Conference on Product-Focused Software Process Improvement & 1 \\
25 & International Conference on Software Architecture Companion (ICSA-C) & 1 \\
26 & International Requirements Engineering Conference Workshops & 1 \\
27 & International Workshop on Empirical Software Engineering in Practice (IWESEP) & 1 \\
28 & Workshop on Machine Learning and Systems & 1 \\
29 & World Wide Web Conference & 1 \\
30 & International Conference on Emerging Technologies and Computing (ICETC) & 1 \\
31 & Brazilian Symposium on Software Components, Architectures, and Reuse & 1 \\
32 & International Conference on Program Comprehension (ICPC) & 1 \\
33 & IEEE Access & 1 \\
34 & IEEE Annual Computing and Communication Workshop and Conference (CCWC) & 1 \\
\hline
\multicolumn{2}{|c|}{\textbf{Total}} & \textbf{60} \\
\hline
\end{tabular}
\label{tab:venue-distribution}
\end{table*}

\subsection{Search terms}
AI-enabled systems share certain forms of technical debt with traditional software systems; however, they also introduce unique types of technical debt that arise from their data-driven architectures, learning components, adaptive behaviors and feedback loops, and emergence. Consequently, it is essential to focus on literature that explicitly addresses technical debt in AI-based systems. Based on this distinction, the following terms were identified as most relevant for constructing the search queries.

\begin{itemize}
    \item AI Terms: Artificial Intelligence, AI, Machine Learning, ML, Deep Learning, DL, Natural Language Processing, NLP, Generative Artificial Intelligence, GenAI, Large Language Model, LLM, Agentic AI. 
    \item Technical Debt Terms: Technical Debt, TD, AITD, Anti-patterns, Self-Admitted Technical Debt, SATD, Smell.
\end{itemize}

\subsection{Search Strings}

Search queries were formulated using appropriate literal and semantic synonyms to capture a wide range of relevant results. Table~\ref{tab:electronicsearch} presents the search strings applied to each data source:

\begin{table*}[ht]
\small
\centering
\caption{Restricted Electronic Search Results for AI and Technical Debt}
\begin{tabular}{p{1.5cm} p{7cm} p{4cm} p{1cm}}
\toprule
\textbf{Data Source}       & \textbf{Search String}                                                                                                                                                          & \textbf{Filters Applied}                                                       & \textbf{Results} \\ \hline
ACM Digital Library        &  [[Abstract: artificial intelligence] OR [Abstract: ai] OR [Abstract: machine learning] OR [Abstract: ml] OR [Abstract: deep learning] OR [Abstract: dl] OR [Abstract: natural language processing] OR [Abstract: nlp] OR [Abstract: generative artificial intelligence] OR [Abstract: genai] OR [Abstract: large language model] OR [Abstract: llm] OR [Abstract: agentic ai]] AND [[Abstract: technical debt] OR [Abstract: td] OR [Abstract: artificial intelligence technical debt] OR [Abstract: aitd] OR [Abstract: anti-patterns] OR [Abstract: self-admitted technical debt] OR [Abstract: satd] OR [Abstract: smell]] & Research articles only, and date                           & 330             \\ \hline
IEEE Xplore                & (("Abstract":"Artificial Intelligence" OR "Abstract":"AI" OR "Abstract":"Machine Learning" OR "Abstract":"ML" OR "Abstract":"Deep Learning" OR "Abstract":"DL" OR "Abstract":"Natural Language Processing" OR "Abstract":"NLP" OR "Abstract":"Generative Artificial Intelligence" OR "Abstract":"GenAI" OR "Abstract":"Large Language Model" OR "Abstract":"LLM" OR "Abstract":"Agentic AI") AND ("Abstract":"Technical Debt" OR "Abstract":"TD" OR "Abstract":"Artificial Intelligence Technical Debt" OR "Abstract":"AITD" OR "Abstract":"Anti-patterns" OR "Abstract":"Self-Admitted Technical Debt" OR "Abstract":"SATD" OR "Abstract":"Smell") )                                                                                                     & Journals and conferences only, and date             & 558             \\ 

\hline

Scopus             & ( "Artificial Intelligence" OR "AI" OR "Machine Learning" OR "ML" OR "Deep Learning" OR "DL" OR "Natural Language Processing" OR "NLP" OR "Generative Artificial Intelligence" OR "GenAI" OR "Large Language Model" OR "LLM" OR "Agentic AI" ) AND ( "Technical Debt" OR "TD" OR "Artificial Intelligence Technical Debt" OR "AITD" OR "Anti-patterns" OR "Self-Admitted Technical Debt" OR "SATD" OR "Smell" )
                                    & 
                                    Document Type: Article and Conference paper;
                                 Subject area: Computer Science and Engineering; Language: English only; and date                     & 203             \\ \hline

Springer                   & "Artificial Intelligence" OR "Machine Learning" OR "Deep Learning" AND "Technical Debt" OR "Anti-patterns"                                                                                                                                     & Research articles and conference papers only; Discipline: Computer Science; Subject area: Software Engineering, Software Testing; Language: English only; and date                                & 260              \\ \hline

\textbf{Total}&  & & \textbf{1,351}    \\ 
\bottomrule
\end{tabular}
\label{tab:electronicsearch}

\end{table*}

\subsection{Search Criteria}
\label{sec:searchcriteria}
Studies were included based on their direct relevance to AI-technical debt. The inclusion and exclusion criteria were as follows:

\subsubsection{Inclusion criteria:}

\begin{itemize} 
    \item \textbf{IC1.} The study must explicitly discuss or analyze technical debt within AI-enabled systems and provide direct insights relevant to one or more of the research questions formulated in this review.
   
    \item \textbf{IC2.} The publication must be a peer-reviewed journal article, conference paper, workshop, or symposium contribution.
    
    \item \textbf{IC3.} The publication must have been released between 2015 and November 2025. This period begins with the introduction of the concept of technical debt in AI and machine learning systems by Sculley et al. \cite{sculley2015hidden} in 2015.
   
    \item \textbf{IC4.} The paper must exceed three pages in length to ensure adequate depth of discussion and analysis. 
\end{itemize}


\subsubsection{Exclusion criteria:}

\begin{itemize} 
    \item \textbf{EC1.} Non-original research articles were excluded. This includes review papers, Ph.D. dissertations, secondary studies (e.g., SLRs and SMSs), posters, editorials, and magazine articles.

    \item \textbf{EC2.} Articles not written in English were excluded.

    \item \textbf{EC3.} Studies without accessible full-text versions were excluded.

    \item \textbf{EC4.} Studies that do not explicitly address technical debt in AI-based systems were excluded.

    \item \textbf{EC5.} Duplicate publications were excluded. These include instances where the same study appeared in multiple databases or was published in more than one venue (e.g., journal, conference, or workshop).

\end{itemize}

\subsection{Study Selection}
The study selection process was conducted in three phases: (1) automatic search restriction and duplicate removal, followed by (2) initial screening of titles and abstracts for relevance, and (2) a full-text review to confirm eligibility based on the predefined inclusion and exclusion criteria.

\begin{figure*}[t]
    \centering
    \includegraphics[width=1.0\textwidth]{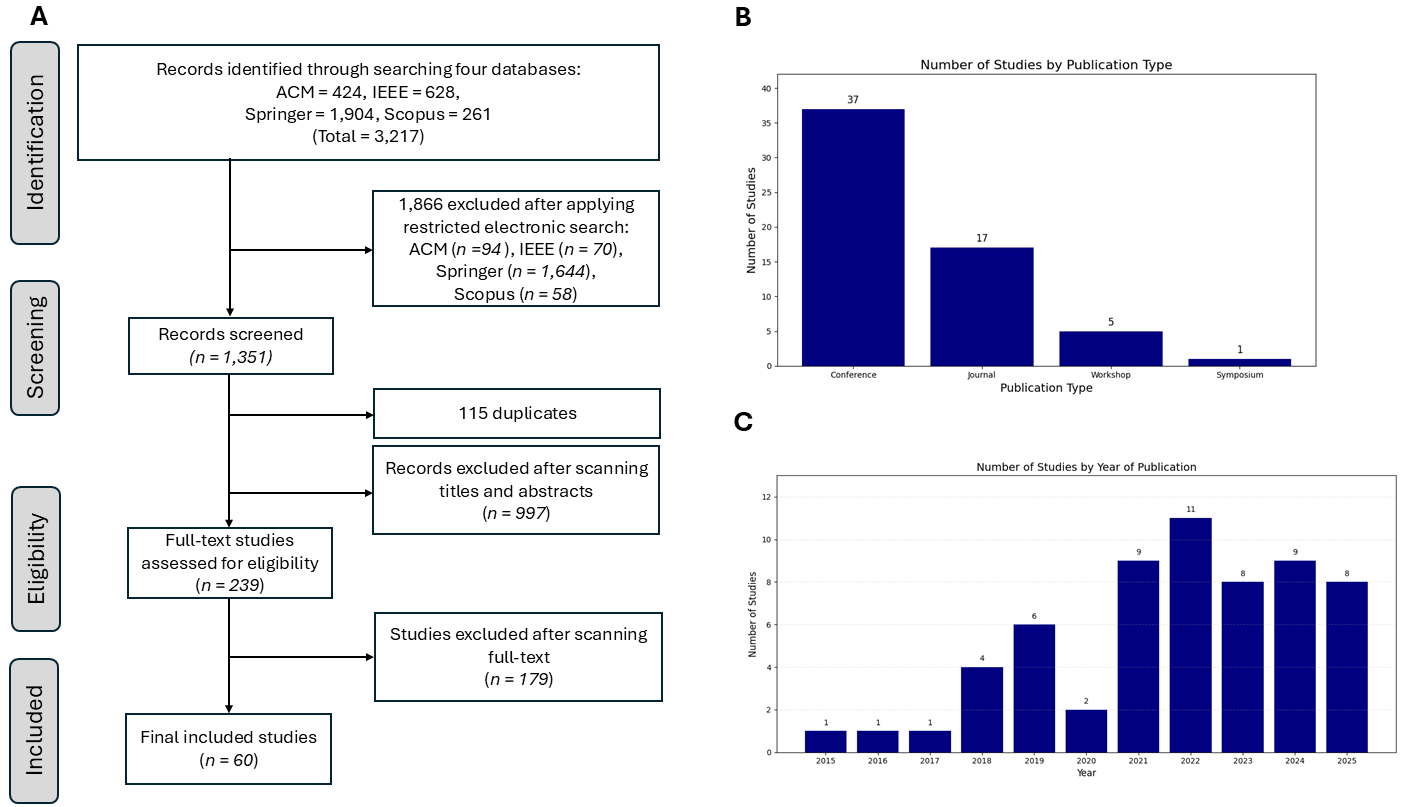}
    
    \caption{Methodology charts: (A) PRISMA chart of the included studies; (B) publication type of the
selected papers. (C) The distribution of studies over the years. 
}\label{fig:prismachart}

\end{figure*}
\subsubsection{Automatic search restriction and duplicate removal}

In this phase, we included papers published between January 2015 and October 2025, as Sculley et al. \cite{sculley2015hidden} were the first to introduce the concept of technical debt in AI and machine learning systems in the year 2015. In addition to the time restriction, we applied additional filters based on the options available in each digital library. For example, the Scopus Digital Library allows filtering by document type, subject area, language, and publication date. 
Table~\ref{tab:electronicsearch} outlines the filters applied and the number of studies retrieved from each digital database. In total, 1,351 studies were initially retrieved, and after removing 115 duplicates, 1,236 unique papers with distinct titles and abstracts were considered for further analysis. Further information and replication package is provided in the \textit{Search Results} folder of the \href{https://drive.google.com/drive/u/1/folders/1qABURByVGpORXmiKOkQTFdj3bcB3gg3O}{Supplementary
 Material} for additional details.

 \subsubsection{Screening based on title and abstract} 
 In this phase, studies were screened based on their titles, abstracts, and the availability of full texts. This resulted in the exclusion of 997 studies due to irrelevance or inaccessible full-texts, leaving 239 unique full-text studies for further evaluation.
 
 \subsubsection{Screening based on full-text}

The remaining studies were subjected to a full-text review, where the inclusion and exclusion criteria outlined in Sec.~\ref{sec:searchcriteria} were rigorously applied. Studies that only discussed technical debt without specifically addressing AI-based systems, or those that did not link technical debt to AI-enabled systems, were excluded. Following a comprehensive review, 60 studies were identified as highly relevant and selected as primary sources. The multidisciplinary nature of this review is evident in the diverse range of publication venues, as outlined in the references (c.f. ~\ref{tab:venue-distribution}. Figure~\ref{fig:prismachart}.B illustrates the types of publications included, while Figure~\ref{fig:prismachart}.C shows the distribution of the selected studies over the past decade.



\subsection{Grounded Theory Coding Procedure}

To systematically identify and categorize the 31 AI Technical Debts (AITDs), we adopted a grounded theory methodology structured around the three classical phases of Corbin and Strauss \cite{corbin1990grounded}: \textit{open coding}, \textit{axial coding}, and \textit{selective coding}. During the open coding phase, two co-authors independently analyzed the full-text content of the 60 primary studies, extracting and labeling recurring technical challenges indicative of debt-like behavior. This stage produced 101 initial codes, with each refined AITD supported by at least three underlying conceptual indicators. In the axial coding phase, these codes were iteratively compared, clustered, and refined to identify conceptual relationships and remove redundancies. Through this constant comparison process, the analysis converged into 31 distinct AITDs, each representing a higher-level abstraction of related debt patterns. In the selective coding phase, these 31 AITDs were organized into seven main root-cause-oriented categories. This classification underwent a structured assessment involving the co-authors. Agreement was achieved through multiple consensus-building sessions, with disagreements resolved via structured deliberation until full consensus was reached. The analysis achieved theoretical saturation after three iterations, at which point no new debt concepts emerged. This rigorous multi-stage approach ensured the reliability, transparency, and saturation of the resulting taxonomy, as summarized in Tables~\ref{tab:AITDranking} and further detailed in Section~\ref{sec:AITDs}. Additional details are provided in the \href{https://drive.google.com/drive/u/1/folders/1qABURByVGpORXmiKOkQTFdj3bcB3gg3O}{Supplementary Material}.  

\section{Overview of the identified AITDs and their taxonomy (RQ1)}

\label{sec:AITDs}
This section offers a comprehensive definition and discussion of the thirty-one (31) AI Technical Debts (AITDs) identified in the primary studies, each illustrated with relevant use cases. The AITDs are systematically categorized, and a detailed analysis is conducted, with each debt ordered by frequency of occurrence using grounded theory as the analytical approach \cite{corbin1990grounded} (see Tables~\ref{tab:AITDranking} and the \href{https://drive.google.com/drive/u/1/folders/1qABURByVGpORXmiKOkQTFdj3bcB3gg3O}{Supplementary Material}). Additionally, an in-depth examination of the overall impact of AITDs on the quality of AI-based systems is provided.


\subsection{Root-Cause-Oriented AITD Taxonomy}
To develop a unified and conceptually coherent understanding of the diverse forms of AI Technical Debt (AITD), the thirty-two identified debts were systematically classified according to their root causes—that is, the underlying technical, organizational, or ethical conditions that give rise to debt accumulation within AI-enabled systems. This root-cause-oriented perspective emphasizes the intrinsic source of each debt rather than its surface manifestation, providing a more holistic view of how AITDs emerge, propagate, and persist throughout the lifecycle of intelligent systems. 
The resulting taxonomy consolidates related technical debts into seven principal categories, each representing a distinct causal domain that influences the introduction and long-term impact of debt:
\begin{enumerate}
    \item Data \& Library–Related Debts 
    \item Model \& Code–Related Debts (Implementation Debts)
    \item Algorithm–Related Debts
    \item Design \& Architecture Debts (Structural Debts)
    \item Operational \& Lifecycle Debts 
    \item Documentation \& Communication Debts
    \item Testing \& Quality Assurance Debts
\end{enumerate}

\begin{figure*}[t]
    \centering
    \includegraphics[width=1.0\textwidth]{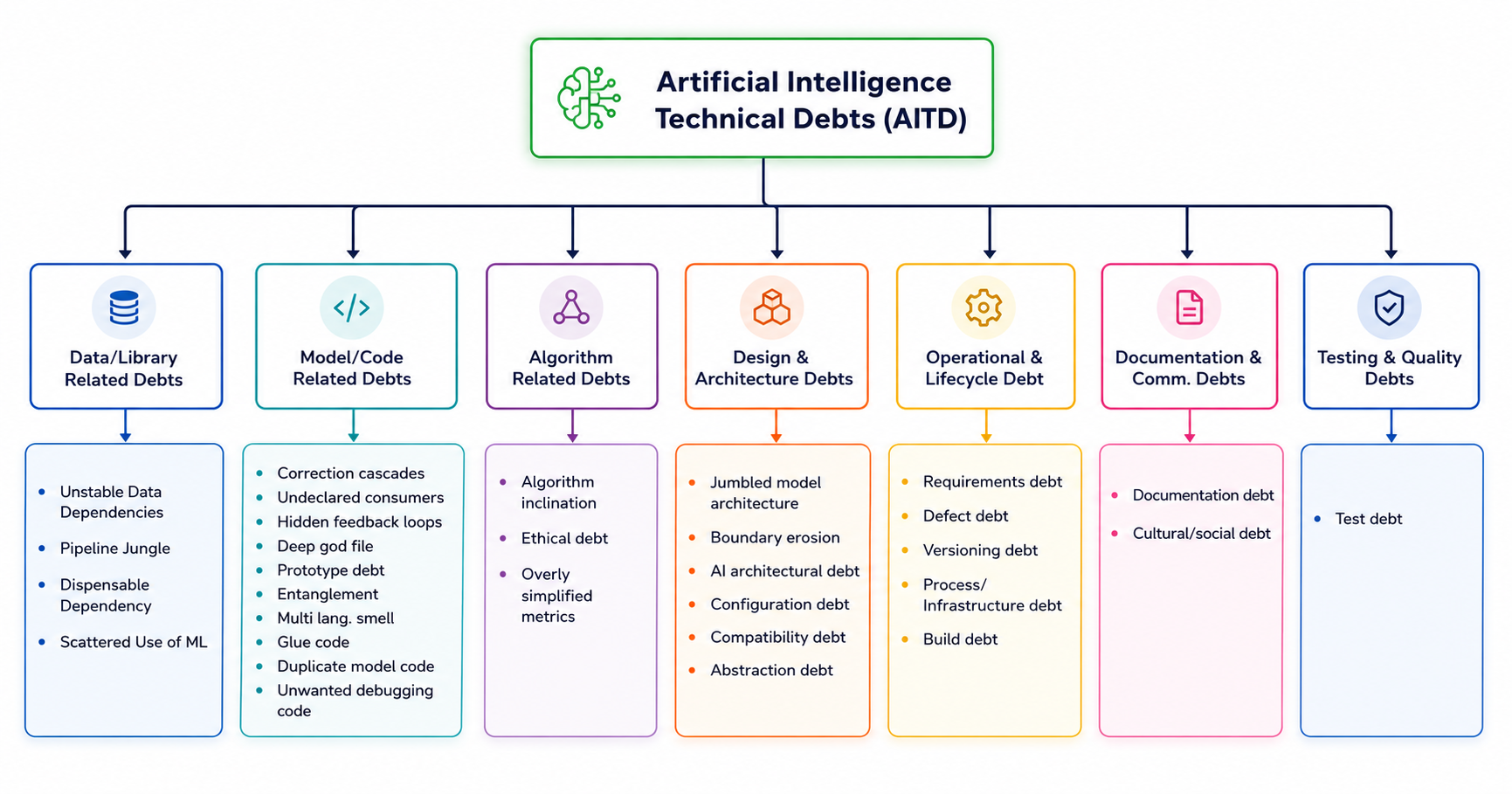}
    \caption{AITD taxonomy showing seven main categories—each divided into subcategories with corresponding debt types.}\label{fig:AITD_Taxonomy}

\end{figure*}

Each category groups together debt types that share a common origin, engineering failure mode, or decision-making trade-off, enabling more precise reasoning about how and why these debts arise. This structure also facilitates targeted mitigation by linking each debt to its underlying technical or managerial driver. The classification process followed an iterative grounded-theory procedure, in which the co-authors independently reviewed and validated the conceptual boundaries of each identified debt type. Disagreements were resolved through structured, consensus-based discussions, resulting in a stable and mutually agreed-upon taxonomy. The hierarchical relationships among the seven categories, their constituent subcategories, and the individual technical debt types are illustrated in Figure~\ref{fig:AITD_Taxonomy}. This taxonomy provides an integrated and extensible representation of the AITD landscape, synthesizing perspectives from software engineering, machine learning systems, responsible AI, and operational practices. The resulting categorizations are presented in the subsections that follow, each accompanied by detailed descriptions, example scenarios, and a justification for its assignment to the respective category.


\subsubsection{Data \& Library-Related Debts}

These debts arise from shortcomings in data governance, feature-pipeline stability, and dependency management—factors that form the foundation of every AI-based system. Because AI models rely heavily on the quality, consistency, and traceability of their training and inference inputs, even minor irregularities in data or library configurations can propagate significant degradation in system reliability, reproducibility, and long-term maintainability. These debts collectively reflect failures to manage the data ecosystem and software dependencies that support the learning pipeline.


\renewcommand{\labelenumi}{\roman{enumi}.}

\begin{enumerate}
    \item \textbf{Data Debt}
     This emerges when datasets used for training, validation, or inference suffer from quality issues such as missing values, biased distributions, poor metadata, untracked schema evolution, or inconsistent preprocessing procedures. In AI systems, where model performance is intrinsically tied to input quality, such deficiencies can severely distort learned representations and weaken generalization. Data Debt often accumulates gradually as datasets evolve without proper versioning, monitoring, or documentation. \textit{Impact:} It introduces instability, model drift, fairness violations, reduced robustness, and increased retraining overhead. Production models may fail unpredictably, degrade silently, or exhibit harmful behavior—such as biased decision-making—when new data deviates from historical patterns.
    \textit{Use Case:} In a real-time fraud-detection system, a financial institution updates its transaction schema by renaming fields and adding new categorical indicators. Because these changes are untracked, the deployed model interprets inputs incorrectly, resulting in a surge of false positives and customer complaints. \textit{Justification (Why):}	Data Debt is classified under Data \& Library-Related Debts because its root cause is inherently tied to data governance failures, unstable data pipelines, and unmanaged dataset evolution—issues central to the foundational data layer of AI systems.
    
     \par \textit{Source(s):} \cite{sculley2015hidden,  alahdab2019empirical, zhang2022code, tang2021empirical,  arpteg2018software, polyzotis2018data, foidl2022data, lenarduzzi2021software, foidl2019technical, breck2017ml, hutchinson2021towards, moreschini2024towards, nahar2022collaboration, shivashankar2022maintainability, roselli2019managing, cote2024quality, belani2019requirements, sas2023architectural, wang2023technical, khanvilkar2025automated, recupito2024unmasking,	cunha2020investigating,	de2025software,			akgul2025aligning,	ximenes2025investigating,			shome2022data, moldovan2024python}.

    \item \textbf{Pipeline Jungle:} This refers to preprocessing or feature-engineering pipelines that have grown organically—often through rapid experimentation—resulting in deeply nested, ad-hoc, and poorly documented chains of transformations. As components evolve, the pipeline becomes opaque, fragile, and difficult to reproduce or debug. AI systems are particularly vulnerable since the learning process depends critically on deterministic and transparent data transformations.~\textit{Impact:} It undermines reproducibility, slows development, and increases the risk of hidden data leaks or inconsistent training-inference paths. Even small upstream changes can cause cascading failures or silent performance degradation.~\textit{Use Case:} A computer-vision team incrementally adds new image augmentations for training. Over time, the pipeline incorporates dozens of conditional operations spread across multiple scripts. During deployment, a mismatch between training and inference transformations leads to lower detection accuracy in real-world images. 
    ~\textit{Justification (Why):}	Pipeline Jungle is grouped under this category because its core deficiencies relate to data pipeline management, transparency, and structural oversight—problems rooted in data engineering rather than model architecture or code structure.

    \par \textit{Source(s):} \cite{recupito2024technical, sculley2015hidden,  alahdab2019empirical,  washizaki2019studying, foidl2022data, moreschini2024towards, shivashankar2022maintainability, belani2019requirements, wang2023technical}.

    \item	\textbf{Scattered Use of ML Libraries (SML):}	
    This debt arises when an AI system relies on multiple machine learning libraries, framework versions, or overlapping APIs without clear standardization. Examples include mixing TensorFlow 1.x and 2.x, combining incompatible PyTorch modules, or using different versions of NumPy for different components. Such inconsistency introduces brittleness, dependency conflicts, and unpredictable behavior. \textit{Impact:} SML leads to nondeterministic outputs, runtime incompatibilities, serialization failures, deployment instability, and increased maintenance cost—especially when model components must interoperate in production. 
    \textit{Use Case:} A research team trains a model in PyTorch but uses TensorFlow scripts for feature extraction. Minor version differences between development and production environments cause silent discrepancies in tensor shapes, breaking the deployment pipeline.
    \textit{Justification (Why):}	SML is rooted entirely in dependency mismanagement and inconsistency within the library ecosystem—precisely the type of foundational support issue that characterizes Data \& Library-Related Debts.
    
    \par \textit{Source(s):}  \cite{recupito2024technical,  10628360}.

    \item \textbf{Dispensible Dependency:}	
    It refers to unused, obsolete, or redundant software packages and libraries that remain in the environment even though they are no longer required by the AI system. These dependencies increase the risk of security vulnerabilities, version conflicts, and inflated environments that are difficult to reproduce or audit. \textit{Impact:} This debt increases attack surfaces, slows build times, complicates dependency resolution, and introduces uncertainty into model reproducibility. It may also hinder deployment on resource-constrained platforms.
    \textit{Use Case:} A legacy feature-extraction toolkit remains installed in a production environment despite being replaced months earlier. A vulnerability scanner later identifies a critical security flaw in the unused library, forcing emergency remediation.
    \textit{Justification (Why):} Dispensable Dependency is placed under Data \& Library-Related Debts because its origin lies in library ecosystem mismanagement and poor dependency hygiene—both fundamental aspects of the AI data-processing environment.
    
    \par \textit{Source(s):}
     \cite{chaudhary2018review,  10628360, breck2017ml}.
     
\end{enumerate}

All the debts in this category share a common root cause: poor control, monitoring, and governance of the data pipelines and software dependencies that constitute the backbone of AI systems.
These debts hinder reproducibility, undermine trust in model behavior, and increase operational friction across the entire AI lifecycle.

    \subsubsection{Model/Code Related Debts}
These originate from deficiencies in implementation practices, modularization, interface management, and the structural integrity of the code that supports AI components. AI systems are particularly susceptible to these debts because model behavior is deeply intertwined with the structure of the underlying code artifacts—such as dataflow logic, model wrappers, feature utilities, and runtime inference pathways. Unlike traditional software components, AI models introduce additional complexity stemming from non-deterministic behaviors, hidden state, and tight coupling between learned parameters and computational code. The debts in this category collectively degrade maintainability, reliability, and integration stability across the model development and deployment lifecycle.

\begin{enumerate}
    \item  \textbf{Correction Cascades (CC):}	Occur when modifications made to fix a model issue inadvertently introduce new defects elsewhere in the system. In AI pipelines, where components depend heavily on learned representations or preprocessing logic, even minor adjustments—such as threshold tuning, feature normalization changes, or loss-function updates—may ripple through downstream modules. \textit{Impact:} This debt increases maintenance effort, leads to unpredictable regressions, and complicates quality assurance. Over time, the system becomes brittle, with every fix posing a risk of degrading other functionalities.  \textit{Use Case:} A development team adjusts the classification threshold of a churn-prediction model to reduce false negatives. The change unexpectedly affects marketing automation logic that relies on threshold-based customer segmentation, resulting in misaligned promotional campaigns. 
    \textit{Justification (Why):}	Correction Cascades are categorized as Model \& Code–Related Debts because their root cause lies in tight coupling, insufficient isolation, and ad-hoc code modifications that propagate unintended side effects across model components.
    \par  \textit{Source(s):} \cite{recupito2024technical, sculley2015hidden,   belani2019requirements, shukla2022challenges}.

    \item \textbf{Undeclared Consumers:}
    This arises when model outputs are consumed by downstream services or processes unknown to the development team. These hidden dependencies often emerge in fast-paced AI development environments where APIs evolve quickly and internal teams reuse AI components without formal registration or interface governance.~\textit{Impact:} They create operational risk, unauthorized data exposure, brittle integration pathways, and unpredictable behavior when model contracts or inference formats change. 
    \textit{Use Case:} A sentiment-analysis model originally designed for customer-support triage is silently reused by a marketing analytics team. When the model is updated with a new tokenization scheme, the downstream system silently begins classifying sentiment incorrectly, altering business reports.
    \textit{Justification (Why):}	The debt originates from undocumented and unmanaged model interfaces, making it fundamentally a model/code-level integration debt rather than a design or architectural issue.
    
    \par \textit{Source(s):} \cite{recupito2024technical, sculley2015hidden,  chaudhary2018review,  washizaki2019studying, belani2019requirements, wang2023technical}.

    
    \item	\textbf{Hidden Feedback Loops:}
    This occurs when model outputs unintentionally influence their future inputs, creating self-reinforcing behavior. Common examples include ranking systems, recommender pipelines, and content-moderation models, where predictions directly affect the distribution of new training data. \textit{Impact:} These loops amplify bias, degrade robustness, and may lead to degenerative behavior (e.g., popularity bias or echo-chamber effects). Additionally, they complicate retraining, making it difficult to disentangle learned behavior from past model outputs.  \textit{Use Case:} A recommender system suggests articles to users based on previous engagement. Over time, the system continues presenting similar content, narrowing topic diversity and reducing exploratory signals in training data.
    \textit{Justification (Why):}	Because feedback loops manifest through runtime coupling between model outputs and model inputs, they are inherently rooted in model-code interactions rather than algorithm design or pipeline configuration.
    \par \textit{Source(s)}  \cite{sculley2015hidden,  arpteg2018software, menshawy2024navigating, moreschini2024towards, shivashankar2022maintainability, roselli2019managing, khritankov2021hidden, wang2023technical, shukla2022challenges}.

    \item \textbf{Deep God File (DG):}	
    This denotes excessively large or monolithic files—often originating from early experimentation—that contain intertwined logic for data preprocessing, model training, evaluation, and deployment~\cite{recupito2024technical,gesi2022code}. Such files hinder modularization, reuse, debugging, and testing. \textit{Impact:} They reduce maintainability, complicate onboarding, hinder automated testing, and increase the likelihood of introducing defects during refactoring.  \textit{Use Case:} A Jupyter notebook used for initial research is promoted directly into production. It contains data cleaning, feature engineering, model training, and inference logic in a single script, making any modification risky.
    Another example is an AI application where data preprocessing, model training, and testing logic are all implemented within the same file, leading to a cluttered codebase. \textit{Justification (Why):}	DG arises from poor code organization, a foundational model-implementation issue, and is therefore grouped under Model \& Code–Related Debts.~\footnote{It is important to note that Deep God File extends the traditional God Class concept to AI contexts, where large monolithic scripts or notebooks combine data preprocessing, model design, training, and evaluation in a single file. Unlike the classic design-level anti-pattern, this debt arises from experimental ML workflows and the lack of modularization typical in AI development. Furthermore, the key difference is the motivation. A classic God Class often results from a lack of architectural discipline. In ML, the "Deep God File" often arises naturally from the exploratory nature of the work, where experimentation takes priority over software engineering best practices. The focus is on getting a working model, not on creating reusable and modular components.}
    \par  \textit{Source(s):}  \cite{recupito2024technical, cunha2020investigating}.

    \item \textbf{Prototype Debt/Dead Experimental Code Paths:}	
    Prototype Debt accumulates when experimental prototypes or proof-of-concept scripts transition into production without refinement, optimization, or appropriate architectural restructuring. While common in AI research settings, such code is fragile and poorly suited for operational use.
    \textit{Impact:} It leads to runtime inconsistencies, missing error handling, reduced scalability, and challenges in debugging or extending the model.
    \textit{Use Case:} A prototype hyperparameter-tuning script is used as the production training pipeline. When data volume increases, the script fails due to hard-coded assumptions and lack of batching logic.
    \textit{Justification (Why):}	Its origin lies in moving research artifacts directly into production, making it fundamentally a model/code-related implementation debt.
    \par \textit{Source(s):} \cite{sculley2015hidden, tang2021empirical,   washizaki2019studying, li2023debtviz, li2023automatic, obrien202223, perez2021technical,  moreschini2024towards, belani2019requirements, wang2023technical, li2022identifying}.

    \item \textbf{Entanglement:}	
    It arises when multiple AI components or feature pathways become tightly interwoven, such that modifying one element requires changing others. This reduces modularity and makes isolated improvements or debugging nearly impossible. \textit{Impact:} It creates high modification cost, restricts model evolution, and slows experimentation cycles. Errors propagate more easily, and testing becomes more complex.  \textit{Use Case:} A set of models share a common feature vector generated through a single, monolithic script. Updating any part of the vector requires revisiting all dependent models.
    \textit{Justification (Why):}	Entanglement reflects poor separation of concerns within model code, justifying its placement within this category.
    
    \par \textit{Source(s):}  \cite{sculley2015hidden,  menshawy2024navigating, belani2019requirements, wang2023technical}.

    \item \textbf{Multiple Language Smells (MLS):}
    MLS occurs when a system incorporates multiple programming languages or framework ecosystems without clear boundaries or tooling support. Though sometimes necessary for performance (e.g., C++ extensions), unmanaged heterogeneity creates build-time and runtime fragility. \textit{Impact:} It increases integration complexity, complicates containerization, introduces serialization challenges, and hinders collaborative development. It also adds complexity and requires diverse expertise for maintenance. \textit{Use Case:} A deep-learning team mixes PyTorch (Python), CUDA kernels (C++/CUDA), and custom Java-based serving infrastructure. Differences in serialization formats cause inference failures. \textit{Justification (Why):}	MLS is rooted in cross-language integration problems that directly affect model code and runtime interactions.
    
    \textit{Source(s):}  \cite{recupito2024technical, sculley2015hidden,  alahdab2019empirical, tang2021empirical, washizaki2019studying, moreschini2024towards, wang2023technical}.

    \item	\textbf{Duplicate Model Code:}	
    This debt occurs when identical or near-identical model functions, utilities, or architectural components are replicated across codebases or modules. It often results from parallel experimentation or copy-paste development practices. \textit{Impact:} Leads to inconsistent behavior, duplicate bugs, and increased maintenance effort. Updating a single model component requires tracking and modifying multiple copies.
    \textit{Use Case:} Two research teams independently copy the same LSTM layer implementation and introduce different small changes, resulting in inconsistent behavior across models. Redundant model code across various analytics models within an organization, requiring changes to be made in multiple locations.
    \textit{Justification (Why):}	The core issue—replication of model logic across code—is a model/code-level maintainability debt.
    \par \textit{Source(s):}  \cite{tang2021empirical,   albuquerque2022comprehending, van2021prevalence, li2023debtviz, li2023automatic, obrien202223, perez2021technical,  jebnoun2022clones, khanvilkar2025automated, li2022identifying}.

    \item	\textbf{Unwanted Debugging Code (UDC):}	
    UDC refers to temporary print statements, debug logs, or ad-hoc instrumentation left in production code. In AI systems, where inputs often include sensitive data, such artifacts may inadvertently expose information or distort performance measurements. \textit{Impact:} It degrades performance, increases log noise, risks leaking confidential information, and complicates monitoring pipelines. \textit{Use Case:} Residual debug logs unintentionally record user-provided medical symptoms in server logs, violating privacy obligations. Temporary code snippets used to debug a recommendation engine that, if left in production, could expose sensitive data paths. 
    \textit{Justification (Why):}	This debt arises from failure to clean up experimental instrumentation, clearly falling within model and code practices.
    
    \textit{Source(s):} \cite{recupito2024technical, li2023debtviz, li2023automatic, obrien202223, perez2021technical}. 
    
    \item \textbf{Glue Code (GC):}
    GC refers to fragile, hand-crafted code used to connect components that were not designed to interoperate. In AI systems, this often appears in bridging incompatible frameworks, data structures, or inference formats. This debt arises when large amounts of code are written specifically to integrate libraries or components that were not initially designed to work together. Glue code often lacks proper organization, which complicates future maintenance. \textit{Impact:} It increases brittleness, reduces portability, complicates debugging, and introduces hidden dependencies that hinder scaling and refactoring. \textit{Use Case:} Developers write custom JSON converters to translate between two ML microservices instead of using a shared schema or API contract. \textit{Justification (Why):}	GC is inherently a code-level integration shortcut, belonging naturally to the Model \& Code–Related Debts category.

    \par \textit{Source(s):}  \cite{recupito2024technical, sculley2015hidden,  alahdab2019empirical,  10628360, tang2021empirical,  washizaki2019studying, van2021prevalence, arpteg2018software, li2023debtviz, li2023automatic, lenarduzzi2021software, obrien202223, perez2021technical,  moreschini2024towards, shivashankar2022maintainability, belani2019requirements, bavota2016large, wang2023technical}. 

\end{enumerate}

All debts in this category arise from implementation-level deficiencies—including poor modularization, ad-hoc experimentation, fragile integrations, runtime coupling, and unmanaged code evolution. While these debts are not exclusive to AI-based systems and may also occur in traditional software, they are particularly prevalent and impactful in AI-enabled systems due to their tight coupling between data, models, and code. As a result, these debts weaken maintainability, increase operational risk, and hinder the scalability of AI-based systems

\subsubsection{Algorithm-Related Debts}
These arise from deficiencies in the conceptual design, ethical alignment, optimization objectives, or evaluation strategies of AI models. Unlike Model \& Code–Related Debts, which stem from implementation and structural issues, Algorithm-Related Debts originate in the fundamental reasoning logic encoded in algorithmic choices, objective functions, and evaluation protocols. These debts influence how AI systems learn, generalize, and behave under real-world conditions, often with long-term consequences for fairness, explainability, robustness, and trustworthiness.

Because algorithmic decisions shape the system’s predictive behavior and risk profile, debts in this category propagate into downstream models, user interactions, and sociotechnical ecosystems. The following debts fall under this category.

\begin{enumerate}
    \item \textbf{Algorithm Inclination Debt (Human Bias Debt):}
    Occurs when bias is inadvertently embedded within the model design, objective functions, or feature selection strategies. Unlike Data Debt, which concerns bias in the training corpus, this debt reflects bias introduced by model developers—through choices such as skewed class weights, inappropriate regularization terms, or the omission of fairness-aware modeling techniques. These decisions influence who benefits or is disadvantaged by the model’s outputs. It also refers to a bias introduced by over-reliance on familiar algorithms, even when they are suboptimal for the problem. 
    \textit{Impact:} It leads to discriminatory outcomes, lower fairness, undermined user trust, and legal or ethical liabilities. Such bias often snowballs in production systems, particularly in high-stakes applications such as credit scoring, hiring, policing, and healthcare. \textit{Use Case:} A loan-approval classifier is optimized solely for overall accuracy on an imbalanced dataset. Because the cost of false negatives is not penalized for minority applicants, the model disproportionately rejects their applications—a consequence stemming from the unfair optimization objective, not the raw data itself. A recommendation system using outdated algorithms that fail to adapt to user behavior, causing dissatisfaction.
    \textit{Justification (Why):}	This debt is categorized under Algorithm-Related Debts because the root cause is the algorithmic formulation—not code quality or data defects. It emerges from modeling choices, loss-function configuration, class balancing strategies, and inductive biases directly encoded into the learning process.

     \par \textit{Source(s):} \cite{chaudhary2018review, liu2020using,  chen2023toward, arpteg2018software, liu2021exploratory, simon2023algorithm, wang2023technical, nikanjam2021design,	li2022identifying,	de2025software}.

    \item \textbf{Ethical Debt:} Ethical Debt accumulates when the development process neglects moral, societal, or human-centric considerations—such as user consent, privacy preservation, transparency, or fairness. Ethical Debt differs from Algorithm Inclination Debt in that it encompasses broader governance-level harms beyond algorithmic optimization, including neglect of ethical guidelines, insufficient documentation of value trade-offs, and absence of bias-impact assessments. \textit{Impact:} It increases the risk of harmful or discriminatory system behavior, erodes public trust, exposes organizations to regulatory penalties, and can lead to reputational damage. Ethical Debt often becomes visible only after deployment when real-world impacts emerge. 
    \textit{Use Case:} A facial-recognition model is deployed in a public environment without transparency documentation or ethical risk analysis. Later, it is found to perform poorly on underrepresented demographic groups, leading to public backlash and regulatory scrutiny. A healthcare AI system that exhibits racial bias due to lack of diverse data, resulting in unequal treatment recommendations.
    \textit{Justification (Why):}	Ethical Debt belongs to this category because its origin is algorithmic governance and moral oversight. It reflects decisions made at the conceptual design and alignment stage, rather than issues of code structure or operational infrastructure.
    
    \par \textit{Source(s):} \cite{menshawy2024navigating, chang2022understanding, roselli2019managing, petrozzino2021pays}.

    	
    \item \textbf{Overly Simplified Metrics:}
    refers to the use of oversimplified, misaligned, or insufficient evaluation metrics that fail to capture the real-world performance, safety, or societal impacts of AI models. Focusing solely on metrics such as accuracy, precision, or F1-score may obscure system vulnerabilities—such as sensitivity to out-of-distribution data, robustness under adversarial conditions, or fairness disparities. It can also refer to the use of metrics that are generic or misaligned with specific problem requirements. 
    \textit{Impact:} This debt leads to misleading validation results, a false sense of model reliability, and unanticipated failures when deployed in dynamic environments. Models optimized for incomplete or Overly Simplified Metrics may perform well in controlled experiments but fail dramatically in real-world contexts.
    \textit{Use Case:} A medical-diagnosis model is evaluated exclusively using overall accuracy. Rare but critical conditions (minority classes) are under-detected because the metric does not capture the cost of false negatives. Deployment then results in missed diagnoses and potential patient harm.
    \textit{Justification (Why):}	The root cause lies in evaluation-design inadequacy, an algorithmic-level failure to define meaningful and context-aware success metrics. Therefore, Overly Simplified Metrics Debt is rightfully placed under Algorithm-Related Debts rather than testing or data categories.
    
    \par \textit{Source(s):}~\cite{chaudhary2018review, khanvilkar2025automated, ximenes2025investigating}.
    
\end{enumerate}

Algorithm-Related Debts arise from conceptual design choices that govern how AI systems learn, reason, and behave.
These debts are epistemic rather than structural—stemming from flawed optimization objectives, inadequate evaluation strategies, or ethical misalignment rather than software engineering deficiencies.
If left unaddressed, they compromise fairness, safety, trustworthiness, and long-term societal impact, making them critical to responsible AI development and governance.

 \subsubsection{Design/Architecture Debts}
   These debts arise from deficiencies in the structural organization, interface boundaries, and configuration management of AI-based systems. These debts reflect weaknesses in how system components are designed, how architectural layers interact, and how configurations are specified, tracked, and maintained. Unlike Model \& Code–Related Debts, which emerge from implementation artifacts, these debts originate earlier in the system’s conceptual and structural design. They compromise long-term extensibility, maintainability, and reproducibility—attributes especially critical in AI systems, where models depend on complex pipelines, configurable hyperparameters, and evolving architectural patterns. The debts in this category collectively undermine internal cohesion, architectural integrity, and configuration reliability.

\begin{enumerate}
    \item \textbf{Jumbled Model Architecture (JMA):}    JMA arises when AI models—particularly deep learning architectures—lack coherent structural organization. Examples include inconsistent layering schemes, irregular activation patterns, or ad-hoc combinations of architectural modules. Such models often originate from rapid experimentation or incremental patching during prototyping. \textit{Impact:} It reduces interpretability, increases training instability, complicates debugging, and diminishes the model’s ability to generalize. Since architecture heavily influences representational learning, fragmented structures can introduce vanishing gradients, bottlenecks, or redundant paths. \textit{Use Case:} A computer vision team alternates between different convolutional block types across layers due to experimentation. The resulting architecture exhibits sporadic gradient explosions during training, making optimization highly unstable; An AI system developed with a mix of different neural network architectures without clear separation between components, leading to difficulties in maintenance.  \textit{Justification (Why):}	JMA belongs to this category because it reflects a design-phase structural deficiency rather than implementation or code-level issues. The root cause is architectural inconsistency originating from poor design discipline.
     \par \textit{Source(s):}  \cite{recupito2024technical,   albuquerque2022comprehending, li2023automatic, foidl2019technical, perez2021technical,  sas2023architectural, li2022identifying,		perez2019proposed, cunha2020investigating}.

    \item	\textbf{Boundary Erosion:}	
    This occurs when architectural boundaries—such as module interfaces, layer abstractions, or service contracts—are violated. In AI systems, this may manifest when data preprocessing modules directly interact with model internals, or when models bypass APIs to access raw system components. Over time, the boundaries between components in an AI system may erode due to complex interdependencies, weakening the modularity of the system and increasing maintenance difficulty. \textit{Impact:} It increases coupling, reduces modularity, and exposes internal components to misuse. Eroded boundaries introduce security risks, complicate dependency management, and restrict the ability to update or replace components independently.
     \textit{Use Case:} In a multi-model AI platform, as models become interdependent, changes in one affect others, eroding the clear separation between components; A feature-engineering script directly injects intermediate tensors into a model’s internal layers for debugging. Later changes to the model break the pipeline, as external components were never meant to interact with these layers; 
     \textit{Justification (Why):}	Boundary Erosion’s root cause is architectural-layer violation, making it conceptually distinct from code smells or operational issues and thus appropriately classified here.
    
     \par \textit{Source(s):}  \cite{sculley2015hidden, chaudhary2018review, tang2021empirical}.

 \item \textbf{AI Architectural Debt}
    Refers to architectural or structural decisions that trade long-term quality for short-term speed, simplicity, or expedience. In AI systems, this often appears as simplistic model wrappers, hard-coded hyperparameters, or missing abstraction layers that prevent future scaling or generalization.  \textit{Impact:} It limits reusability, slows feature development, and introduces fragility into evolving codebases. Poor design also impedes systematic monitoring, experimentation, and integration—key aspects of continuous AI deployment. 
    \textit{Use Case:} A research team wraps a model inside a single script without separating preprocessing, inference, and monitoring logic. As production requirements evolve, each modification forces large-scale refactoring.
    \textit{Justification (Why):}	Design Debt is inherently a structural design failure, justifying its assignment to this category rather than implementation or operational domains.
    
    \par \textit{Source(s):}  \cite{liu2020using,   albuquerque2022comprehending, li2023debtviz, li2023automatic, perez2021technical,  jebnoun2022clones, bavota2016large, yan2018automating, liu2021exploratory, nikanjam2021design,	li2022identifying, de2025software, sutoyo2024satdaug}.

    \item	\textbf{Configuration Debt:}
    This debt arises when hyperparameters, environment variables, or system settings are poorly documented, inconsistently defined, or not version-controlled. In AI systems—where configuration governs behavior as much as code—such inconsistencies cause unpredictability and hinder reproducibility.
    Over time, configuration settings in an AI system become overly complex or difficult to manage, increasing the risk of errors and misconfiguration. \textit{Impact:} It produces experimental inconsistencies, failed model replications, training divergence, or mismatched behavior between training and inference. Configuration Debt frequently leads to irreproducible results, a central challenge in machine learning research. 
    \textit{Use Case:} A deep learning model fails to reproduce validation accuracy because random seeds, batch sizes, and learning schedules were not logged during earlier experiments; An AI system with numerous hyperparameter settings, making it difficult for operators to manage configurations accurately across deployment environments.
    \textit{Justification (Why):}	This debt reflects configurational mismanagement, a structural rather than implementation issue, and is therefore appropriately grouped under this category.
    
    \par \textit{Source(s):}  \cite{sculley2015hidden,  alahdab2019empirical, zhang2022code,  tang2021empirical,   chen2023toward, foidl2019technical, breck2017ml,  jebnoun2022clones, belani2019requirements, khanvilkar2025automated, nikanjam2021design,	ximenes2025investigating}.

    \item	\textbf{Compatibility Debt:} 
    This emerges when AI components rely on outdated, incompatible, or conflicting frameworks, model formats, or serialization schemes. This often occurs when transitioning between versions of TensorFlow, PyTorch, ONNX, or other tooling ecosystems. \textit{Impact:} It hinders system evolution, complicates deployment across environments, increases integration overhead, and may prevent models from being exported or reused across platforms; Incompatibilities among system components lead to inefficient workarounds, raising system complexity and maintenance challenges.  \textit{Use Case:} A model trained in an older TensorFlow version cannot be exported to ONNX without extensive patching, delaying migration to a faster inference engine. 
    \textit{Justification (Why):}	Its origin lies in architectural and design decisions regarding toolchain selection, framework versions, and dependency constraints—not in code-level implementations.
    
    \par \textit{Source(s):}  \cite{liu2020using,  chen2023toward, lenarduzzi2021software, liu2021exploratory}.

    \item	\textbf{Abstraction Debt:}	
    Occurs when interfaces are overly generic, insufficiently defined, or inconsistently structured. Poor abstraction design leads to convoluted inheritance hierarchies, unclear responsibilities, and tightly coupled components. \textit{Impact:} It makes model components harder to extend, test, and reuse. Abstraction Debt often introduces hidden coupling and forces developers to modify low-level details even for high-level changes.  \textit{Use Case:} A base “Model” class defines ambiguous abstract methods, causing inconsistent implementations across subclasses. Adding a new training mode requires modifying nearly all subclassed models.
    \textit{Justification (Why):}	This debt’s root cause is inadequate abstraction in design, making it a prototypical structural debt falling under Design, Architecture \& Configuration.
    
    \par \textit{Source(s):}  \cite{sculley2015hidden,  10628360, tang2021empirical,  washizaki2019studying}.

  \end{enumerate}

  Design, Architecture \& Configuration Debts reflect structural and conceptual weaknesses introduced during the design and configuration stages of AI system development. These debts degrade modularity, hinder reproducibility, and amplify technical friction across the lifecycle. They differ from implementation-level debts because their origins lie in poor architectural foresight, inconsistent configuration practices, and inadequate abstraction mechanisms rather than the code itself.

\subsubsection{Operational \& Lifecycle Debts}
These emerge from weaknesses in the processes, infrastructure, and versioning practices that support the continuous development, deployment, and maintenance of AI systems. Unlike debts rooted in design or implementation, these reflect failures in operational discipline and lifecycle governance, including inadequate CI/CD automation, untracked evolution of model artifacts, and ambiguous requirements that lead to rework.
Because AI systems require tight synchronization across model, data, and environment versions, even routine operational lapses can manifest as long-term technical liabilities, particularly in production environments.

\renewcommand{\labelenumi}{\roman{enumi}.}
\begin{enumerate}

    \item \textbf{	Requirement Debt:}	
    This accumulates when functional or quality attribute specifications are ambiguous, incomplete, or inconsistently communicated~\cite{ernst2021technical}. AI systems are particularly vulnerable because their performance depends on well-defined metrics, target thresholds, and operational constraints. Poor requirement alignment often leads to misinterpretation of expected outcomes or inappropriate model objectives. \textit{Impact:} This debt results in rework, misaligned performance goals, wasted experimentation cycles, and delayed delivery. It also increases the risk of deploying models that fail to meet stakeholder expectations or regulatory requirements.  \textit{Use Case:} A healthcare model is trained to maximize accuracy, but stakeholders later clarify that minimizing false negatives is critical for patient safety. The model must be retrained, causing significant delays; An AI-based e-commerce system where user experience requirements were insufficiently defined, leading to an incomplete recommendation feature.
    \textit{Justification (Why):}	Because the root cause is inadequate requirements engineering within the lifecycle, Requirements Debt is classified as an Operational \& Lifecycle Debt rather than an algorithmic or design-level issue.
    
    \par \textit{Source(s):}  \cite{liu2020using,   albuquerque2022comprehending, li2023debtviz, li2023automatic, obrien202223, perez2021technical,  moreschini2024towards, nahar2022collaboration, belani2019requirements, bavota2016large, liu2021exploratory, li2022identifying,						de2025software,	sutoyo2024satdaug}.

    \item	\textbf{	Defect Debt:}	
    This arises when known bugs or system flaws are deferred for future resolution, often due to time pressure or resource limitations. These defects may include incorrect preprocessing logic, mislabeled data, unstable model behaviors, or unhandled exceptions. \textit{Impact:} It reduces system reliability, increases the frequency of runtime failures, and demands greater developer time to diagnose and correct issues later in the lifecycle.  \textit{Use Case:} A feature extractor intermittently outputs NaN values during training. The issue is noted but deprioritized. Over time, the intermittency increases, leading to complete training failure during a critical deployment phase.
    \textit{Justification (Why):}	The root cause is post-development defect tolerance, an operational choice that directly affects lifecycle maintenance.
    \par \textit{Source(s):}  \cite{liu2020using,   albuquerque2022comprehending, li2023automatic, obrien202223, perez2021technical,  bavota2016large, yan2018automating, liu2021exploratory, li2022identifying}.

    \item 	\textbf{	Versioning Debt:}
    This occurs when model artifacts, datasets, code modules, or environment configurations are not appropriately version-controlled. AI systems heavily depend on deterministic versioning to ensure reproducibility, auditability, and rollback capability. Poor version control practices lead to challenges in tracking changes, maintaining compatibility, and reproducing past results. \textit{Impact:} It obstructs experiment tracking, prevents reproducibility of results, hinders debugging, and complicates compliance audits. Models may behave unpredictably because different environment or dataset versions inadvertently enter the pipeline. \textit{Use Case:} A new model underperforms during A/B testing. Investigation reveals that the training dataset version differs from the dataset used in validation, but no version metadata was tracked.
    \textit{Justification (Why):}	Because the root cause lies in lifecycle artifact management, this debt belongs to the Operational \& Lifecycle category.
    
    \par \textit{Source(s):} \cite{washizaki2019studying, albuquerque2022comprehending, perez2021technical,  cote2024quality, shukla2022challenges}.

    \item	\textbf{	Process/Infrastructure Debt:}	
    This debt arises from outdated, inconsistent, or manual operational workflows, including CI/CD pipelines, orchestration processes, hardware provisioning, or monitoring infrastructure. AI lifecycles often rely on complex tooling (e.g., model registries, deployment orchestrators, monitoring dashboards), making process deficiencies especially harmful. \textit{Impact:} It slows deployment cycles, increases operational risk, reduces system scalability, and introduces failure points due to manual interventions.   
    \textit{Use Case:} A model deployment pipeline requires manual execution of several scripts. When an urgent patch is needed, a step is executed incorrectly, causing service downtime; An AI-based recommendation system with an outdated build pipeline lacking automation.
    \textit{Justification (Why):}	Since the underlying issue is lack of automation and process maturity, it is appropriately placed under Operational \& Lifecycle Debts.
    
    \textit{Source(s):} \cite{lenarduzzi2021software, perez2021technical, nahar2022collaboration}.

    \item	\textbf{ Build Debt:} 
    This accumulates when build scripts, Dockerfiles, environment specifications, or compilation pipelines become fragmented, non-standardized, or poorly maintained. This often results from ad-hoc experimentation during early development stages. \textit{Impact:} It causes inconsistent builds, extends debugging time, creates deployment fragility, and increases onboarding complexity. Build Debt is a common source of environment drift between development, testing, and production; Results in unreliable deployment processes and system instability.  \textit{Use Case:} A production inference container fails because a dependency version differs from development. The mismatch originates from an outdated Dockerfile with ambiguous version specifications. 
    \textit{Justification (Why):}	The root cause is build environment instability, making this a lifecycle and operational concern rather than a code-level one.
    
    \par \textit{Source(s):}  \cite{li2023automatic, perez2021technical, li2022identifying}.

\end{enumerate}
These debts all stem from operational misalignment, inadequate automation, and poor lifecycle management practices.
If unaddressed, they significantly undermine the reliability, reproducibility, and agility of AI development and deployment.

\subsubsection{Documentation and Communication Debt}
These debts originate from insufficient documentation, weak communication channels, and inadequate knowledge transfer practices across teams. AI systems rely on complex interactions between data engineers, ML researchers, software developers, and domain experts.
When documentation is incomplete or communication is fragmented, system behavior becomes opaque, maintenance becomes slower, and organizational risk increases. These debts reflect socio-technical deficiencies rather than purely technical ones.

\begin{enumerate}
    \item \textbf{Documentation Debt:}
   This arises when key artifacts—such as model behavior, preprocessing steps, data assumptions, evaluation methodologies, or architectural decisions—are insufficiently documented, outdated, or missing entirely~\cite{ernst2021technical}. \textit{Impact:} It slows onboarding, increases maintenance effort, complicates audits, and leads to miscommunication during handovers. For AI systems, lack of documentation also undermines explainability, transparency, and regulatory compliance; Poor documentation creates obstacles for developers to maintain and extend AI systems effectively, resulting in reduced system reliability.  \textit{Use Case:} A deployment engineer attempts to reproduce a training run but cannot locate information on normalization strategies used in the original experiment, resulting in inconsistent model behavior; Sparse documentation in a complex AI model for disease diagnosis, which complicates onboarding new developers and troubleshooting.
    \textit{Justification (Why):}	This debt’s root cause is deficiency in documentation and knowledge management, making it a socio-technical rather than algorithmic or architectural debt.

    \textit{Source(s):}  \cite{liu2020using,   albuquerque2022comprehending, li2023debtviz, li2023automatic, hutchinson2021towards, obrien202223, perez2021technical,  nahar2022collaboration, chang2022understanding, cote2024quality, bavota2016large, liu2021exploratory, wang2023technical, li2022identifying,								sutoyo2024satdaug}.

    \item	\textbf{Cultural/People/Social Debt:}	 
    This debt arises when teams have misaligned communication practices, insufficient collaboration, or conflicting expectations. In AI settings, teams often operate in silos—data engineers manage schemas, ML researchers iterate on models, and software engineers maintain deployments—causing integration friction. It also refers to the differences in culture, values, or priorities between teams, such as research and engineering, leading to misalignments~\cite{ernst2021technical}. \textit{Impact:} It leads to coordination failures, inconsistent workflows, misconfigured models, duplicated effort, and erroneous assumptions about system behavior. It also increases risk in safety-critical or regulated contexts.  \textit{Use Case:} An ML team updates a model’s input schema but does not communicate the change to the data engineering team. As a result, the data pipeline continues to produce the old schema, causing runtime failures and misaligned features.
    \textit{Justification (Why):}	The root cause is organizational misalignment, not technical implementation defects. Thus, it aligns naturally with Documentation \& Communication Debts.
    
    \par \textit{Source(s):} \cite{arpteg2018software, lenarduzzi2021software, perez2021technical, menshawy2024navigating, moreschini2024towards, nahar2022collaboration,mailach2023socio, annunziata2025uncovering, de2025software,	akman2025people}. 

\end{enumerate}

Documentation \& Communication Debts highlight the human and organizational dimension of technical debt.
These debts impair transparency, collaboration, and maintainability, especially in interdisciplinary AI teams where knowledge silos are common.

    \subsubsection{Testing and Quality Assurance Debt}
    These debts emerge from insufficient verification, limited test coverage, and overlooked quality-assurance practices. AI systems require testing not only of code correctness but also of data pipelines, model behavior under distribution shifts, fairness metrics, and robustness to adversarial conditions. Inadequate testing accelerates technical debt accumulation, reduces trustworthiness, and increases the likelihood of harmful failures in production.

\begin{enumerate}
    \item  \textbf{Test Debt:}	
     Test Debt accumulates when testing suites are incomplete, outdated, or insufficient to assess AI system behavior. This includes lack of unit tests for preprocessing logic, missing integration tests, absence of fairness or robustness evaluations, and insufficient testing of edge-case inputs~\cite{ernst2021technical}. \textit{Impact:} It leads to undetected regressions, silent model degradation, vulnerability to distribution shifts, and reduced confidence in deployments. Test Debt often manifests only when unexpected failures occur in real-world usage. \textit{Use Case:} A sentiment-analysis model fails to handle emerging slang and social-media phrases because its test suite only covers standard dictionary terms. A predictive analytics model deployed without comprehensive testing, resulting in undetected errors in real-time use.
    \textit{Justification (Why):}	The debt originates from insufficient QA practices, making it distinct from data, architecture, and operational deficiencies.

    \par \textit{Source(s):}  \cite{liu2020using, tang2021empirical,   albuquerque2022comprehending, arpteg2018software, li2023debtviz, li2023automatic, lenarduzzi2021software, breck2017ml, obrien202223, perez2021technical,  shivashankar2022maintainability, bavota2016large, liu2021exploratory, wang2023technical, li2022identifying, de2025software, sutoyo2024satdaug}.

    
\end{enumerate}

\begin{tcolorbox}
\textbf{RQ1.1:} 
The analysis revealed that AI Technical Debts (AITDs) are best explained through a root-cause–oriented perspective comprising seven overarching categories: 
(1) Data \& Library-Related Debts, (2) Model \& Code-Related Debts, 
(3) Algorithm-Related Debts, (4) Design, Architecture \& Configuration Debts, 
(5) Operational \& Lifecycle Debts, (6) Documentation \& Communication Debts, 
and (7) Testing \& Quality Assuarance Debts. 
By grounding the taxonomy in the underlying engineering sources—such as data instability, structural design flaws, algorithmic misalignment, inadequate lifecycle processes, and socio-technical communication gaps—the classification provides a holistic and integrative view of the diverse mechanisms through which technical debt accumulates in AI-based systems.
\end{tcolorbox}

\begin{tcolorbox}
\textbf{RQ1.2:} 
Through grounded theory analysis, we identified and refined a total of 31 distinct AITDs reported across the 60 primary studies. 
These debts span all seven root-cause categories and collectively reflect the multifaceted nature of technical debt in AI-enabled systems. 
Their prevalence varies considerably across the literature, with \textit{Data Debt (45.00\%)}, \textit{Glue Code (30.00\%)}, and \textit{Test Debt (28.33\%)} emerging as the most frequently reported issues. 
Table~\ref{tab:AITDranking} summarizes the occurrence frequencies and provides a detailed overview of how often each debt type appears across empirical studies.
\end{tcolorbox}

\begin{table*}[ht]
\centering
\caption{Overview of AI Technical Debts (AITDs) as part of the systematic review}
\label{tab:AITDranking}
\begin{tabular}{c l c c l}
\toprule
\textbf{\#} & \textbf{AITD} & \textbf{Freq(n=60)} & \textbf{\%} &
\textbf{Category}\\ \hline

1  & Data Debt/Unstable Data Dependencies  & 27 & 45.00 & Data/Library Related Debts \\ \hline
2  & Glue Code (GC) & 18 & 30.00 &Model/Code Related Debts \\ \hline
3  & Test Debt & 17 & 28.33 & Testing \& Quality Assurance Debts\\ \hline
4  & Documentation Debt & 15 & 25.33 & Documentation \& Communication Debts \\ \hline
5  & Requirement Debt & 14 & 23.33 & Operational \& Lifecycle Debts\\ \hline

6  & AI Architectural Debt & 13 & 21.67 & Design \& Architecture Debts \\ \hline

7  & Configuration Debt & 12 & 20.00 & Design \& Architecture Debts\\ \hline
8 & Dead Experimental Code Paths/Prototype Debt & 11 & 18.33 & Model/Code Related Debts \\ \hline

9 & Algorithm Debt/Inclination/Human Bias & 10 & 16.67 & Algorithm Related Debts \\ \hline

10 & Duplicate Model Code & 10 & 16.67  & Model/Code Related Debts \\ \hline

11 & Cultural/People/Social Debt & 10 & 16.67 & Documentation \& Communication Debts \\ \hline
12  & Pipeline Jungle & 9 & 15.00  & Data/Library Related Debts \\ \hline
13 & Jumbled Model Architecture (JMA) & 9 & 15.00 & Design \& Architecture Debts \\ \hline

14 & Hidden Feedback Loops & 9 & 15.00 & Model/Code Related Debts \\ \hline

15  & Defect Debt & 9 & 15.00 & Operational \& Lifecycle Debts \\ \hline

16 & Multiple Language Smells (MLS) & 7  & 11.67 & Model/Code Related Debts\\ \hline

17 & Undeclared consumers & 6  & 10.00 & Model/Code Related Debts \\ \hline
18 & Ethical Debt & 6  & 10.00 & Algorithm Related Debts\\ \hline

19 & Unwanted Debugging Code (UDC) & 5  & 8.33 &Model/Code Related Debts\\ \hline

20 & Versioning Debt & 5  & 8.33 & Operational \& Lifecycle Debts\\ \hline

21 & Correction Cascades (CC) & 4  & 6.67 & Model/Code Related Debts \\ \hline
22 & Entanglement & 4  & 6.67 & Model/Code Related Debts\\ \hline

23 & Compatibility Debt & 4  & 6.67 & Design \& Architecture Debts \\ \hline

24 & Abstraction Debt & 4  & 6.67 & Design \& Architecture Debts \\ \hline

25 & Dispensable Dependency & 3  & 5.00 & Data/Library Related Debts \\ \hline
26 & Boundary Erosion & 3  & 5.00 & Design \& Architecture Debts \\ \hline

27 & Overly Simplified Metrics & 3  & 5.00 & Algorithm Related Debts \\ 
\hline

28 & Process/Infrastructure Debt & 3 & 5.00 & Operational \& Lifecycle Debts \\ \hline
29 & Build Debt & 3  & 5.00 & Operational \& Lifecycle Debts\\ \hline
30 & Scattered Use of ML Libraries (SML) & 2  & 3.33 & Data/Library Related Debts\\ \hline

31 & Deep God File (DG) & 2  & 3.33 & Model/Code Related Debts \\

\bottomrule
\end{tabular}

\end{table*}


\subsection{Analysis of the Identified AITDs and Their Implications Through the Lens of AI TRiSM}

Table~\ref{tab:AITDranking} summarizes the 31 AI Technical Debts (AITDs) identified across the 60 primary studies and ranks them by frequency and percentage occurrence. The distribution reveals several dominant forms of debt that consistently appear across diverse AI system development contexts.

\textbf{Data Debt} is the most frequently reported AITD, appearing in \textbf{45.00\%} of the studies. This debt category reflects persistent data governance and data-quality challenges, including unstable data dependencies, schema inconsistencies, missing metadata, and untracked transformations. Given that AI systems rely on continuously evolving and heterogeneous data sources, these deficiencies lead to model drift, silent performance degradation, and reproducibility failures. Data Debt remains the most systemic and foundational problem recorded in the literature.

\textbf{Glue Code} is the second-most prevalent, reported in \textbf{30.00\%} of the studies. It captures the ad-hoc integration logic used to connect disparate data pipelines, libraries, model components, and deployment tools. Although Glue Code enables rapid experimentation—particularly in early development phases—it produces fragile, highly coupled systems that are difficult to extend or debug. This debt reflects the rapid, tool-centric, and exploratory nature of typical AI workflows.

\textbf{Test Debt} ranks third, appearing in \textbf{28.33\%} of the included studies. AI-based systems require extensive testing not only at the code level but also across data transformations, model behavior, fairness, robustness, and performance under distribution shift. The literature indicates that teams frequently deprioritize testing due to the difficulty of specifying expected ML behavior or the iterative nature of model development. This leads to insufficient test coverage, poor monitoring, and higher risk of failure in deployment settings.

\textbf{Documentation Debt} follows closely at \textbf{25.33\%}. As AI projects involve interdisciplinary teams (AI researchers, ML engineers, domain experts), inadequate documentation compromises maintainability, reproducibility, and onboarding. Missing or outdated dataset descriptions, model cards, configuration files, feature-engineering rationales, and pipeline diagrams hinder long-term system understanding and evolution.

\textbf{Requirement Debt} appears in \textbf{23.33\%} of the studies, illustrating how evolving stakeholder needs and ambiguous system goals affect AI development. Requirement Debt often arises when model behavior is not fully specified or when performance and fairness requirements evolve during experimentation. This leads to incomplete implementations and rework.

\textbf{AI Architectural Debt} and \textbf{Configuration Debt} were reported in \textbf{21.67\%} and \textbf{20.00\%} of the papers respectively. Design Debt arises from rushed architectural decisions and insufficient modularity—often exacerbated by early experimentation culture. Configuration Debt reflects the complexity of hyperparameters, environment variables, orchestration specifications, and deployment parameters, which, when poorly managed, cause reproducibility gaps and environment-dependent failures.

\textbf{Prototype Debt}appear in \textbf{18.33\%} of the studies. Prototype Debt stems from experimental code that is prematurely reused in production. It illustrate the tension between research-driven iteration and production engineering.

Following these, a second tier of moderately frequent debts (reported in 15–17\% of studies) includes: \textit{Algorithm Inclination/Human Bias (16.67\%)}, \textit{Duplicate Model Code (16.67\%)}, \textit{Cultural/People/Social Debt (16.67\%)}, \textit{Pipeline Jungle (15.00\%)}, \textit{Jumbled Model Architecture (JMA) (15.00\%)}, \textit{Hidden Feedback Loops (15.00\%)}, \textit{Defect Debt (15.00\%)}, \textit{Multiple Language Smells (MLS) (11.67\%)}, \textit{Undeclared Consumers (10.00\%)}, \textit{Ethical Debt (10.00\%)}. These debts highlight structural, behavioral, and socio-technical issues such as unplanned architecture evolution (JMA), unintended system dynamics (Hidden Feedback Loops), implementation redundancy (Duplicate Model Code), ethically problematic decision pathways (Ethical Debt), and misalignment across teams (Cultural/People/Social Debt). 

The remaining AITDs, each appearing in fewer than \textit{10\%} of the studies, include Unwanted Debugging Code (8.33\%), Versioning Debt (8.33\%), Correction Cascades (6.67\%), Entanglement (6.67\%), Compatibility Debt (6.67\%), Abstraction Debt (6.67\%), Dispensable Dependency (5.00\%), Boundary Erosion (5.00\%), Overly Simplified Metrics (5.00\%), Process/Infrastructure Debt (5.00\%), Build Debt (5.00\%), Scattered Use of ML Libraries (3.33\%), and Deep God File (3.33\%).

Although individually less frequent, these debts are still significant. Many reflect emerging or specialized risks unique to AI complexity, such as: unstable component borders (Boundary Erosion), cascading dependency breakages (Correction Cascades), flawed or manipulated evaluation criteria (Overly Simplified Metrics), architectural monoliths (Deep God File), tool fragmentation (Scattered ML Libraries). Their lower frequency does not imply low severity; several represent high-impact failure modes that are under-investigated in current research.

The prevalence distribution reinforces the inherently multifaceted nature of AITDs. High-impact debts (Data Debt, Glue Code, Test Debt) correlate strongly with foundational AI engineering workflows—data management, model integration, and verification—while medium-frequency and low-frequency debts highlight structural, algorithmic, operational, and socio-technical weaknesses. In summary, the landscape of AITDs reveals a complex interplay between AI-specific engineering challenges and long-standing software quality concerns. The prevalence analysis indicates that AI-intensive systems demand specialized management strategies addressing data governance, integration practices, rigorous testing, reproducibility, and responsible algorithmic design to ensure sustainable and trustworthy AI development.


\paragraph{\textbf{Implications of Prevalent AITDs Through the Lens of AI TRiSM}} 
Taken together, the ten most prevalent AITDs identified in this review—Data Debt/Unstable Data Dependencies (Data/Library-Related Debts); Glue Code, Prototype Debt, and Duplicate Model Code (Model/Code-Related Debts); Test Debt (Testing \& Quality Assurance Debts); Documentation Debt (Documentation \& Communication Debts); Requirement Debt (Operational \& Lifecycle Debts); AI Architectural Debt and Configuration Debt (Design \& Architecture Debts); and Algorithmic Inclination/Human Bias (Algorithm-Related Debts)—exhibit clear, mechanism-level intersections with AI safety and AI security concerns.

Data and library–related weaknesses (e.g., unstable data dependencies) increase the likelihood of unsafe behavior under data drift and reduce confidence in input validity. Model and code–related debts (e.g., glue code, duplicate model code, and prototype paths) introduce fragile integration layers and hidden dependencies that broaden attack surfaces and make failures harder to isolate. Testing and documentation deficits reduce fault and vulnerability discoverability and weaken auditability and incident response capabilities, while operational and architectural debts—such as requirements ambiguity, configuration drift, and architectural shortcuts—can embed unsafe assumptions, misconfigurations, and weak boundary controls into production systems. Finally, algorithm-related bias debt directly contributes to harmful outcomes and safety violations through systematic misbehavior across specific populations or operational contexts.

This analysis demonstrates how the most frequently reported AITDs consistently map to concrete safety- and security-relevant risk pathways, thereby supporting an AI-TRiSM–aligned perspective on AITD management. This mapping, however, should be regarded as an initial analytical step; Sections~\ref{sec:secsafeconcerns} and~\ref{sec:secsafeguidelines} provide a deeper examination of how these AITDs propagate and compound risks related to security and safety, as well as their management -the foundational pillars of AI TRiSM—thereby directly addressing \textbf{RQ2–RQ3}.

\section{AITDs - Safety and Security Concerns (RQ2)}
\label{sec:secsafeconcerns}

\subsection{Safety Concerns}
The increasing integration of AI in high-stakes domains has amplified the need to address critical safety concerns that extend beyond technical debt and performance limitations. As detailed in prior AI safety literature, particularly in \cite{salhab2024systematic, 10.1007/978-3-030-83906-2_20, Gyevnár2025531, bucaioni2025checklist}, key safety risks emerge due to the opaque, non-deterministic, and dynamic nature of AI systems. These concerns are especially pronounced in systems leveraging deep learning and reinforcement learning models, which are often sensitive to environmental noise, distributional shifts, and adversarial manipulation. Table~\ref{tab:safety-concerns-aitds} presents a comprehensive mapping between these safety concerns and their related AITDs, providing an explanatory context for each relationship.

\subsubsection{Explainability and Interpretability}

AI models - particularly deep learning architectures - often operate as black boxes, making it difficult for developers and stakeholders to understand, validate, or justify system decisions. In addition to this opacity, many AI models exhibit inherently stochastic behavior due to probabilistic inference, non-deterministic training processes, and adaptive updates. Such stochasticity reduces predictability, which has traditionally been a cornerstone of safety and security engineering. Together, limited interpretability and non-deterministic behavior pose significant safety risks in domains such as healthcare, finance, and autonomous systems, where accountability, traceability, and reliable error analysis are essential~\cite{Gyevnár2025531, Xu2019563, sheh2021explainable, Albahri2023156, 8466590, krajna2022explainable, chamola2023review}.


\par Related AITDs:

\begin{itemize}
    \item \textit{Jumbled Model Architecture}: Poorly structured and undocumented model architectures obscure the internal logic of AI systems, making it difficult to interpret model behavior or trace decision flows.
    \item \textit{Overly Simplified Metrics}: When models are evaluated using high-level, generic metrics (e.g., accuracy), without alignment to context-specific safety goals (e.g., false negatives in medical diagnosis), the model’s performance becomes misleading and less interpretable.
    \item \textit{Documentation Debt}: Lack of sufficient documentation about models, features, and training processes severely impairs the ability of users and auditors to interpret system logic, creating blind spots in safety validation efforts.
\end{itemize}

\subsubsection{Robustness and Reliability}

Robustness is the ability of AI systems to remain stable and accurate under perturbations, noisy data, or adversarial conditions  \cite{Gyevnár2025531, wang2023distribution, girard2022caisar, zhao2021detecting, akram2022stadre, tarchoun2022investigating,  arnez2021improving, mziou2019safety, choi2022argan, carlini2017towards}. Reliability ensures consistent performance over time and across various deployment environments. A lack of robustness and reliability can cause erratic behavior and system failures, undermining trust and safety in AI operations \cite{fisher2021towards, schwartz2020regularization, maabreh2022robustness, kshetry2019safety, zhao2020safety, werner2023assurance}.

\par Related AITDs:

\begin{itemize}
    \item \textit{Data Debt}: Poor quality, inconsistent, or evolving training data leads to brittle models that fail when exposed to real-world or shifted inputs.
    \item \textit{Defect Debt}: Known but unresolved bugs in the system compromise reliability, especially as they accumulate and interact unpredictably with evolving components.
    \item \textit{Test Debt}: Inadequate testing, particularly in edge-case scenarios, increases the likelihood of system crashes or incorrect outputs under stress conditions.
\end{itemize}

\subsubsection{Transparency and Trustworthiness}

Transparency refers to the ability to inspect, understand, and verify AI processes \cite{cooper2022believe,krajna2022explainable,sokol2019counterfactual,zhao2020safety,sheh2021explainable, buczak2022explainable, chamola2023review, samadi2023safe}. Trustworthiness builds upon transparency by assuring users that the system behaves ethically, consistently, and as intended \cite{bravo2022human, srinivasan2019understanding, he2021challenges, chamola2023review, steimers2021sources, hagendorff2021linking}. In safety-critical settings, lack of transparency can result in hidden vulnerabilities, while eroded trust may prevent responsible AI adoption.

\par Related AITDs:

\begin{itemize}
    \item \textit{Documentation Debt}: Insufficient documentation prevents clear understanding of system behavior, hindering transparency and the ability to audit the model pipeline.
    \item \textit{Versioning Debt}: Poor version tracking of datasets, models, or parameters makes it difficult to reproduce results, undermining system transparency and regulatory traceability.
    \item \textit{Ethical Debt}: Failing to explicitly consider ethics or fairness in system design leads to opacity in value alignment, undermining user trust.
\end{itemize}

\subsubsection{Bias and Fairness}

Bias in AI systems arises from imbalanced datasets, unrepresentative sampling, or discriminatory design choices. These biases compromise fairness and can result in systemic harm to individuals or groups, particularly in sensitive domains such as hiring, lending, and criminal justice \cite{kim2020fair, srinivasan2019understanding,jaipuria2022deeppic, kim2020fair}.

\par Related AITDs:

\begin{itemize}
    \item \textit{Algorithmic Inclination / Human Bias}: Over-reliance on familiar or convenient algorithms, rather than evaluating fairness implications, can encode human or institutional bias into the system.
    \item \textit{Ethical Debt}: Failure to integrate ethical design principles allows biased data or assumptions to propagate unchecked, resulting in unfair or unsafe model decisions.
    \item \textit{Overly Simplified Metrics}: Metrics that fail to account for fairness dimensions (e.g., equal opportunity, demographic parity) obscure potential bias and mask harmful outcomes.
\end{itemize}

\subsubsection{Adversarial and Poisoning Attacks}

Adversarial attacks involve input manipulations crafted to mislead AI systems, while data poisoning injects malicious data into training pipelines to degrade model integrity. These threats compromise system safety by causing incorrect outputs that may go undetected in critical environments \cite{he2021challenges, carlini2017towards, maabreh2022developing, maabreh2022robustness, 9256597}.

\par Related AITDs:

\begin{itemize}
    \item \textit{Algorithmic Inclination / Human Bias}: Use of predictable or widely known algorithms increases susceptibility to adversarial manipulation, especially if security hardening is not prioritized.
    \item \textit{Correction Cascades}: Inter-model dependencies make it possible for adversarial changes in one component to cascade through the system, amplifying unsafe behaviors.
    \item \textit{Entanglement}: Tight coupling between components means that adversarially targeted faults in one model can trigger ripple effects, leading to systemic vulnerabilities.
\end{itemize}

\subsubsection{Out-of-Distribution (OOD) Generalization}

AI systems are often trained in controlled settings and may fail when deployed in real-world environments that differ from their training data distributions. OOD generalization failures can result in unsafe decisions, especially when the system is unaware of its own limitations \cite{10.1007/978-3-030-83906-2_20, Kamoi2020OutofDistributionDW,10.1007/978-3-030-83906-2_17, rossolini2022increasingconfidencedeepneural, girard2022caisar, bravo2022human,carlini2017towards, Haider2023851}.

\par Related AITDs:

\begin{itemize}
    \item \textit{Pipeline Jungle}: Overly complex and interdependent data pipelines hinder the ability to adapt to new data distributions, leading to failures in OOD settings.
    \item \textit{Dead Experimental Code Paths / Prototype Debt}: Experimental or prototype components may lack robustness and are often not validated against diverse real-world data, making them vulnerable to distribution shifts.
    \item \textit{Overly Simplified Metrics}: Over-reliance on aggregate metrics during training can hide poor generalization performance on unseen inputs, leading to unsafe or unexpected system behavior post-deployment.
\end{itemize}

\begin{table*}[ht]
\centering
\caption{Mapping Safety Concerns to Related AI Technical Debts (AITDs)}
\label{tab:safety-concerns-aitds}
\begin{tabular}{p{0.3cm} p{2.5cm} p{4.0cm} p{7.0cm}}
\toprule
\textbf{S/N} & \textbf{Safety Concern} & \textbf{Related AITDs} & \textbf{Explanation} \\
\hline
1 & Explainability and Interpretability & 
Jumbled Model Architecture, Overly Simplified Metrics, Documentation Debt &
Opaque model structures, vague or misaligned evaluation metrics, and poor documentation reduce the ability to understand or justify AI decisions—critical in high-stakes scenarios like healthcare and law. \\
\hline
2 & Robustness and Reliability & 
Data Debt, Defect Debt, Test Debt &
Poor data quality, unresolved bugs, and insufficient testing undermine model stability and cause erratic or unreliable performance in unpredictable real-world environments. \\
\hline
3 & Transparency and Trustworthiness & 
Documentation Debt, Versioning Debt, Ethical Debt &
Incomplete documentation, weak version tracking, and disregard for ethical considerations hinder traceability and reduce stakeholder confidence in the system’s outputs. \\
\hline
4 & Bias and Fairness & 
Algorithmic Inclination / Human Bias, Ethical Debt, Overly Simplified Metrics &
Systems trained on biased data or developed without ethical oversight can produce unfair or discriminatory results, particularly when using simplistic or poorly aligned performance metrics. \\
\hline
5 & Adversarial and Poisoning Attacks & 
Algorithmic Inclination / Human Bias, Correction Cascades, Entanglement &
Overused algorithms, coupled components, and cascading model dependencies increase the attack surface and enable faults or adversarial perturbations to propagate across the system. \\
\hline
6 & Out-of-Distribution (OOD) Generalization & 
Pipeline Jungle, Dead Experimental Code Paths, Overly Simplified Metrics &
Complex or disorganized pipelines, unvetted prototype code, and overly generic evaluation metrics result in models that perform poorly or dangerously when encountering unfamiliar input distributions. \\
\bottomrule
\end{tabular}
\end{table*}

\begin{figure*}[t]
    \centering
    \includegraphics[width=1.0\textwidth]{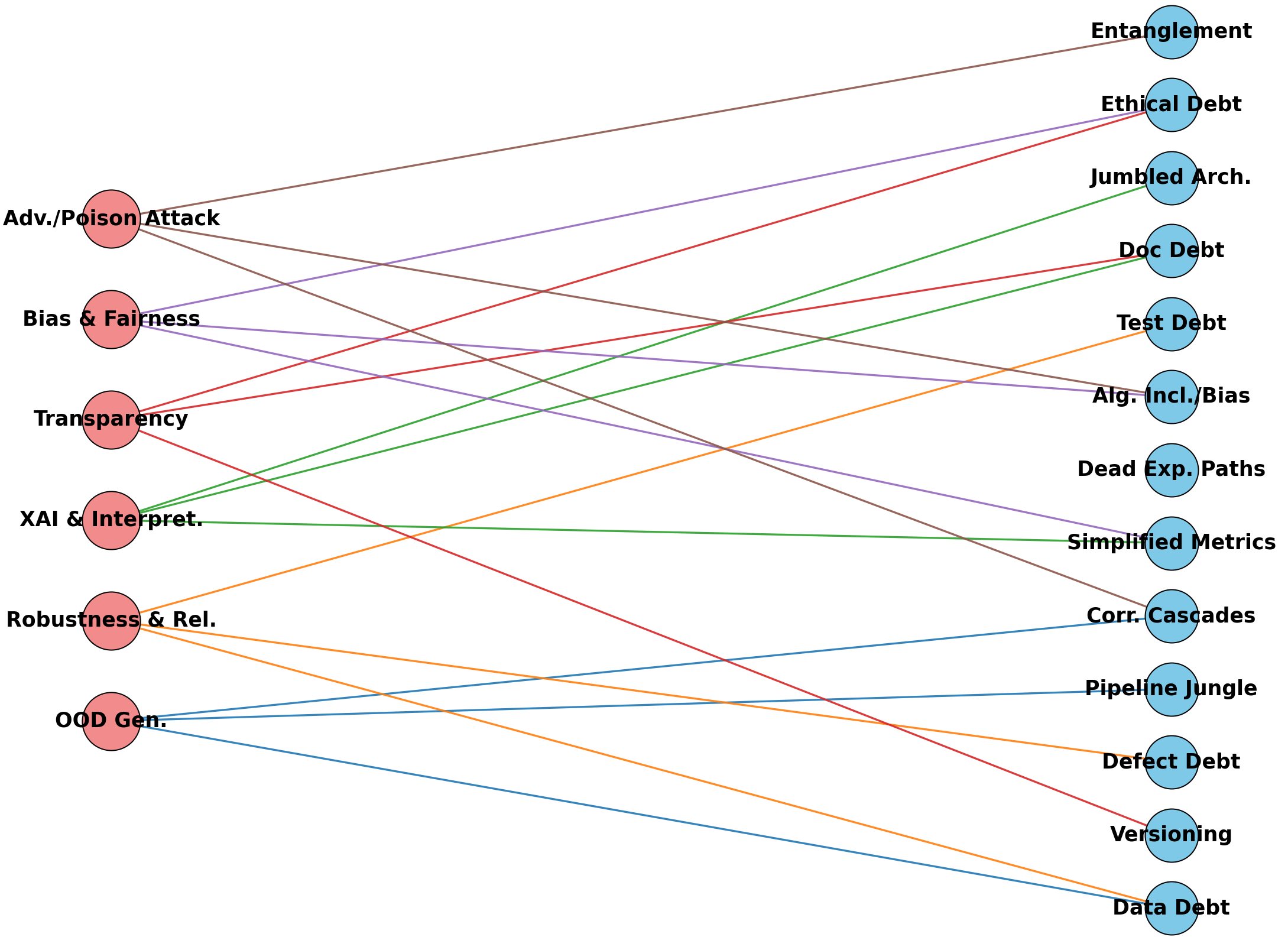}
    
    \caption{\textbf{Visual Mapping of Safety Concerns to AI Technical Debts:} Red circles represent safety concerns, while blue circles represent AITDs. Each safety concern is linked to its associated technical debts using color-coded edges, illustrating the multidimensional impact of technical debt on AI system safety.} \label{fig:safety-aitd}

\end{figure*}

To provide a visual representation of the relationships between identified safety concerns and the corresponding AI technical debts, we present Figure~\ref{fig:safety-aitd}. This mapping helps to clarify how various safety attributes are undermined by specific AITDs and reinforces the interconnectedness between safety assurance and technical debt management in AI systems.

\begin{tcolorbox}[colframe=gray!60, colback=gray!5, coltitle=black, title=\textbf{RQ2.1 – Safety Concerns in AI Systems}]
AI Technical Debts (AITDs) undermine the safety of AI systems in high-stakes environments. Key safety risks include:

\begin{itemize}
    \item \textbf{Low Explainability:} Complex and poorly structured models reduce interpretability.
    \item \textbf{Poor Robustness:} Fragile models fail under noisy or adversarial inputs.
    \item \textbf{Opaque Behavior:} Documentation and versioning issues erode transparency and trust.
    \item \textbf{Bias Risks:} Algorithmic shortcuts and Overly Simplified Metrics conceal unfair outcomes.
    \item \textbf{OOD Failures:} Models trained on limited data generalize poorly to real-world inputs.
\end{itemize}

\textit{Note: These risks demand proactive design, validation, and monitoring to ensure safe AI deployment.}
\end{tcolorbox}

\subsection{Security Concerns}

AI systems, similar to conventional software systems, are exposed to dynamic threats and evolving adversarial attacks. However, these challenges are often amplified in AI-based systems due to their reliance on data-driven learning, probabilistic decision-making, and adaptive behaviors, which introduce additional uncertainty and complexity. Over time, these characteristics contribute to increased maintenance burdens, expanded attack surfaces, and greater difficulty in anticipating and mitigating security risks. As a result, these systems are vulnerable to a range of security threats that can introduce technical debt which can in turn introduce other security risks if not mitigated early.
This section examines the security concerns associated with identified AI technical debts (AITDs). For each concern, a clear definition is provided, followed by a discussion of the relevant AITDs and their implications. Table~\ref{tab:security-aitd} presents a comprehensive mapping between these security concerns and their related AITDs, providing an explanatory context for each relationship. It is important to note that many of these security concerns are not exclusive to AI-intensive systems; rather, consistent with our earlier observation that ML systems inherit the maintenance challenges of traditional software while introducing additional AI-specific issues~\cite{sculley2015hidden}, these concerns also arise in conventional systems but are often intensified in AI-enabled contexts due to data dependence, adaptive behavior, and model-driven decision-making.


\subsubsection{Authentication and Authorization}
This concern arises when AI systems lack robust authentication and authorization mechanisms, which are essential for controlling who can access the system or its data \cite{recupito2024technical}. Without these controls, unauthorized users could gain access to sensitive model outputs or even manipulate system behavior. Inadequate authorization systems can also allow unauthorized users to alter critical system parameters, increasing the risk of security breaches \cite{sculley2015hidden}. This is particularly dangerous in safety-critical applications where malicious modifications or unauthorized access could lead to severe consequences, such as in healthcare \cite{thota2018centralized} or autonomous systems \cite{katzenbeisser2019security}.
\par Related AITDs:
\begin{itemize}
    \item Undeclared Consumers: This AITD is directly related, as it arises from the failure to implement proper access controls and authentication mechanisms, leading to unauthorized usage of the model’s outputs.
    \item Configuration Debt: Misconfigurations can weaken or bypass authentication and authorization mechanisms, increasing security risks.
\end{itemize}

\subsubsection{Data Integrity and Validation}
When AI systems do not adequately validate their input data or ensure the integrity of the data they rely on, they become vulnerable to security risks. Data integrity debt occurs when a system begins to process inaccurate, incomplete, or manipulated data, leading to unsafe or unreliable results \cite{foidl2019risk, recupito2024technical}. For example, in AI systems handling financial transactions, if input data isn't properly validated, attackers might exploit the system by feeding it fraudulent or corrupt data. This could lead to incorrect predictions or actions, such as approving unauthorized transactions or making unsafe decisions based on flawed information.
\par Related AITDs:
\begin{itemize}
    \item Unstable Data Dependencies/Data Debt: This directly relates to data integrity issues, as unreliable data sources can introduce incorrect or outdated information into the model, leading to unsafe or insecure outcomes.
    \item Hidden Feedback Loops: Feedback loops can interfere with the integrity of the system’s decision-making process, making it more prone to errors due to invalid data inputs.
\end{itemize}

\subsubsection{Security Patch and Update}
This concern arises when security patches and updates for the underlying software, libraries, or dependencies in AI systems are not applied regularly. Over time, unpatched vulnerabilities expose the system to attacks that exploit known security flaws. Systems relying on outdated or vulnerable libraries are especially prone to exploitation \cite{wang2021patchrnn}. In safety-critical applications like healthcare or autonomous vehicles, failure to apply security patches can lead to catastrophic outcomes, as attackers might leverage these vulnerabilities to compromise the system's functionality, cause data breaches, or trigger unsafe behaviors.
\par Related AITDs:
\begin{itemize}
    \item Dispensable Dependency: Dependencies that are no longer necessary but not removed create vulnerabilities due to missed security patches and updates.
    \item Inconsistent Use of ML Libraries: Scattered library use makes it difficult to apply security patches consistently across the system, leaving some parts vulnerable to attack.
\end{itemize}

\subsubsection{Data Privacy and Confidentiality}
AI systems often handle sensitive data, and failing to implement adequate privacy and confidentiality safeguards creates debt. This includes insufficient encryption, anonymization, or access control for data stored or processed by the system. As a result, unauthorized access to personal or sensitive data can occur, leading to privacy breaches, legal liabilities, and loss of trust \cite{al2021quality}. For example, in AI-driven healthcare systems, improper protection of patient records could expose confidential medical information, putting both patients and the organization at risk of legal penalties and reputational damage \cite{nankya2024security}.
\par Related AITDs:
\begin{itemize}
    \item Undeclared Consumers: Failure to control access to model outputs can lead to privacy breaches where sensitive data is exposed to unauthorized parties.
    \item Glue Code: Poorly written integration code may fail to protect data, exposing it to potential leaks or unauthorized access.
\end{itemize}

\subsubsection{Adversarial Attack Resilience}
This concern is significant, especially when AI systems are not designed or hardened against adversarial attacks, where input is intentionally manipulated to trick the system into making incorrect or harmful decisions. Attackers may subtly alter input data (such as images or text) in ways that are imperceptible to humans, but cause AI to make erroneous predictions \cite{dai2021artificial}. For instance, an adversarial attack on an autonomous vehicle's AI could involve subtly altering road signs to cause the vehicle to misinterpret a stop sign as a speed limit sign, leading to unsafe driving behaviors \cite{girdhar2023cybersecurity}.
\par Related AITDs:
\begin{itemize}
    \item Algorithmic Inclination/Human Bias: AI systems that rely too heavily on specific algorithms may be vulnerable to adversarial attacks that exploit known weaknesses in those algorithms.
    \item Correction Cascades (CC): Complex dependencies between models may create opportunities for adversarial manipulation, where altering input to one model can cascade through the system, resulting in unexpected and unsafe outcomes.
\end{itemize}

\subsubsection{Fail-Safe Mechanisms}
Fail-safe mechanisms are critical for ensuring that AI systems can safely handle unexpected errors, failures, or external threats. 
In safety-critical environments such as industrial robotics or autonomous driving, the lack of fail-safe mechanisms could result in hazardous conditions when an error occurs, as the system may continue to operate in an unsafe manner instead of shutting down or reverting to a safe state \cite{farrell2021evolution}.
\par Related AITDs:
\begin{itemize}
    \item Dead Experimental Code Paths/Prototype Debt: Experimental or prototype systems often lack robust fail-safe mechanisms, leading to unpredictable behavior when they encounter unexpected situations in production environments.
    \item Boundary Erosion: As clear separations between components weaken, fail-safe mechanisms may fail to isolate issues, allowing errors in one component to propagate through the system.
\end{itemize}

\subsubsection{Model Interpretability and Transparency}
There is concern when AI models are too opaque to allow users to understand how decisions are made, particularly in complex, black-box models like deep neural networks \cite{nallakaruppan2024explainable, hoangexplainable}. Without model interpretability, it becomes difficult to detect biases, vulnerabilities, or potential failures, especially in safety-critical applications \cite{hoangexplainable}. For example, in AI-driven medical diagnosis, a lack of transparency in how AI arrives at its recommendations could prevent healthcare professionals from identifying flaws in the model, potentially leading to incorrect treatments or adverse outcomes \cite{mohammed2024developing, kumarml}.

\par Related AITDs:
\begin{itemize}
    \item Jumbled Model Architecture: Disorganized model architectures can make it difficult to interpret or understand the behavior of the system, increasing the risk of undetected vulnerabilities or unsafe decisions.
    \item Overly Simplified Metrics: AI systems optimized for abstract performance metrics may lack transparency, making it harder to trace decisions and assess model safety in critical situations.
\end{itemize}

\subsubsection{Real-Time Security Monitoring}
AI systems often operate in dynamic environments where real-time threats and security risks must be continuously monitored \cite{chirra2020ai}. An uncertain consequence can arise as a result of insufficient real-time security monitoring and incident response infrastructure in place. Without robust monitoring, security breaches or failures can go unnoticed for long periods, allowing attackers to exploit vulnerabilities or manipulate system behavior undetected \cite{chirra2020ai}. For instance, in AI-controlled energy grids, the absence of real-time security monitoring could allow a cyberattack to disrupt operations without being detected until it is too late.
\par Related AITDs:
\begin{itemize}
    \item Hidden Feedback Loops: Without proper real-time monitoring, feedback loops may go unnoticed, leading to system degradation or vulnerabilities that can be exploited over time.
    \item Dead Experimental Code Paths: Leftover experimental code that is not actively monitored in production can introduce vulnerabilities that are difficult to detect without proper real-time oversight.
\end{itemize}

\subsubsection{Access Control and Data Protection}
A lack of sufficient controls over who can access sensitive data or model outputs in AI systems is a significant concern in AI systems \cite{tibebuframework}. If access controls are weak or nonexistent, unauthorized users could gain access to data that should be protected, such as proprietary model outputs or sensitive customer data. This increases the risk of data breaches and compliance violations, particularly in highly regulated industries such as finance \cite{hernandez2019data} or healthcare \cite{de2023guide}. Furthermore, in safety-critical systems, unauthorized access to model outputs could lead to unsafe actions if the model predictions are used inappropriately \cite{ruland2018access}.

\par Related AITDs:
\begin{itemize}
    \item Undeclared Consumers: As this AITD involves lack of control over who accesses model outputs, it directly contributes to data protection issues, making it difficult to secure sensitive information.
    \item Deep God File: A large, monolithic file containing multiple components makes it harder to enforce access controls effectively, increasing the risk of data leaks.
\end{itemize} 

\subsubsection{Complexity-Induced Vulnerabilities}
AI systems with complex architectures or pipelines often hide vulnerabilities that are difficult to detect and address \cite{sculley2015hidden}. This occurs when the system's complexity makes it challenging to perform thorough security assessments or safety analyses. For example, in systems where multiple models or components are interdependent, it may be difficult to track how data flows through the system or identify weak points where security measures are inadequate. This complexity can lead to vulnerabilities that attackers can exploit or cause system failures in safety-critical applications.

\par Related AITDs:
\begin{itemize}
    \item Pipeline Jungle: The complexity of AI pipelines increases the likelihood of hidden vulnerabilities, making it harder to secure the system and identify potential weaknesses.
    \item Entanglement: Tightly coupled components make it difficult to manage security risks, as vulnerabilities in one part of the system can affect other interconnected components.

\end{itemize}

\subsubsection{Dependency-Related Vulnerabilities}
AI systems often rely on external libraries, tools, and frameworks to function. Whenever outdated or unused dependencies are not properly managed, it leads to creating vulnerabilities that can be exploited by attackers \cite{chaudhary2018review}. Unused dependencies that remain in the system without regular updates can serve as backdoor for malicious activity \cite{10628360}. In safety-critical applications such as autonomous driving or healthcare, dependency-related vulnerabilities can lead to unpredictable system behavior or integration issues, potentially causing unsafe outcomes or system failures.
\par Related AITDs:
\begin{itemize}
    \item Dispensable Dependency: Unnecessary dependencies in the system introduce security vulnerabilities, as they are often not updated or patched regularly.
    \item Compatibility Debt: Dependency on outdated or insecure libraries can result in vulnerabilities that compromise system security and safety.
\end{itemize}

\subsubsection{Ethical and Bias-Induced Safety and Security}
AI systems that are developed without adequate consideration for ethical principles, such as fairness, accountability, and transparency, accumulate ethical debt \cite{petrozzino2021pays} can lead to biased decisions that result in unsafe or unethical outcomes \cite{roselli2019managing}. In safety-critical environments, such as healthcare, biased AI models could deny certain demographic groups access to critical treatments, resulting in harm \cite{sujan2023looking}. Similarly, from a security perspective, attackers could exploit biased algorithms to influence the system’s behavior in ways that compromise both safety and security \cite{chaudhary2018review}.

\par Related AITDs:
\begin{itemize}
    \item Ethics Debt: This is directly related, as biases in AI models can lead to unethical decisions, which may also introduce security risks if attackers exploit those biases.
    \item Algorithmic Inclination/Human Bias: Over-reliance on specific algorithms may introduce systemic bias, leading to unsafe decisions in critical applications and increasing susceptibility to exploitation.
\end{itemize}

\begin{figure*}[t]
    \centering
    \includegraphics[width=1.0\textwidth]{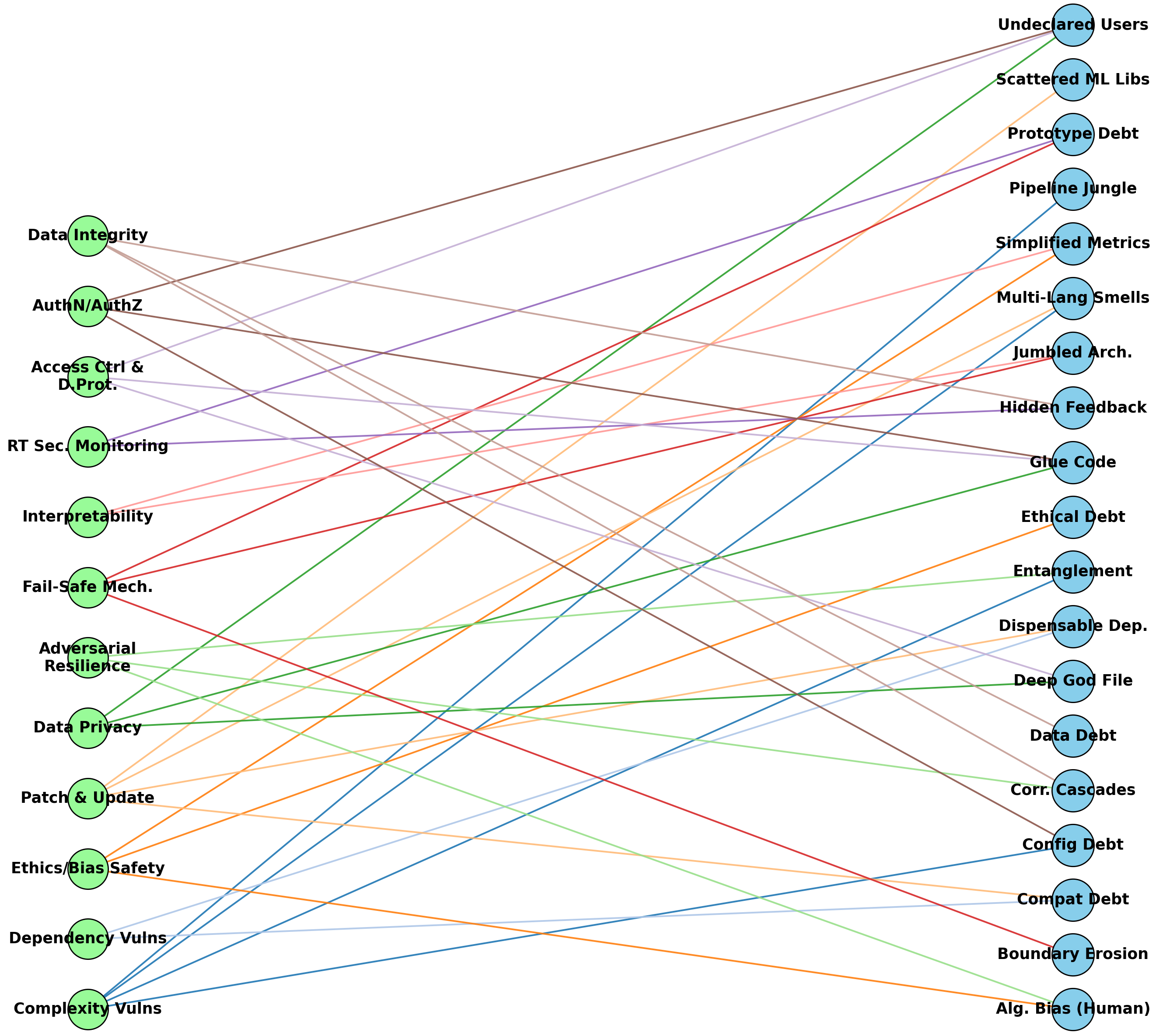}

    \caption{Bipartite mapping between security concerns and AI Technical Debts (AITDs), showing how identified AITDs contribute to or exacerbate specific security risks in AI-enabled systems. Green circles represent security concerns, while blue circles represent AITDs.}


    \label{fig:security-aitd}

    \end{figure*}

These security concerns are deeply interwoven with technical debt manifestations across AI-enabled systems. Figure~\ref{fig:security-aitd} visually consolidates these relationships, highlighting their interconnections, providing a graphical representation of how each AITD contributes to broader risk categories.

\begin{tcolorbox}[colframe=gray!60, colback=gray!5, coltitle=black, title=\textbf{RQ2.2 – Security Concerns in AI Systems}]

The identified AITDs, such as Undeclared Consumers, Data Debt, Algorithmic Bias, and Ethical Debt, significantly impact the security of AI-based systems. These debts can lead to vulnerabilities, including unauthorized access, data breaches, and biased outcomes. Critical risks identified include:
\begin{itemize}
    \item Authentication and Authorization: Weak controls can lead to unauthorized access to sensitive AI model outputs.
    \item Data Integrity: Poor validation processes make systems vulnerable to manipulated or inaccurate data.
    \item Adversarial Attacks: Systems without resilience mechanisms are prone to exploitation by adversarial inputs, jeopardizing both security and safety.
    \item Ethical and Bias-Induced: Failing to address fairness and accountability in AI systems can lead to harmful or unethical outcomes.
\end{itemize}
Impact on High-Stakes Domains: The safety-critical nature of sectors like healthcare, finance, and autonomous systems magnifies these concerns, requiring robust mitigation strategies.
\end{tcolorbox}

\begin{table*}[htbp]
\centering 
\caption{Mapping Security  Concerns to Related AI Technical Debts (AITDs)}
\begin{tabular}{c p{3.5cm} p{4.0cm} p{6.2cm}}
\toprule
\textbf{S/N} & \textbf{Security Concern} & \textbf{Related AITDs} & \textbf{Explanation} \\
\hline
1 & Authentication and Authorization & Undeclared Consumers, Configuration Debt, Glue Code & Lack of access control and misconfigurations allow unauthorized access to model outputs and parameters. \\
\hline
2 & Data Integrity and Validation & Data Debt (Unstable Dependencies), Hidden Feedback Loops, Correction Cascades & Inadequate validation of input data or changes in source data can compromise prediction accuracy and trustworthiness. \\
\hline
3 & Security Patch and Update & Dispensable Dependency, Scattered Use of ML Libraries, Compatibility Debt, Multiple Language Smells & Outdated, scattered, or heterogeneous codebases increase difficulty in applying consistent security updates. \\
\hline
4 & Data Privacy and Confidentiality & Undeclared Consumers, Glue Code, Deep God File & Poor control of model outputs and monolithic code structures may expose sensitive data. \\
\hline
5 & Adversarial Attack Resilience & Algorithmic Inclination / Human Bias, Correction Cascades, Entanglement & Lack of algorithm diversity and complex model interdependencies increase susceptibility to adversarial manipulation. \\
\hline
6 & Fail-Safe Mechanisms & Prototype Debt, Boundary Erosion, Jumbled Model Architecture & Weak architectural modularity and experimental code can prevent safe fallback during system failure. \\
\hline
7 & Model Interpretability and Transparency & Jumbled Model Architecture, Overly Simplified Metrics & Opaque model design and inappropriate evaluation metrics hinder understanding and safety validation. \\
\hline
8 & Real-Time Security Monitoring & Hidden Feedback Loops, Prototype Debt & Insufficient oversight allows critical issues (e.g., feedback drift) to go unnoticed in production.. \\
\hline
9 & Access Control and Data Protection & Undeclared Consumers, Glue Code, Deep God File & Lack of modularity and loosely managed outputs can expose the system to unauthorized use or data leakage. \\
\hline
10 & Complexity-Induced Vulnerabilities & Pipeline Jungle, Entanglement, Configuration Debt, Multiple Language Smells & Interdependent modules and use of multiple languages increase brittleness and vulnerability to subtle configuration flaws. \\
\hline
11 & Dependency-Related Vulnerabilities & Dispensable Dependency, Compatibility Debt & Poor management of software/library dependencies increases exposure to known exploits. \\
\hline
12 & Ethical and Bias-Induced Safety & Ethical Debt, Algorithmic Inclination / Bias, Overly Simplified Metrics & Neglecting fairness, transparency, and appropriate evaluation criteria can lead to unsafe, discriminatory outcomes. \\
\bottomrule
\end{tabular}
\label{tab:security-aitd}
\end{table*}

\section{Safety and Security guidelines for mitigating the identified AITDs (RQ3)}
\label{sec:secsafeguidelines}

\subsection{Safety Guidelines}
To mitigate the impact of AI Technical Debts (AITDs) on system safety, this study identifies a set of strategic guidelines derived from established AI safety research and best practices. These guidelines address core safety concerns, including robustness, explainability, fairness, transparency, and operational assurance. Each guideline is explicitly linked to specific AITDs that introduce safety risks, as identified in Section~\ref{sec:AITDs}. Our contribution lies in systematically consolidating, synthesizing, and mapping these established guidelines to the concrete forms of AITD uncovered in this review.

\begin{table*}[ht]
\centering
\caption{Mapping of Safety Guidelines to Related AI Technical Debts (AITDs)}
\label{tab:safety-guidelines-to-aitds}
\begin{tabular}{p{0.4cm} p{3.0cm} p{4.4cm} p{6.5cm}}
\toprule
\textbf{S/N} & \textbf{Safety Guideline} & \textbf{Related AITDs} & \textbf{Explanation} \\
\hline
1 & Explainable AI (XAI) & Jumbled Model Architecture, Overly Simplified Metrics, Documentation Debt & Improves model transparency and enables users to understand and justify AI decisions, enhancing trust and accountability. \\
\hline
2 & Fairness Assessment and Bias Mitigation & Algorithmic Inclination / Human Bias, Ethical Debt, Overly Simplified Metrics & Detects and corrects biases in data and model behavior to ensure equitable and non-discriminatory AI outcomes. \\
\hline
3 & Adversarial Training & Algorithmic Inclination / Human Bias, Correction Cascades, Entanglement & Increases model robustness against malicious inputs and reduces vulnerability propagation through tightly coupled components. \\
\hline
4 & OOD Detection & Pipeline Jungle, Prototype Debt, Overly Simplified Metrics & Prevents unsafe decisions on unfamiliar data by identifying distribution shifts and handling anomalies effectively. \\
\hline
5 & Formal Verification & Defect Debt, Test Debt, Jumbled Model Architecture & Ensures correctness of AI systems via mathematical proofs, addressing unverified bugs and untestable complex models. \\
\hline
6 & Domain Adaptation and Transfer Learning & Data Debt, Pipeline Jungle, Dead Experimental Code Paths & Enhances model generalization across diverse contexts by adapting to new domains and discarding obsolete or fragile components. \\
\hline
7 & Black Box Auditing and Logging & Documentation Debt, Versioning Debt  & Facilitates traceability and compliance through systematic logging and monitoring of AI decision-making processes. \\
\hline
8 & Secure and Federated Learning & Configuration Debt, Dispensable Dependency, Compatibility Debt & Preserves data privacy and system integrity by securely distributing training without centralizing sensitive information. \\
\bottomrule
\end{tabular}
\end{table*}

\subsubsection{Explainable AI (XAI)}
Explainable AI encompasses methods and techniques that make AI system decisions understandable to humans. This includes model-agnostic tools like Local Interpretable Model-agnostic Explanations (LIME) and SHapley Additive exPlanations (SHAP), inherently interpretable models like decision trees, and visualizations like saliency maps. By enhancing transparency, XAI helps end-users, developers, and regulators to audit and trust AI outputs \cite{Dağlarli20, Xu2019563, sheh2021explainable, Albahri2023156, 8466590, krajna2022explainable, chamola2023review}.
\par \textbf{Related AITDs:} \textit{Jumbled Model Architecture} obscures the internal structure and decision pathways of models, complicating explanation. \textit{Overly Simplified Metrics} do not convey interpretable performance indicators. \textit{Documentation Debt} limits external understanding of design choices and model logic.

\subsubsection{Fairness Assessment and Bias Mitigation}
This guideline ensures that AI systems treat individuals and groups equitably. Techniques include fairness-aware training algorithms \cite{shi2023towards}, pre-processing data balancing \cite{parmar2023review}, and post-hoc fairness evaluations using metrics like demographic parity or equal opportunity \cite{10.5555/3157382.3157469}. The goal is to detect and reduce both data and algorithmic biases \cite{ferrara2023fairness,  10.5555/3157382.3157469}.
\par \textbf{Related AITDs:} \textit{Algorithmic Inclination / Human Bias} encodes unchecked human biases into model logic. \textit{Ethical Debt} stems from ignoring fairness and social implications during development. \textit{Overly Simplified Metrics} may fail to capture nuanced biases in model outputs.

\subsubsection{Adversarial Training}
Adversarial training improves robustness by exposing models to adversarial examples during training. This enables the system to learn resilience against malicious perturbations. It is crucial for security-critical applications where robustness is a primary safety concern \cite{10.5555/3600270.3600944, a15080283}.
\par \textbf{Related AITDs:} \textit{Algorithmic Inclination / Human Bias} may prioritize performance over resilience. \textit{Correction Cascades} allow adversarial perturbations in one module to affect downstream outputs. \textit{Entanglement} creates interdependencies that amplify security vulnerabilities.

\subsubsection{Out-of-Distribution (OOD) Detection}
OOD detection allows AI systems to recognize when they are presented with inputs that significantly deviate from their training data. Strategies include confidence scoring, uncertainty modeling, and Bayesian inference. By deferring or flagging these cases, the system avoids unsafe decisions \cite{Kamoi2020OutofDistributionDW, Haider2023851}.
\par \textbf{Related AITDs:} \textit{Pipeline Jungle} complicates systematic integration of OOD detectors. \textit{Prototype Debt} results in fragile components not validated for distributional robustness. \textit{Overly Simplified Metrics} often miss edge-case or anomaly behavior.

\subsubsection{Formal Verification}
Formal verification applies mathematical proofs and logical reasoning to ensure that AI systems meet predefined specifications. This is particularly useful for verifying safety properties, such as invariants and boundary conditions in control systems \cite{Meng2022AdversarialRO,pmlr-v161-corsi21a}.
\par \textbf{Related AITDs:} \textit{Defect Debt} includes unaddressed errors that may invalidate verification results. \textit{Test Debt} results in insufficient test coverage for verifying formal properties. \textit{Jumbled Model Architecture} complicates reasoning due to its opacity and lack of modularity.

\subsubsection{Domain Adaptation and Transfer Learning}
These approaches allow models trained on one domain to adapt effectively to another. They involve fine-tuning or augmenting training data to generalize across varying distributions \cite{9134370, Kamath2019}. These are essential for deploying models in dynamic or diverse environments.
\par \textbf{Related AITDs:} \textit{Data Debt} limits generalization due to insufficient diversity. \textit{Pipeline Jungle} obstructs the integration of domain adaptation pipelines. \textit{Dead Experimental Code Paths} leave legacy components that are poorly adapted to new domains.

\subsubsection{Black Box Auditing and Logging}
This involves systematically recording the inputs, outputs, and internal states of AI systems to enable retrospective analysis \cite{8466590}. Logging enables debugging, monitoring, and compliance with regulatory standards. It is essential for safety transparency \cite{Chennam2023}.

\par \textbf{Related AITDs:} \textit{Documentation Debt} restricts the reproducibility of logs and audit trails. \textit{Versioning Debt} impairs traceability across system updates. 

\subsubsection{Secure and Federated Learning}
These are privacy-preserving learning paradigms where models are trained collaboratively without sharing raw data \cite{Zhang2022}. Federated learning improves security and trust in multi-agent environments. Encryption and differential privacy techniques are often employed \cite{Shejwalkar20221354}.

\par \textbf{Related AITDs:} \textit{Configuration Debt} creates exposure due to insecure settings. \textit{Dispensable Dependency} brings in third-party components that may not be vetted for privacy. \textit{Compatibility Debt} hinders integration of federated architectures into existing systems.

\begin{figure*}[t]
    \centering
    \includegraphics[width=1.0\textwidth]{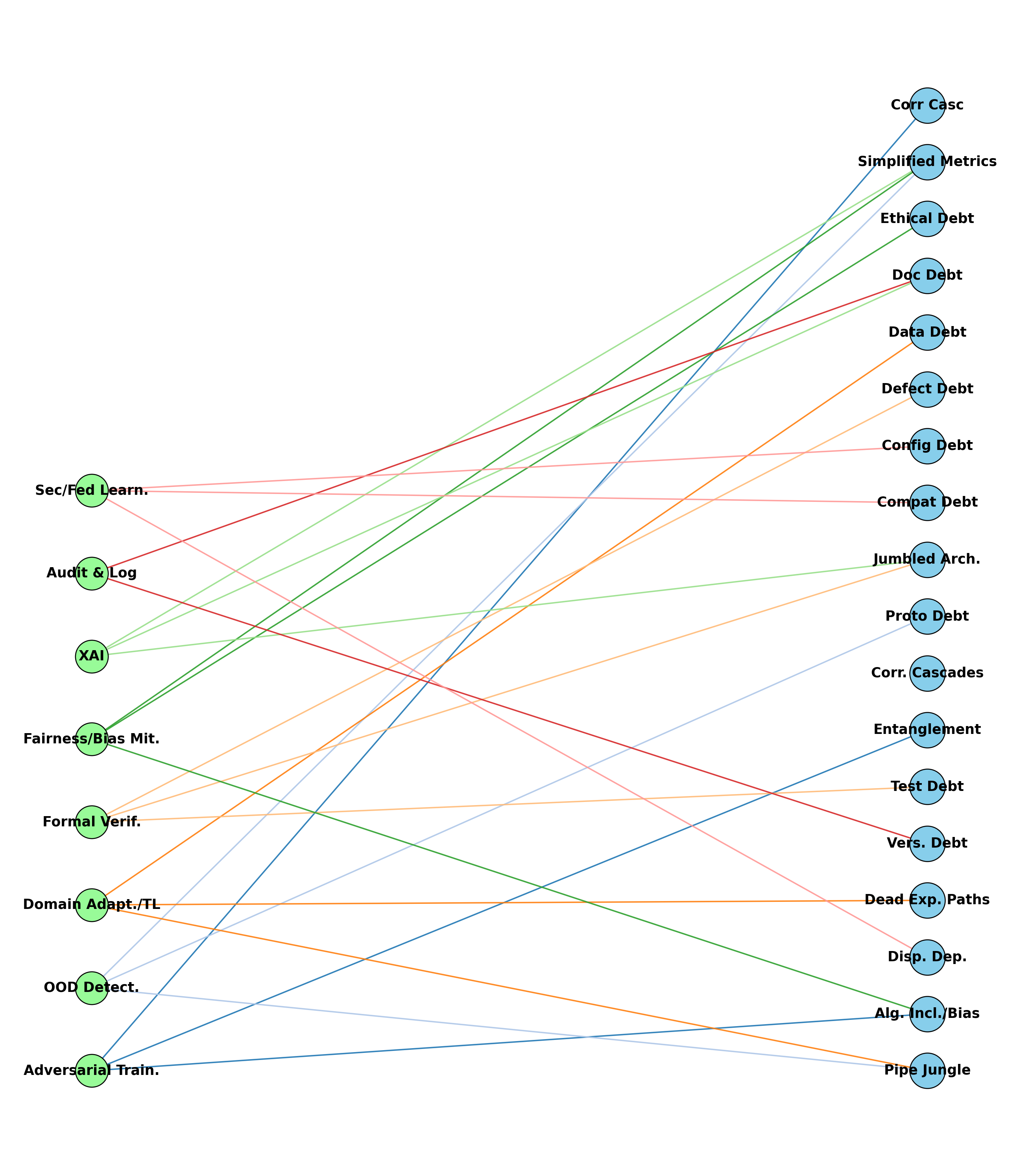}   
    \caption{\textbf{Mapping of Safety Guidelines to Related AI Technical Debts (AITDs).} This figure visualizes the relationships between key AI safety guidelines and the specific technical debts they help mitigate. Each guideline (green nodes) is connected to one or more AITDs (blue nodes) that pose risks to safety attributes such as explainability, robustness, fairness, and security. } \label{fig:safeguide-aitd}
\end{figure*}

\noindent These guidelines provide a multidimensional defense-in-depth strategy that enhances the safety of AI-based systems while also facilitating sustainable debt management. As summarized in Table~\ref{tab:safety-guidelines-to-aitds}, each safety guideline is explicitly linked to one or more AI Technical Debts (AITDs), highlighting the targeted impact of these practices. Additionally, Figure~\ref{fig:safeguide-aitd} provides a visual mapping that illustrates how each guideline contributes to the mitigation of specific AITDs, supporting a comprehensive safety assurance strategy.

\begin{tcolorbox}[colframe=gray!80!black, colback=gray!5, coltitle=black, title=\textbf{RQ3.1: Safety Guidelines for Mitigating AITDs}]
To enhance the safety of AI-based systems, the following guidelines are recommended for addressing AI Technical Debts (AITDs):

\begin{itemize}
    \item \textbf{Explainable AI (XAI):} Enhances system transparency, mitigating debts related to interpretability and trust.
    \item \textbf{Robustness Testing:} Detects failure points under adversarial or noisy conditions.
    \item \textbf{Continuous Monitoring:} Ensures early detection of drift, decay, or bias during deployment.
    \item \textbf{Ethical and Fairness Audits:} Mitigates bias, ethical, and societal debts through regular assessments.
    \item \textbf{Modular and Versioned Pipelines:} Reduces architecture and documentation debt by enabling traceability and maintainability.
    \item \textbf{OOD Detection Mechanisms:} Handles novel inputs and prevents unsafe behavior from unrecognized data.
\end{itemize}

\textit{Adopting these guidelines supports resilient, fair, and transparent AI systems in safety-critical domains.}
\end{tcolorbox}

\subsection{Security Guidelines}
\label{sec:secguidelines}
\begin{figure*}[t]
    \centering
    \includegraphics[width=0.9\textwidth]{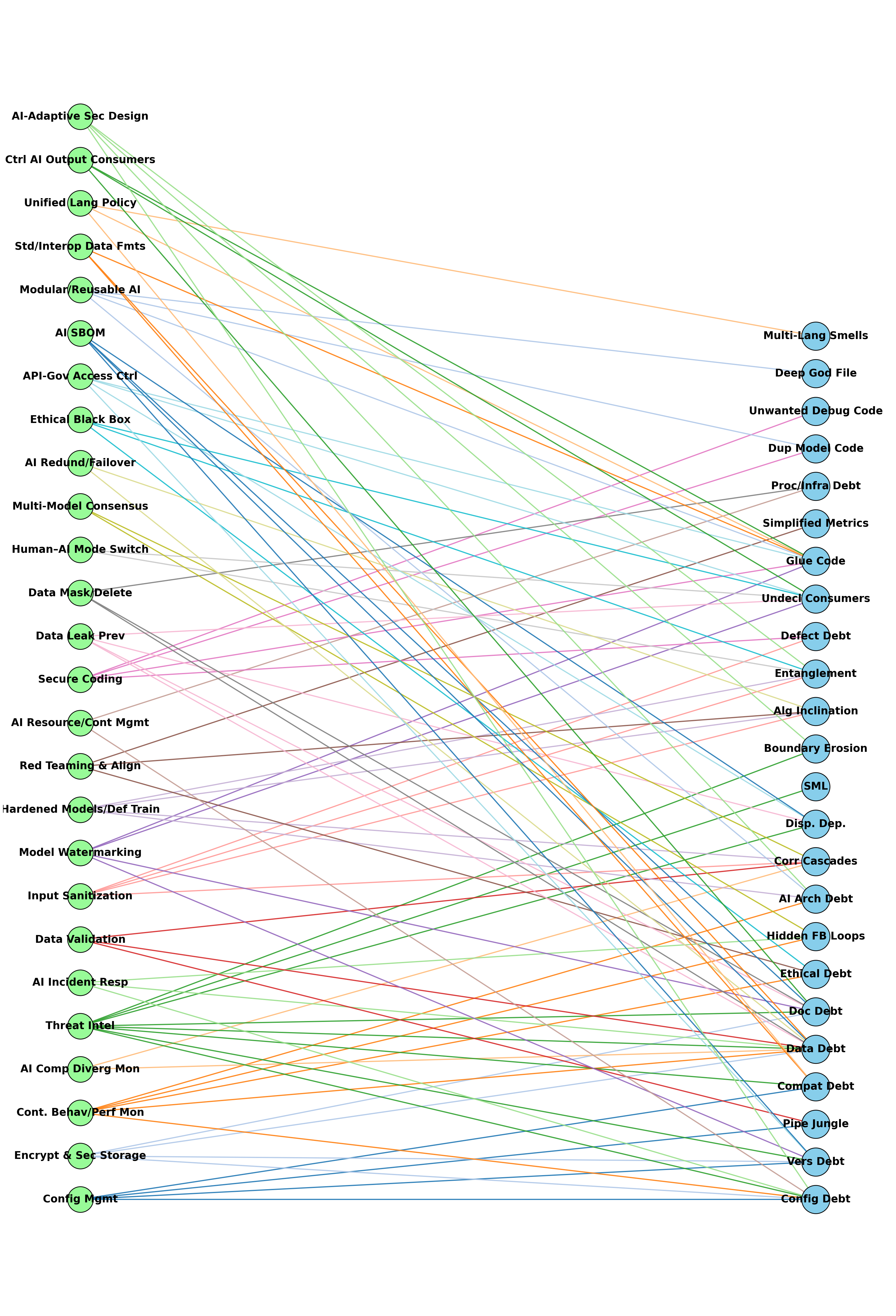}
    \caption{\textbf{Mapping of Security Guidelines to Related AI Technical Debts (AITDs)}. This figure visualizes the relationships between key AI security guidelines and the specific technical debts they help mitigate. Each guideline (green node) is connected to one or more AITDs (blue nodes) that introduce vulnerabilities or risks to system security and reliability. The connections illustrate how implementing these guidelines can address multiple forms of AI Technical Debt—such as data, model, and configuration-related debts—thereby enhancing the overall resilience and trustworthiness of AI-enabled systems.}
    \label{fig:secguides-aitd}

\end{figure*}

This section presents a comprehensive set of security guidelines designed to effectively mitigate the AI Technical Debts (AITDs) identified in this review. Our contribution lies in systematically consolidating, synthesizing, and mapping established guidelines to the specific forms of AITD uncovered in this study. The selected guidelines reflect the most relevant and impactful practices for addressing the root causes of AITDs, spanning critical areas such as secure data handling, adversarial resilience, configuration discipline, system monitoring, dependency governance, and organizational preparedness. In the following subsections, each guideline is presented with: (i) a concise definition, (ii) a practical example illustrating its application in real-world AI deployments, and (iii) an explanation of the specific AITDs it helps mitigate.
To enhance traceability and support evidence-based mitigation planning, Table~\ref{tab:sec-guidelines-to-aitd} provides a structured mapping between the security guidelines and the AITDs they address, while Figure~\ref{fig:secguides-aitd} visually illustrates their interconnections. Together, these artifacts enable practitioners and researchers to understand how targeted security practices can reduce the accumulation of technical debt across the AI lifecycle and strengthen the robustness, trustworthiness, and sustainability of AI-enabled systems.

\subsubsection{Configuration Management}
Robust configuration management ensures that all AI components—models, pipelines, hyperparameters, dependencies, and deployment environments—are consistently versioned, controlled, and auditable throughout the lifecycle~\cite{ISOIEC27002, simonetta2024iso, soundararajansecure}. In AI-enabled systems, where small changes in data sources, hyperparameters, or library versions can significantly alter model behavior, disciplined configuration practices are essential to maintain reproducibility, traceability, and secure rollback in case of failures or regressions. Centralizing configuration artifacts (e.g., as code, in registries, or configuration databases) also supports governance, compliance, and cross-team collaboration in complex MLOps settings.

\begin{itemize}
\item Example: An MLOps pipeline stores configuration snapshots (hyperparameters, model versions, data schemas) in a managed registry before every retraining cycle, ensuring reproducibility.
\item Related AITDs: Configuration Debt (uncontrolled parameter drift), Versioning Debt (untracked changes across updates), Pipeline Jungle (inconsistent configurations across stages), Compatibility Debt (misaligned module configurations).
\end{itemize}

\subsubsection{Encryption and Secure Data Storage}

Encryption and secure data storage are foundational security practices for AI systems, ensuring that all sensitive data—during collection, preprocessing, model training, evaluation, and deployment—is protected from unauthorized access, tampering, or interception. This guideline encompasses two complementary aspects: encryption of data at rest and in transit and the secure storage of training, validation, and retraining datasets, especially those containing personally identifiable or confidential information. Compliance-focused AI systems must implement industry-grade cryptographic standards (e.g., AES-256 for stored data, TLS 1.2+ for data in transit), robust key-management procedures, and controlled access to secure storage environments. These safeguards align with regulatory requirements such as GDPR and ISO 27002 by ensuring confidentiality, integrity, and long-term protection of sensitive data throughout the AI lifecycle \cite{shahriar2023survey, hamon2024three, schneider2024designing, Sujatha2023, Hjerppe2019265}.

\begin{itemize}
\item Example:
In a healthcare diagnostic AI system, patient records uploaded for model inference are transmitted over encrypted channels (TLS), while historical datasets used for retraining are stored in encrypted repositories (AES-256) with strict access controls. This prevents unauthorized access, accidental exposure, or inference-time data leakage.

\item Related AITDs:
This guideline mitigates several data- and configuration-related debts, including: \textit{Data Debt / Unstable Data Dependencies} — securing data flows reduces exposure to compromised or inconsistent data sources. \textit{Configuration Debt} — enforcing proper cryptographic settings, key rotation, and secure storage parameters prevents misconfigurations that weaken system security. \textit{Documentation Debt} — requiring well-documented encryption policies and data-handling procedures enhances traceability and audit readiness.  \textit{Versioning Debt} — secure key management and versioning of encrypted assets prevent stale or misaligned cryptographic states.

\end{itemize}

\subsubsection{Continuous Behavioral \& Performance Monitoring}
This involves the systematic, real-time observation of AI system behavior to ensure that models operate within expected ethical, functional, and performance boundaries. This unified guideline integrates two complementary monitoring practices:  
(i) continuous behavioral monitoring, which evaluates adherence to ethical, fairness, and security policies; and  
(ii) performance monitoring, which tracks accuracy, drift, degradation, and operational reliability over time.  
Together, these monitoring practices enable early detection of anomalies, misconfigurations, emerging biases, model drift, or hidden system interactions that may compromise system integrity or safety \cite{lu2024responsible, hamon2024three, bogner2021characterizing, ISOIEC27002, Feng2022, Guo2025197}.  
Continuous monitoring is especially critical for AI systems deployed in dynamic or high-stakes environments where data distributions evolve and decision quality must remain consistently reliable.
\begin{itemize}
    \item Example:
    In a real-time fraud detection system, continuous monitoring evaluates whether model predictions remain fair and unbiased across customer groups, while performance monitoring checks for drops in detection accuracy that may indicate emerging fraud patterns or model drift. Alerts are triggered when behavioral or performance deviations exceed thresholds, prompting retraining, recalibration, or human review.
    
    \item Related AITDs:  
    This guideline mitigates multiple AITDs, including:  
    \emph{Ethical Debt} — by continuously checking for fairness, bias, or ethical violations in model outputs;  
    \emph{Configuration Debt} — by detecting misconfigurations or runtime deviations that affect behavior or safety;  
    \emph{Unstable Data Dependencies (Data Debt)} — by identifying drift or changes in upstream data that degrade model performance;  
    \emph{Hidden Feedback Loops} — by surfacing recurrent or cyclic interactions that cause unexpected model behaviors; and  
    \emph{Performance or Design Debt} — by detecting accuracy drops, latency spikes, or reliability issues early in the AI lifecycle.
\end{itemize}

\subsubsection{Data Validation}
This guideline requires that all input data feeding AI pipelines undergo strict validation checks for correctness, format consistency, completeness, and authenticity~\cite{sheeba2025decentralized, hamon2024three, ranjitsingh2025establish}. Strong validation reduces the risk of corrupted, noisy, or malicious data introducing instability.

\begin{itemize}
    \item Example: Before retraining a recommendation system, automated validators check for missing fields, incorrect data types, and out-of-distribution samples, preventing corrupted data from degrading model accuracy.
    
    \item Related AITDs: Data Debt (unvetted data introduces drift and errors), Pipeline Jungle (complex pipelines lacking validation checkpoints), Correction Cascades (invalid data propagating errors downstream). 
\end{itemize}

\subsubsection{Secure Coding Practices}
This guideline emphasizes applying standard secure-software engineering principles—such as static analysis, dependency scanning, code reviews, and avoiding hard-coded secrets—to AI pipelines and model-serving infrastructure~\cite{ISOIEC27002, simonetta2024iso, loncar2024secure}.

\begin{itemize}
    \item Example: During deployment, static code analysis flags insecure API endpoints and outdated dependencies used in the inference server, prompting corrective refactoring.
    
    \item Related AITDs: Glue Code Debt (quick fixes introducing vulnerabilities), Duplicate Model Code (security flaws replicated across modules), Unwanted Debugging Code (leftover logs leaking sensitive info), Defect Debt (known issues not addressed).

\end{itemize}

\subsubsection{Incident Response for AI Systems} Incident response involves establishing structured procedures to detect, investigate, contain, and recover from security incidents affecting AI workflows~\cite{jawhar2024ai, matsuda2019cyber, hamon2024three, jones2025analysing}. This includes AI-specific events such as adversarial data poisoning, unauthorized model access, or misconfiguration-induced failures.

\begin{itemize}
    \item Example: In a fraud-detection AI system, if an attacker injects adversarial inputs that manipulate prediction thresholds, an incident response workflow automatically detects anomalies, isolates compromised components, and triggers human review.
    
    \item Related AITDs: Hidden Feedback Loops (undetected abnormal system behavior), Unstable Data Dependencies (data poisoning events), Configuration Debt (incidents caused by misconfigurations). 
\end{itemize}

\subsubsection{Threat Intelligence} It involves the continuous collection, analysis, and application of information about emerging vulnerabilities, attack vectors, system misuses, and adversarial behaviors relevant to AI-enabled systems. In the context of AI security, threat intelligence extends beyond traditional cybersecurity practices by incorporating AI-specific risks such as data poisoning, model evasion, adversarial perturbations, prompt injection, and manipulation of training or inference pipelines. By proactively monitoring threat landscapes—including AI model repositories, dependency ecosystems, upstream data sources, and adversarial research communities—organizations can anticipate misconfigurations, model weaknesses, and infrastructural exposures that could manifest as AI Technical Debt (AITD).  
Integrating actionable threat intelligence into AI development pipelines enhances preparedness, supports risk-informed decision-making, and ensures that AI components evolve in alignment with emerging security requirements~\cite{ISOIEC27002, camilo2024ai, sarker2024introduction}.

\begin{itemize}
    \item {Example:}  
    A financial-services AI platform incorporates threat intelligence feeds that track newly discovered vulnerabilities in ML libraries (e.g., TensorFlow, PyTorch), recent data poisoning campaigns detected in public datasets, and adversarial evasion techniques emerging from academic research. When a critical vulnerability is reported in a dependency, the system automatically flags relevant components, triggering audits, dependency upgrades, and revalidation of affected models.

    \item {Related AITDs:}  
    Threat Intelligence mitigates a range of debt types, including:  
    \emph{Dispensable Dependency} — by detecting obsolete or vulnerable dependencies that should be removed or replaced;  
    \emph{Scattered Use of ML Libraries (SML)} — by identifying fragmented or inconsistent library usage that increases exposure to security flaws;  
    \emph{Configuration Debt} — by uncovering unsafe or outdated configuration settings tied to evolving threat patterns;  
    \emph{Data Debt / Unstable Data Dependencies} — by detecting compromised or manipulated upstream data sources;  
    \emph{Boundary Erosion and Compatibility Debt} — by highlighting insecure integrations or outdated interfaces that may become attack vectors;  
    \emph{Documentation and Versioning Debt} — by reinforcing the need for traceability of updates, patches, and threat responses across AI lifecycle artifacts.
\end{itemize}

\subsubsection{Red Teaming and AI Alignment (RLHF)}
Red teaming systematically tests AI systems using adversarial, stress-testing, or probing techniques to expose vulnerabilities. RLHF (Reinforcement Learning from Human Feedback) aligns model behavior with safe and acceptable outcomes, reducing misuse and unintended behaviors~\cite{spelda2025security, ouyang2022training, bai2022training, walter2024red, feffer2024red}.

\begin{itemize}
    \item Example: Before deployment, a large language model undergoes red-teaming sessions where experts test harmful prompts, manipulations, and jailbreak attempts. RLHF is then used to retrain the model and correct unsafe behaviors.
    
    \item Related AITDs: Ethical Debt (lack of alignment with societal norms), Algorithmic Inclination / Bias (unchecked bias requiring feedback alignment), Overly Simplified Metrics (models optimized with insufficient safety metrics).

\end{itemize}

\subsubsection{Data Leak Prevention (DLP)}
Data Leak Prevention involves monitoring, filtering, and controlling the flow of sensitive data inside AI pipelines to prevent accidental exposure or exfiltration~\cite{ISOIEC27002, simonetta2024iso, jaeyalakshmi2023self, alneyadi2016survey}.

\begin{itemize}
    \item Example: A model-training cluster blocks outbound file transfers containing sensitive patterns (e.g., credit card numbers), preventing unintended dataset exposure.
    \item Related AITDs: Data Debt (poorly governed sensitive data), Documentation Debt (missing data-handling guidelines), Dispensable Dependency (unnecessary third-party components leaking data), Undeclared Consumers (unmonitored data usage).
\end{itemize}

\subsubsection{Data Masking \& Information Deletion}
Data masking anonymizes sensitive data used in training or inference. Information deletion ensures obsolete or risky data is securely removed, minimizing retention-related risk~\cite{ISOIEC27002, simonetta2024iso, gaddam2024ai, bilakantisecure}.
\begin{itemize}
    \item Example: An AI system masks personally identifiable information (PII) before performing analytics and automatically deletes expired training logs after a retention window.
    
    \item Related AITDs: Data Debt (unmasked or outdated data remains in the pipeline), Unstable Data Dependencies (stale data causing unexpected behavior), Process/Infrastructure Debt (missing secure-deletion procedures), Documentation Debt (unclear retention policies).
\end{itemize}

\subsubsection{Controlled AI Output Consumers}
This guideline ensures that outputs produced by AI models—predictions, scores, embeddings, or generated content—are only accessed by explicitly authorized systems, services, or users \cite{bogner2021characterizing, recupito2024technical, martinez2022software}. Enforcing strict consumption policies prevents downstream components from silently depending on AI outputs, which can introduce hidden couplings, regulatory violations, and unmonitored propagation of errors or sensitive information. Controlled output consumption also supports auditability by requiring explicit logging, authentication, and authorization checks whenever AI outputs are requested or consumed. This reduces the risk of unintended reuse, shadow pipelines, or unauthorized integrations that frequently lead to systemic vulnerabilities in AI-driven architectures.
\begin{itemize}
\item Example:
In an AI-driven cybersecurity monitoring platform, only vetted detection modules and SOC (Security Operations Center) tools can access threat-classification outputs through authenticated API calls. Unauthorized services attempting to query the AI’s outputs are automatically blocked and logged, preventing silent dependencies or accidental leakage of sensitive threat intelligence.

\item Related AITDs:
\emph{Undeclared Consumers} — prevents hidden or unauthorized modules from silently consuming AI outputs; 
\emph{Glue Code Debt} — reduces ad-hoc integrations that bypass approved consumption pathways;  
\emph{Documentation Debt} — requires clear documentation of permitted consumers and access rules.

\end{itemize}

\subsubsection{Model Watermarking (Model Integrity Controls)}
Model watermarking embeds identifiers or verification signatures into AI models to detect unauthorized replication, tampering, or misuse~\cite{pooyandeh2022cybersecurity, banerjee2025securing, mekhfioui2025optimized, narula2025exploring}. This strengthens model integrity and provides forensic traceability.


\begin{itemize}
    \item Example: A commercial NLP model embeds a proprietary watermark into its weight matrices. If leaked or cloned, developers can verify fingerprints to identify unauthorized deployments.
    
    \item Related AITDs: Undeclared Consumers (untracked or unauthorized model usage), Versioning Debt (lack of traceability enabling misuse), Documentation Debt (missing model-ownership records), Glue Code Debt (uncontrolled integrations facilitating illicit replication).
\end{itemize}

\subsubsection{AI System Software Bill of Materials (AI-SBOM)}
An AI System Software Bill of Materials (AI-SBOM) is a machine-readable inventory that lists all datasets, model versions, libraries, configurations, and dependencies used across the AI lifecycle \cite{rajbahadur2025building, xia2025operationalising, sood2025malicious}. It provides end-to-end traceability of how models were trained, what data they relied on, and which components they depend on, thereby strengthening transparency, auditability, and supply-chain security. AI-SBOMs also support rapid vulnerability assessment when flaws or dataset integrity issues arise \cite{Xia20232630}.

\begin{itemize}
\item Example:
In a healthcare AI system, an AI-SBOM records training datasets, preprocessing steps, model checkpoints, and library versions, enabling auditors to verify compliance with GDPR and medical-device regulations.
\item Related AITDs:
\emph{Data Debt / Unstable Data Dependencies} — through clear provenance tracking.  
\emph{Dispensable Dependency} — by identifying outdated or unnecessary components.  
\emph{Versioning Debt} — by documenting model and dependency histories.  
\emph{Documentation Debt} — by formalizing system lineage and configuration details.
\end{itemize}

\subsubsection{Modular \& Reusable AI Components}
This guideline emphasizes designing AI systems using modular, self-contained components with clearly defined interfaces and separation of concerns \cite{schneider2024designing, Martínez-Fernández2022, riggio2021ai, celepija2025towards}. Modular architectures minimize unnecessary coupling between data pipelines, model logic, and service layers, making AI systems easier to test, secure, maintain, and evolve. Reusable components also promote architectural consistency across projects, reduce duplication, and lower the risk of integrating ad-hoc patches or experimental code that later becomes technical debt.

\begin{itemize}
\item Example:
In an AI-based fraud detection platform, independently developed modules for feature extraction, model inference, and case scoring can be reused across credit card fraud, loan fraud, and identity-theft scenarios. This avoids rewriting redundant logic and enables secure updates without impacting unrelated components.
\item Related AITDs:  
\emph{Glue Code Debt} — modular interfaces reduce the need for brittle integration scripts.  
\emph{Deep God File} — decomposition prevents monolithic files accumulating multiple responsibilities.  
\emph{Design Debt} — modularity enforces cleaner architectural boundaries.  
\emph{Duplicate Model Code} — reusable components reduce redundant implementations across the system.
\end{itemize}

\subsubsection{Standardized \& Interoperable Data Formats}
This guideline promotes the use of consistent, well-defined, and interoperable data formats across all stages of the AI pipeline \cite{bogner2021characterizing, gujar2025data, zhi2024algorithm, naeem2021trends}. Standardization minimizes the need for ad-hoc data conversions and reduces the likelihood of mismatches between feature schemas, model inputs, and downstream services. By enforcing shared data conventions—such as unified schemas, typed data structures, consistent encodings, and versioned dataset specifications—organizations can prevent integration failures, enhance security by reducing transformation-related vulnerabilities, and improve long-term maintainability of AI workflows.
\begin{itemize}
\item Example:
In a smart city AI platform aggregating data from traffic sensors, public transit feeds, and environmental monitors, enforcing interoperable formats (e.g., standardized timestamps, unified geospatial encodings, consistent JSON schemas) ensures that downstream prediction models can reliably process data without error-prone conversion scripts.

\item Related AITDs:  
\emph{Glue Code Debt} — reduces the need for excessive data wrangling or conversion scripts used to bridge incompatible formats;  
\emph{Unstable Data Dependencies (Data Debt)} — ensures predictable, consistent data flows across heterogeneous systems;  
\emph{Compatibility Debt} — avoids failures originating from mismatched data structures between models, services, or pipeline components.

\end{itemize}

\subsubsection{API-Governed AI Access Control}
This guideline promotes accessing AI models and components exclusively through well-governed, authenticated, and rate-limited APIs rather than distributing models for local execution \cite{siriwardena2019advanced, paidy2024securing, lu2024responsible, hamon2024three}. Centralizing access through APIs strengthens security by enforcing fine-grained permissions, monitoring usage patterns, applying request validation, and preventing unauthorized or uncontrolled consumption of AI outputs. It also reduces the attack surface by avoiding model proliferation across uncontrolled environments, enabling organizations to maintain better visibility, governance, and lifecycle control over deployed AI capabilities.
In addition, API-mediated access supports compliance and auditability, since all interactions with the AI system can be logged, attributed, and reviewed, ensuring that AI predictions and model updates remain traceable and accountable.


\begin{itemize}
\item Example:
An AI-based weather prediction model is deployed behind a secure API gateway. External partners in agriculture, aviation, and logistics can query predictions through authenticated API calls, without receiving direct access to model weights or internal logic. This prevents model theft, unauthorized reuse, and uncontrolled distribution, while enabling centralized monitoring and permission enforcement.

\item Related AITDs:
\emph{Undeclared Consumers} — by ensuring AI outputs are accessed only by authorized systems and recorded in access logs.  
\emph{Dispensable Dependency} — by reducing the need to embed or replicate models locally, avoiding unnecessary third-party integrations that create security and maintenance burdens.  
\emph{Glue Code Debt} — by minimizing ad-hoc local integrations and promoting standardized, well-defined API interfaces.  
\emph{Versioning Debt} — by enabling centralized version control of models and ensuring clients always access the correct model version through the API.

\end{itemize}

\subsubsection{Adequate AI Resource \& Continuity Management} This focuses on ensuring that AI systems are provisioned with adequate computational resources—such as CPU/GPU capacity, memory, storage, and network bandwidth—while also maintaining organizational readiness for operational continuity in the event of outages, failures, or unexpected load surges. This guideline combines two complementary security and reliability practices: (i) resource sufficiency, which supports stable and timely AI model execution, and (ii) ICT readiness for business continuity, which ensures fallback mechanisms, redundancy, and systematic recovery procedures are in place to maintain secure and safe AI operation even under adverse conditions \cite{bogner2021characterizing, shneiderman2020bridging, shahriar2023survey, ISOIEC27002}.

\begin{itemize}
    \item Example:  
    In an autonomous driving system, ensuring sufficient GPU acceleration and memory bandwidth allows AI models to process multimodal sensor inputs (e.g., LiDAR, radar, cameras) with millisecond latency. Simultaneously, redundancy in compute nodes and a business continuity plan (e.g., automatic failover to a secondary control module) ensures that vehicle perception and decision-making continue uninterrupted during hardware faults or peak load conditions.
    
    \item Related AITDs:  
    This guideline mitigates several AITDs, including:  
    \emph{Configuration Debt} — by ensuring proper resource provisioning and avoiding misconfigured system parameters that lead to performance bottlenecks;   and 
    \emph{Process/Infrastructure Debt} — by strengthening operational readiness, redundancy, and failover strategies.
\end{itemize}

\subsubsection{AI Component Divergence Monitoring}
This guideline focuses on the continuous detection and analysis of behavioral discrepancies between redundant AI components or between an AI system’s expected outputs and its actual runtime behavior \cite{aagaard2024discrepancies, renard2024understanding, lu2024responsible}. Such divergence may signal a variety of underlying issues—including data quality problems, configuration drift, adversarial interference, model degradation, or faulty integration logic. By systematically monitoring for mismatches, organizations can identify anomalies at an early stage, initiate corrective action, and prevent security- or safety-critical consequences. Divergence monitoring is especially important in safety-critical and real-time AI systems, where silent failures or unnoticed drift can propagate downstream and trigger cascading errors.

\begin{itemize}
\item Example: In an AI-driven industrial control system, if two redundant models controlling a robotic arm produce different movement instructions, this discrepancy would trigger an alert for human review. The system may automatically revert to a safe fallback mode or require operator approval before executing actions, thereby preventing accidental equipment damage or unsafe operations.
\item Related AITDs: Helps mitigate \emph{Unstable Data Dependencies} by catching inconsistencies in upstream data or model behavior early, and \emph{Correction Cascades} by preventing erroneous or misaligned outputs from propagating through interconnected components in the AI pipeline.
\end{itemize}

\subsubsection{Multi-Model Consensus Decision Framework}
This guideline advocates deploying multiple AI models in parallel—each trained with different architectures, datasets, feature sets, or optimization strategies—to independently generate decisions that are then compared to derive a consensus \cite{lu2024responsible,schneider2024designing, shahin2025automated, gupta2025optimizing}. By diversifying the decision-making process across multiple models, the system becomes inherently more fault-tolerant, reducing susceptibility to single-model failures, dataset biases, adversarial perturbations, or unexpected distribution shifts.
Consensus-based decision mechanisms strengthen the robustness of AI-driven systems by identifying anomalous outputs early, isolating underperforming models, and enabling ensemble-level validation \cite{Dang2022192}.
Moreover, multi-model consensus supports adaptive system design, allowing continuous improvement by monitoring divergence patterns across models and informing when retraining, recalibration, or model retirement is necessary.

\begin{itemize}
\item Example:
In healthcare diagnostics, separate AI models—such as a CNN, a transformer-based classifier, and a segmentation-based model—can independently analyze radiographic images. If one model generates an output that deviates from the majority consensus, the system automatically flags the case for manual review, ensuring higher reliability and preventing diagnostic errors.

\item Related AITDs:
Mitigates \emph{Correction Cascades} by preventing downstream components from acting on a faulty single-model output. Addresses \emph{Hidden Feedback Loops} by distributing decision-making across diverse model pathways, reducing the risk that self-reinforcing errors or biased predictions propagate through the pipeline without detection.

\end{itemize}

\subsubsection{AI Redundancy \& Failover Mechanisms}
This guideline emphasizes the deployment of multiple identical—or functionally equivalent—AI components operating in parallel to provide resilience, robustness, and continuity under failure conditions \cite{tutuncuoglu2024zero, musunuru2025optimizing, gupta2025gpu}. By maintaining redundant AI modules, the system can seamlessly transition to a backup component whenever the primary model encounters unexpected behavior, hardware degradation, adversarial interference, or operational faults.
Failover mechanisms are essential in high-availability and safety-critical environments, where a single point of AI failure can produce cascading errors or compromise system safety. Redundancy reduces operational risk by ensuring that model predictions remain stable even when one instance becomes compromised, overloaded, or misconfigured. It also enables cross-validation between identical components, improving anomaly detection and strengthening reliability during runtime. 
Beyond resilience, redundancy supports secure and responsible AI deployment by enabling rolling updates, safe model rollbacks, and phased testing of new model versions. This ensures that experimental or updated AI components do not destabilize system behavior, as stable redundant units remain available to maintain uninterrupted operation. Collectively, redundancy and failover strategies enhance system robustness, protect against adversarial failure modes, and uphold service continuity.
\begin{itemize}
\item Example:
In a smart grid energy distribution system, multiple redundant AI controllers continuously monitor load patterns and grid conditions. If the primary controller is compromised—due to a cyberattack, hardware fault, or sensor malfunction—the secondary controller instantly assumes control, preventing blackouts, safety hazards, or cascading failures across the grid.

\item Related AITDs:
Mitigates \emph{Algorithmic Inclination / Human Bias} by reducing dependence on a single algorithmic decision-maker. Addresses \emph{Unstable Data Dependencies} by ensuring that failures in one data pathway do not compromise the entire system, as redundant components can operate on parallel or validated data streams.

\end{itemize}

\subsubsection{Use of Hardened Models and Defensive Training}
This guideline advocates for using models trained with defensive mechanisms such as adversarial training, robust optimization, or certified defenses to increase resilience against malicious perturbations~\cite{hamon2024three, olutimehin2025adversarial, samuel2025adversarial}.

\begin{itemize}
    \item Example: An image classifier used in a biometric access system is trained with adversarially perturbed samples, increasing robustness against evasion attacks that slightly manipulate facial images.
    
    \item Related AITDs: Algorithmic Inclination / Human Bias (models optimized only for accuracy, not robustness), Entanglement (high interdependencies amplifying vulnerabilities), Correction Cascades (unrobust models propagating errors), Design Debt (architectures not built for robust defense).
\end{itemize}

\subsubsection{Input Sanitization (Defense Against Evasion Attacks)} This involves applying filters, normalization techniques, and anomaly detection to remove adversarial perturbations or malicious payloads~\cite{hamon2024three, barlas2022exploiting, kiranbabu2025challenge}. This is critical for mitigating evasion attacks targeting model vulnerabilities.

\begin{itemize}
    \item Example: A vision-based authentication model normalizes pixel distributions and rejects suspiciously manipulated images before inference, reducing exposure to adversarial examples designed to bypass access controls.
    
    \item Related AITDs (not sure - double check): Algorithmic Inclination / Human Bias (models not trained to handle edge cases), Entanglement (tight coupling amplifying adversarial effects), Correction Cascades (malicious inputs causing downstream failures), Defect Debt (unhandled edge cases left unresolved).
\end{itemize}

\subsubsection{Ethical Black Box}
This is a comprehensive logging mechanism that captures AI system inputs, outputs, intermediate signals, model states, and the reasoning (where available) behind decisions \cite{lu2024responsible, schneider2024designing, franzoni2023black, pedreschi2019meaningful}. Inspired by aviation black boxes, this mechanism provides a continuous, tamper-evident record of the AI system’s behavior. Such detailed traceability is essential for post-incident analysis, forensic auditing, compliance verification, and diagnosing failures—particularly in high-stakes domains involving safety, fairness, or accountability \cite{Bélisle-Pipon20231507}.
By enabling stakeholders to reconstruct decision pathways, the Ethical Black Box strengthens transparency and supports ethical AI governance. It also facilitates proactive oversight by making deviations from expected behavior detectable long before they propagate into harmful system outcomes.

\begin{itemize}
\item Example:
In an AI-driven financial system, an ethical black box records every loan-approval decision, including input features (e.g., credit score, income), model-inferred risk factors, confidence scores, and decision rationale. During audits or dispute resolution, this record allows regulators or internal teams to evaluate whether decisions were fair, explainable, and aligned with organizational and legal requirements.
\item Related AITDs:
Mitigates \emph{Ethical Debt} by providing the transparency and traceability needed to assess fairness, detect discrimination, and verify decision legitimacy.  
Addresses \emph{Entanglement} by offering clear logs that disentangle complex interactions between system components.  
Mitigates \emph{Undeclared Consumers} by recording which systems or entities accessed or consumed AI outputs, ensuring accountability and preventing unauthorized use.
\end{itemize}

\subsubsection{Human–AI Control Mode Switching}
This guideline introduces a configurable mechanism that allows users or system operators to dynamically determine whether AI-generated outputs should be executed autonomously or treated as advisory suggestions requiring human confirmation before action \cite{lu2023responsible, lu2022software, ge2024mode, lu2023developing}. By enabling seamless switching between automated and human-in-the-loop modes, this guideline enhances operational transparency, prevents over-reliance on automated decisions, and ensures that critical tasks remain under appropriate levels of human oversight. Such control mechanisms are especially vital in high-stakes or safety-critical environments—such as transportation, healthcare, finance, and industrial automation—where fully autonomous operation may expose the system to unacceptable levels of risk or error propagation.
Human–AI control mode switching also provides an additional safeguard against unexpected system behaviors, model drift, or misinterpretations by ensuring that human judgment can override or validate AI outputs before execution.

\begin{itemize}
\item Example:
In an autonomous vehicle, the mode-switching mechanism allows the driver to specify whether the AI should automatically initiate lane changes or merely suggest them, enabling manual review in dense traffic or hazardous conditions. This ensures that the human operator retains situational awareness and can intervene when contextual nuances exceed the model’s competence.
\item Related AITDs: 
Mitigates \emph{Entanglement} by providing a clear separation between AI-generated recommendations and their execution, reducing the risk of tightly coupled autonomous decisions leading to cascading failures. Also addresses \emph{Undeclared Consumers} by ensuring that AI outputs are not silently consumed or acted upon without explicit human approval, thereby improving transparency and accountability in how AI-driven decisions are used.

\end{itemize}

\subsubsection{Unified Programming Language Policy}
Adopting a single, standardized programming language across the AI development lifecycle reduces fragmentation in implementation practices and streamlines system-wide security, testing, and audit processes \cite{yang2024multi, bogner2021characterizing, Martínez-Fernández2022}. When AI pipelines depend on multiple languages—such as Python for modeling, R for analytics, and Java or C++ for deployment—teams face increased complexity in dependency management, debugging, and vulnerability scanning. A unified programming language policy improves maintainability, enhances developer productivity, supports more consistent security hardening, and reduces integration overhead by minimizing cross-language interoperability challenges.

\begin{itemize}
\item Example:
In an AI-based e-commerce recommendation system, standardizing all components—feature engineering, model training, batch scoring, and API serving—on Python enables consistent dependency management (e.g., pip/conda), uniform static analysis, and streamlined deployment pipelines, lowering the risk of language-specific vulnerabilities or mismatched runtime environments.

\item Related AITDs:  
\emph{Multiple Language Smells (MLS)} — reduces fragmentation and eliminates hard-to-maintain cross-language boundaries;  
\emph{Glue Code Debt} — minimizes the need for connectors or wrappers between mismatched languages;  
\emph{Compatibility Debt} — avoids inconsistencies in runtime environments and library ecosystems caused by heterogeneous language stacks.
\end{itemize}

\subsubsection{AI-Adaptive Secure Design Practices}
This guideline emphasizes the integration of AI-specific ethical, security, and operational requirements into established system design methodologies, such as Unified Modeling Language (UML), SysML, workflow diagrams, or architectural blueprints \cite{shahriar2023survey, ebad2022exploring, spelda2025security, patel2021creating}. Traditional design methods often fall short in representing the unique risks of AI components—such as model drift, data dependency volatility, adversarial susceptibility, or fairness constraints. By adapting these methods to explicitly model AI behaviors, data flows, decision boundaries, and governance constraints, organizations can surface potential vulnerabilities early in the design phase. This ensures that AI-driven components are not treated as black-box add-ons but as first-class citizens whose ethical, safety, and security requirements shape the architecture from its inception.

Embedding AI-aware design practices also supports cross-team alignment by enabling data engineers, ML researchers, and software architects to reason about system interactions holistically. Early modeling of AI-specific design concerns reduces downstream refactoring, minimizes architectural inconsistencies, and strengthens long-term system maintainability.

\begin{itemize}
\item Example:
When architecting an AI-powered traffic management system, UML component diagrams may explicitly represent fairness constraints, model-update workflows, input-validation checkpoints, and monitoring hooks. Including these aspects early ensures the final system adheres to ethical expectations, safety constraints, and regulatory requirements.

\item Related AITDs:  
\emph{Design Debt} — mitigated by embedding robust architectural principles and ethical/security considerations from the outset;  
\emph{Boundary Erosion} — reduced by clearly representing AI component boundaries and interfaces;  
\emph{Configuration Debt} — minimized by documenting AI-specific configuration parameters within formal design artifacts;  
\emph{Entanglement} — avoided by designing modular, well-isolated AI components with explicitly defined interactions.

\end{itemize}

\begin{table*}[htbp]
\small
\centering
\caption{Mapping of Security Guidelines to Related AI Technical Debts (AITDs)}
\begin{tabular}{c p{3.6cm} p{5.2cm} p{7.2cm}}
\toprule
\textbf{S/N} & \textbf{Security Guideline} & \textbf{Related AITDs} & \textbf{Explanation} \\
\midrule

1 & Configuration Management & Configuration Debt, Versioning Debt, Pipeline Jungle, Compatibility Debt & Ensures consistent, auditable, and controlled configurations across models, pipelines, dependencies, and environments. \\
\hline

2 & Encryption and Secure Data Storage & Data Debt, Configuration Debt, Documentation Debt, Versioning Debt & Protects sensitive data via encryption at rest/in transit, secure storage, and proper cryptographic configuration. \\
\hline

3 & Continuous Behavioral \& Performance Monitoring & Ethical Debt, Configuration Debt, Data Debt, Hidden Feedback Loops, Design Debt & Detects drift, misconfigurations, unfair outcomes, and anomalous behaviors. \\
\hline

4 & Incident Response for AI Systems & Hidden Feedback Loops, Unstable Data Dependencies, Configuration Debt, SATD & Enables structured detection, containment, and recovery from AI-specific incidents (e.g., poisoning). \\
\hline

5 & Data Validation & Data Debt, Pipeline Jungle, Correction Cascades & Ensures input data is correct, consistent, authentic, and complete before entering AI pipelines. \\
\hline

6 & Input Sanitization (Defense Against Evasion Attacks) & Algorithmic Inclination, Entanglement, Correction Cascades, Defect Debt & Removes adversarial inputs and malicious payloads to prevent evasion attacks. \\
\hline

7 & Model Watermarking (Model Integrity Controls) & Undeclared Consumers, Versioning Debt, Documentation Debt, Glue Code & Detects tampering, unauthorized use, and illicit replication of models. \\
\hline

8 & Use of Hardened Models \& Defensive Training & Algorithmic Inclination, Entanglement, Correction Cascades, Design Debt & Improves robustness via adversarial training and defense-oriented modeling. \\
\hline

9 & Red Teaming and AI Alignment (RLHF) & Ethical Debt, Algorithmic Inclination, Overly Simplified Metrics, SATD & Identifies vulnerabilities and aligns model behavior with safe and ethical outcomes. \\
\hline

10 & Adequate AI Resource \& Continuity Management & Configuration Debt, Process/Infrastructure Debt & Ensures AI systems have reliable resources and continuity plans for safe operation. \\
\hline

11 & Secure Coding Practices & Glue Code, Duplicate Model Code, Unwanted Debugging Code, Defect Debt & Enforces secure development, dependency scanning, and removal of insecure code artifacts. \\
\hline

12 & Data Leak Prevention (DLP) & Data Debt, Documentation Debt, Dispensable Dependency, Undeclared Consumers & Prevents accidental or malicious exfiltration of sensitive data. \\
\hline

13 & Data Masking \& Information Deletion & Data Debt, Unstable Data Dependencies, Process/Infrastructure Debt, Documentation Debt & Anonymizes and securely deletes sensitive/obsolete data to reduce risk. \\
\hline

14 & Threat Intelligence & Dispensable Dependency, SML, Configuration Debt, Data Debt, Boundary Erosion, Compatibility Debt, Documentation/Versioning Debt & Provides early awareness of AI-specific threats and emerging vulnerabilities. \\
\hline

15 & Human–AI Control Mode Switching & Entanglement, Undeclared Consumers & Enables human oversight, preventing unintended or unsafe autonomous decisions. \\
\hline

16 & Multi-Model Consensus Decision Framework & Correction Cascades, Hidden Feedback Loops & Uses diverse models to detect inconsistencies and avoid cascading failures. \\
\hline

17 & AI Redundancy \& Failover Mechanisms & Algorithmic Inclination, Unstable Data Dependencies & Eliminates single-model or single-data-source failure points. \\
\hline

18 & Ethical Black Box & Entanglement, Undeclared Consumers, Ethical Debt & Enhances traceability and accountability for AI decisions. \\
\hline

19 & AI Component Divergence Monitoring & Correction Cascades, Unstable Data Dependencies & Detects divergence across redundant components. \\
\hline

20 & Controlled AI Output Consumers & Undeclared Consumers, Glue Code, Documentation Debt & Prevents unauthorized or untracked consumption of AI outputs. \\
\hline

21 & API-Governed AI Access Control & Undeclared Consumers, Dispensable Dependency, Glue Code, Versioning Debt & Enforces controlled access to AI components and outputs. \\
\hline

22 & AI System Software Bill of Materials (AI-SBOM) & Data Debt, Dispensable Dependency, Versioning Debt, Documentation Debt & Tracks datasets, libraries, and model lineage for transparency and safety. \\
\hline

23 & Modular \& Reusable AI Components & Glue Code, Deep God File, Design Debt, Duplicate Model Code & Promotes modular designs that reduce coupling and simplify updates. \\
\hline

24 & Standardized \& Interoperable Data Formats & Glue Code, Unstable Data Dependencies, Compatibility Debt & Ensures interoperability and reduces risky conversions. \\
\hline

25 & Unified Programming Language Policy & Multiple Language Smells (MLS), Glue Code, Compatibility Debt & Reduces complexity and lowers cross-language security risks. \\
\hline

26 & AI-Adaptive Secure Design Practices & Design Debt, Boundary Erosion, Configuration Debt, Entanglement  & Embeds ethical and security considerations directly into AI system design. \\
\bottomrule
\end{tabular}
\label{tab:sec-guidelines-to-aitd}
\end{table*}

\begin{tcolorbox}[colframe=gray!80!black, colback=gray!5, coltitle=black, title=\textbf{RQ3.2: Security Guidelines for Mitigating AITDs}]

Through a systematic consolidation of industry best practices, standards, and research insights, the study proposes a comprehensive set of \textbf{26 security-oriented guidelines} designed to mitigate the risks associated with AI Technical Debt (AITD). These guidelines address vulnerabilities across the entire AI development and operational lifecycle, with emphasis on data pipelines, model training and deployment, system integration, and governance structures. The key thematic categories include:

\begin{itemize}
    \item \textbf{Transparency and Accountability Mechanisms}: Ethical Black Boxes, auditable decision logs, and traceability tools to support post-hoc analysis and regulatory compliance.

    \item \textbf{Human--AI Oversight and Control}: Human–AI mode switching, human-on-the-loop and human-in-the-loop procedures for high-stakes decisions, and manual override pathways.

    \item \textbf{Continuous Monitoring and Drift Management}: Behavioral monitoring, data drift and concept drift detection, performance tracking, and anomaly detection pipelines.

    \item \textbf{Resilience and Continuity Engineering}: Multi-model consensus, redundancy and failover mechanisms, fallback policies, and robustness techniques against adversarial or unexpected system behavior.

    \item \textbf{Secure Data Handling and Governance}: Encryption, access control, data masking, secure storage, automated deletion routines, and standardized data formats to reduce data-related AITDs.

    \item \textbf{Model Security and Integrity}: Watermarking, hardened models, adversarial and defensive training, and integrity validation checks across the model lifecycle.

    \item \textbf{Threat Intelligence and Incident Response}: AI-specific threat modeling, red-teaming, alignment testing, and rapid incident response workflows tailored for AI-enabled systems.

    \item \textbf{Configuration, Dependency, and Resource Management}: Version control for AI components, environment reproducibility, AI SBOM management, modularization, and resource continuity planning to reduce operational AITD accumulation.
\end{itemize}

\textbf{Overall Emphasis}: The presented set of 26 guidelines aims to enhance system resilience, reduce vulnerability exposure, and ensure that AI-enabled systems remain secure, transparent, and trustworthy throughout their lifecycle. These guidelines form the foundation for mitigating the diverse forms of AITD identified in this study.
\end{tcolorbox}

\clearpage
\section{Principal findings}
\label{sec:principalfindings}

This systematic review consolidates evidence from 60 primary studies and identifies 31 distinct forms of technical debt as they manifest in AI-enabled systems. Using grounded theory and a root-cause–oriented synthesis approach, these debts were organized into a unified taxonomy consisting of seven major classes: (1) Data \& Library–Related Debts, (2) Model \& Code–Related Debts, (3) Algorithm–Related Debts, (4) Design \& Architecture Debts, (5) Operational \& Lifecycle Debts, (6) Documentation \& Communication Debts, and (7) Testing \& Quality Assurance Debts. This classification shifts the perspective from traditional categorizations toward a cause-driven view, enabling deeper analysis of how AITDs originate, propagate, and affect downstream system behavior, quality, and risk.

\textbf{Root-Cause–Oriented Debt Landscape:}  
The taxonomy reveals that AITDs arise from heterogeneous origins across the AI development pipeline. \textit{Data \& Library–Related Debts}, such as Data Debt and Pipeline Jungle, remain dominant and often triggered by incomplete, biased, outdated, or inconsistently processed datasets. These debts impede reproducibility and increase the likelihood of bias amplification and model drift. 
\textit{Model \& Code–Related Debts} capture issues such as Model Entanglement, Model Complexity, and Glue Code, which introduce maintainability bottlenecks and obscure system behavior.  
\textit{Algorithm–Related Debts}, including Algorithmic Inclination or misaligned objective functions, reflect deeper issues with optimization choices or inadequately specified learning criteria.  
\textit{Design \& Architecture Debts} highlight structural shortcomings such as poor modularity, tightly coupled components, and undocumented dependencies, which impair scalability and evolution capability.  
\textit{Operational \& Lifecycle Debts} underscore challenges in deployment, monitoring, retraining, and model governance—areas increasingly recognized as critical in MLOps environments.  
Finally, \textit{Documentation \& Communication Debts} and \textit{Testing \& QA Debts} reveal persistent gaps in transparency, traceability, and verification, further weakening system accountability and reliability.
Collectively, this taxonomy illustrates that AITDs originate from systemic and interconnected root causes spanning data, model internals, architectural decisions, operational workflows, and organizational practices.

\textbf{Impact on System Quality Attributes:}  
Across the reviewed studies, AITDs were found to substantially degrade  qualities such as maintainability, performance, scalability, robustness, and efficiency. Debts related to data governance and model behavior often introduce latent risks—manifesting only during redeployment, domain shifts, or integration with new pipelines. Model Entanglement, for instance, impedes debugging and retraining, while missing or poor documentation complicates audits and compliance checks.  
The analysis also highlights strong interactions between AITDs and explainability, interpretability, and fairness—quality dimensions that carry unique significance in AI engineering. These impacts extend beyond typical software failures, affecting trustworthiness and societal acceptance of AI systems. Notably, the findings show that many AITDs have a cascading effect: errors in data processing propagate into models, which then propagate into downstream decisions, magnifying technical and ethical risks.

\textbf{Safety and Security Integration:}  
A core contribution of this study is the explicit mapping of the 31 AITDs to \textbf{6 safety} and \textbf{12 security} risks identified in the literature. Certain debts, such as Data Debt, Undeclared Consumers, and Correction Cascades, were repeatedly linked to vulnerabilities including adversarial susceptibility, data leakage, unreliable model reuse, and unstable behavior under distribution shifts.  Ethical and documentation-related debts were associated with failures in accountability, auditability, and safe human oversight. Operational debts further compromise resilience by impairing monitoring, retraining, and incident response workflows.  This mapping confirms that AITDs function not merely as engineering inefficiencies but as systemic vulnerabilities with direct implications for safety, compliance, and risk governance—highlighting the need to integrate AITD assessment with security engineering and safety certification processes.

\textbf{Mitigation Strategies:}  
The study synthesizes a total of \textbf{34 mitigation guidelines} - \textbf{8 safety guidelines} and \textbf{26 security} - designed to address the root causes and propagation pathways of AITDs. These include mechanisms such as:  
\textit{Human–AI Control Mode Switching}, \textit{Ethical Black Boxes for decision logging}, \textit{Continuous Behavioral and Drift Monitoring}, \textit{Model Watermarking and Hardening}, \textit{Redundancy and Failover Architectures}, \textit{Out-of-Distribution Detection}, and \textit{Formal Verification and Model Assurance} for high-risk deployments.  
Each guideline is mapped to one or more debt classes (Tables~\ref{tab:safety-guidelines-to-aitds}–\ref{tab:sec-guidelines-to-aitd}), creating a traceable link between AITD type, underlying cause, associated risks, and appropriate mitigation techniques. By grounding mitigation in root causes, the framework supports proactive debt prevention during design, development, deployment, and continuous operation.

\textbf{AITD-MAP Framework:}  
Building on the taxonomy and mappings, the study introduces the AITD-MAP (Mapping AI Technical Debt: Types, Impact, and Guidelines) framework. AITD-MAP integrates three dimensions: (i) the root-cause-oriented AITD taxonomy, (ii) the identified impacts on software quality, safety, and security, and (iii) the 34 mitigation guidelines. This unified model enables practitioners to diagnose debts based on symptoms, trace their propagation across pipelines, evaluate associated risks, and select targeted mitigation strategies. It also offers a conceptual foundation for future automated tools for AITD detection, monitoring, and refactoring in MLOps ecosystems.

\textbf{Prevalence and Prioritization:}  
The prevalence analysis shows that \textit{Data Debt/Unstable Data Dependencies} is the most frequently reported AITD, appearing in 27 of 60 studies (45\%). This is followed by \textit{Glue Code} (30\%), \textit{Test Debt} (28.33\%), and \textit{Documentation Debt} (25.33\%). Other commonly reported debts include \textit{Requirement Debt} (23.33\%), \textit{Design Debt} (21.67\%), and \textit{Configuration Debt} (20\%). 
These results indicate that data-related and model/code-related debts dominate current AI development challenges, particularly those associated with unstable data pipelines, ad-hoc integrations, prototype code, and insufficient testing practices. The relatively high frequency of documentation, requirement, and design debts further reflects the ongoing struggle to maintain traceability, architectural clarity, and consistent communication in rapidly evolving AI workflows.
Overall, the distribution of AITDs underscores that the most pervasive issues stem from the early stages of the AI pipeline (data and preprocessing), the implementation layer (model/code quality), and system-level coordination (requirements, documentation, and configuration). This prioritization highlights the need for context-aware debt management practices tailored to system maturity, domain criticality, and operational risk profiles.

\textbf{Synthesis and Implications:}  
Overall, this review provides the first root-cause–oriented, security-aware, and mitigation-driven characterization of AI Technical Debt. By connecting technical debt origin, systemic impact, and prescriptive mitigation, the study bridges a significant gap in the AI engineering literature. The findings emphasize that AITD management must be continuous, interdisciplinary, and transparent-spanning data governance, model design, architecture, testing, deployment, and organizational practices. Integrating AITD assessment into DevOps and MLOps workflows is essential for early detection and automated mitigation. This work lays a robust foundation for future empirical validations, tool development, and policy frameworks aimed at promoting sustainable, trustworthy, and resilient AI systems.

\subsection{Strengths and limitations}

This study demonstrates several notable strengths. It systematically identifies and categorizes 31 distinct types of AI Technical Debts (AITDs), offering a comprehensive taxonomy that captures their structural, operational, and ethical implications in AI-based systems. By emphasizing the intersection between AITDs and critical concerns—particularly security and safety—the study provides a nuanced understanding of how these debts affect system attributes such as maintainability, robustness, and resilience. The introduction of AITD-MAP (Mapping AI Technical Debt: Types, Impact, and Guidelines) further strengthens the contribution by organizing insights across taxonomy, quality attributes, and mitigation strategies into a coherent, actionable framework.

Another key strength lies in the proposed mitigation strategies. The study presents 34 security and safety guidelines that are explicitly mapped to the identified AITDs, offering practical pathways to address risks in AI system development and deployment. This integration of mitigation guidance into a structured framework represents a novel step toward responsible and sustainable AI engineering. Additionally, the application of grounded theory ensures a rigorous and data-driven process for categorization, enhancing the validity and reliability of findings. The review draws on 60 primary studies from diverse domains, reinforcing the generalizability of the findings across various AI use cases and development contexts.

However, the study is not without limitations. The reliance on selected academic databases (ACM Digital Library, IEEE Xplore, Scopus, and Springer) may have led to the exclusion of relevant studies from other specialized or emerging sources. The manual nature of the selection, coding, and classification process—though mitigated by cross-checking and author consensus—still poses a risk of human bias or oversight. Furthermore, the study's focus on recent literature (2015–2025) offers a timely synthesis but may omit insights from earlier foundational works on technical debt or AI safety.

While the inclusion of mitigation guidelines is a major strength, their practical applicability remains to be validated through industrial case studies or empirical trials. Additionally, some emerging debts—such as Correction Cascades and Boundary Erosion—appear less frequently in the literature, making their classification provisional and highlighting the need for further empirical investigation. Despite these limitations, this research significantly advances the field by providing a taxonomy-driven, security-aware, and mitigation-oriented perspective on AI Technical Debt. It sets a solid foundation for future empirical studies and practical tool development aimed at improving the quality, accountability, and resilience of AI-based systems.

\section{Conclusion and future work}
\label{sec:conclusion}

As AI-based systems increasingly operate in safety-critical and trust-sensitive environments, the accumulation of AI Technical Debt (AITD) poses a serious challenge to their long-term sustainability, maintainability, and trustworthiness. This study conducted a systematic review of 60 primary studies and identified 31 distinct types of AITDs, offering a detailed taxonomy and analysis of their origin, impact, and mitigation strategies. A key contribution of this work is the adoption of AI TRiSM (AI Trust, Risk, and Security Management) as a conceptual lens to reinterpret AITDs through the dimensions of trust, risk, safety, and security. This perspective consolidates fragmented concerns across technical and governance domains and emphasizes the interdependence between safety and security debt—both of which are often treated in isolation in existing literature.

To operationalize this perspective, we proposed the AITD-MAP framework, which unifies taxonomic classification, impact assessment, and mitigation planning for AITDs. We further introduced 34 actionable guidelines to assist practitioners and researchers in identifying and addressing AITDs—especially those that compromise safety, security, and overall system trustworthiness. In summary, this study offers a structured, multi-dimensional understanding of AITDs and introduces a practical roadmap for addressing them. The proposed taxonomy, impact mappings, and mitigation framework contribute to the growing body of knowledge on responsible and sustainable AI engineering—supporting researchers, developers, and policymakers in building safe, secure, reliable, and maintainable AI systems.

While this study consolidates and maps a comprehensive set of safety and security guidelines to the identified AITDs, future work should focus on empirically validating these guidelines in industrial settings and evaluating their effectiveness within MLOps pipelines. A natural extension of this work involves operationalizing the AITD-MAP framework into automated tools for debt detection, monitoring, and risk-aware refactoring. Such developments can draw on principles of responsible, safe, and secure AI engineering, as discussed in Section~\ref{sec:secsafeguidelines}, with the aim of improving guideline applicability and supporting continuous assurance.


Moreover, although the reviewed debts primarily reflect challenges in foundational AI systems, the insights extend to emerging paradigms such as Agentic-AI~\cite{shavit2023practices, chawla2024agentic} and large language models (LLMs)~\cite{chang2024survey}. These systems introduce increased autonomy, complexity, and dependency on dynamic data sources, which may intensify the accumulation and propagation of AITDs. Investigating how the proposed taxonomy, guidelines, and AITD-MAP framework apply to these new architectures represents a promising direction for future research.

\subsection{Future Challenges and Research Directions}

\label{sec:futurechallenges}

Building upon the findings and risk-informed roadmap proposed in this study, several open challenges and research directions emerge that warrant further exploration to advance the management of AI Technical Debt (AITD) in intelligent systems.

\begin{itemize}

\item \textbf{Standardized Metrics and Measurement Frameworks.}
Although this study provides a taxonomy and qualitative mapping of AITDs, the absence of standardized metrics for assessing the severity, propagation, and resolution of such debts remains a critical gap. Future research should focus on defining quantifiable indicators and benchmark datasets to enable empirical measurement of AITD impact on software quality, maintainability, and trustworthiness.

\item \textbf{Tooling and Automation for Debt Detection.}
While mature automated tools already exist for detecting several forms of technical debt—particularly code- and design-related debts - many AI-specific and cross-cutting AITDs (e.g., data-related debt, pipeline-level debt, and socio-technical debt) still rely heavily on manual qualitative analysis. Developing AI-aware and integrated detection and monitoring approaches-leveraging static analysis, ML-driven pattern recognition, and natural language processing of development artifacts-will therefore be crucial for operationalizing comprehensive AITD management within MLOps and continuous integration pipelines.


\item \textbf{Dynamic Risk Modeling and Simulation.}
As AITDs evolve over time through feedback loops and system updates, static analyses may fail to capture their long-term implications. Future studies should adopt dynamic simulation techniques (e.g., digital twins, system dynamics) to model how debts accumulate, interact, and impact reliability, security, and ethics throughout the system lifecycle.

\item \textbf{Integration with AI Governance and TRiSM Frameworks.}
The mapping with AI TRiSM and related governance models revealed strong conceptual overlap between AITDs and concerns such as fairness, transparency, and accountability. Future work should formalize these relationships by embedding debt-aware risk assessment modules within governance, assurance, and certification frameworks.

\item \textbf{Towards a Debt-Aware Software Engineering Paradigm.}
The long-term vision is to establish a debt-aware AI engineering paradigm where design decisions, testing strategies, and deployment pipelines are continuously informed by AITD risk indicators. Future work should develop integrated frameworks that connect technical debt management with responsible and sustainable AI system development.

\end{itemize}

\noindent By addressing these open challenges, future research can move toward a quantitative, automated, and context-aware understanding of AI Technical Debt. This will strengthen both theoretical foundations and practical interventions, ensuring the responsible and sustainable evolution of intelligent systems.

\bibliographystyle{ACM-Reference-Format}
\bibliography{sample-base}

@String{Computing = "Computing" }

@String{Computer = "{IEEE} Computer" }

@String{Springer = "Springer-Verlag" }

@BOOK{test,
   author = "Donald E. Knuth",
   title = "Seminumerical Algorithms",
   volume = 2,
   series = "The Art of Computer Programming",
   publisher = "Addison-Wesley",
   address = "Reading, MA",
   edition = "2nd",
   month = "10~" # jan,
   year = "1981",
}

@article{tricco2018prisma,
  title={PRISMA extension for scoping reviews (PRISMA-ScR): checklist and explanation},
  author={Tricco, Andrea C and Lillie, Erin and Zarin, Wasifa and O'Brien, Kelly K and Colquhoun, Heather and Levac, Danielle and Moher, David and Peters, Micah DJ and Horsley, Tanya and Weeks, Laura and others},
  journal={Annals of internal medicine},
  volume={169},
  number={7},
  pages={467--473},
  year={2018},
  publisher={American College of Physicians},
  doi={10.7326/M18-0850}
}

@article{sculley2015hidden,
  title={Hidden technical debt in machine learning systems},
  author={Sculley, David and Holt, Gary and Golovin, Daniel and Davydov, Eugene and Phillips, Todd and Ebner, Dietmar and Chaudhary, Vinay and Young, Michael and Crespo, Jean-Francois and Dennison, Dan},
  journal={Advances in neural information processing systems},
  volume={28},
  year={2015}
}

@article{corbin1990grounded,
  title={Grounded theory research: Procedures, canons, and evaluative criteria},
  author={Corbin, Juliet M and Strauss, Anselm},
  journal={Qualitative sociology},
  volume={13},
  number={1},
  pages={3--21},
  year={1990},
  publisher={Springer},
  doi = {https://doi.org/10.1007/BF00988593}
}

@article{lu2024responsible,
  title={Responsible AI pattern catalogue: A collection of best practices for AI governance and engineering},
  author={Lu, Qinghua and Zhu, Liming and Xu, Xiwei and Whittle, Jon and Zowghi, Didar and Jacquet, Aurelie},
  journal={ACM Computing Surveys},
  volume={56},
  number={7},
  pages={1--35},
  year={2024},
  publisher={ACM New York, NY}
}

@inproceedings{bogner2021characterizing,
  title={Characterizing technical debt and antipatterns in AI-based systems: A systematic mapping study},
  author={Bogner, Justus and Verdecchia, Roberto and Gerostathopoulos, Ilias},
  booktitle={2021 IEEE/ACM International Conference on Technical Debt (TechDebt)},
  pages={64--73},
  year={2021},
  organization={IEEE}
}

@article{shahriar2023survey,
  title={A survey of privacy risks and mitigation strategies in the Artificial Intelligence life cycle},
  author={Shahriar, Sakib and Allana, Sonal and Hazratifard, Seyed Mehdi and Dara, Rozita},
  journal={IEEE Access},
  volume={11},
  pages={61829--61854},
  year={2023},
  publisher={IEEE}
}

@inproceedings{schneider2024designing,
  title={Designing Secure AI-based Systems: a Multi-Vocal Literature Review},
  author={Schneider, Simon and Saha, Ananya and Mezzi, Emanuele and Tuma, Katja and Scandariato, Riccardo},
  booktitle={2024 IEEE Secure Development Conference (SecDev)},
  pages={13--19},
  year={2024},
  organization={IEEE}
}

@article{recupito2024technical,
  title={Technical debt in AI-enabled systems: On the prevalence, severity, impact, and management strategies for code and architecture},
  author={Recupito, Gilberto and Pecorelli, Fabiano and Catolino, Gemma and Lenarduzzi, Valentina and Taibi, Davide and Di Nucci, Dario and Palomba, Fabio},
  journal={Journal of Systems and Software},
  volume={216},
  pages={112151},
  year={2024},
  publisher={Elsevier}
}

@article{cunningham1992wycash,
  title={The WyCash portfolio management system},
  author={Cunningham, Ward},
  journal={ACM Sigplan Oops Messenger},
  volume={4},
  number={2},
  pages={29--30},
  year={1992},
  publisher={ACM New York, NY, USA}
}

@article{gesi2022code,
  title={Code Smells in Machine Learning Systems},
  author={Gesi, Jiri and Liu, Siqi and Li, Jiawei and Ahmed, Iftekhar and Nagappan, Nachiappan and Lo, David and de Almeida, Eduardo Santana and Kochhar, Pavneet Singh and Bao, Lingfeng},
  year={2022}
}

@inproceedings{alahdab2019empirical,
  title={Empirical analysis of hidden technical debt patterns in machine learning software},
  author = {Alahdab, Mohannad and Çalıklı, G{\"u}l},
  booktitle={Product-Focused Software Process Improvement: 20th International Conference, PROFES 2019, Barcelona, Spain, November 27--29, 2019, Proceedings 20},
  pages={195--202},
  year={2019},
  organization={Springer}
}

@inproceedings{chaudhary2018review,
  title={A review on hidden debts in machine learning systems},
  author={Chaudhary, Dev Kumar and Srivastava, Sandeep and Kumar, Vikas},
  booktitle={2018 Second International Conference on Green Computing and Internet of Things (ICGCIoT)},
  pages={619--624},
  year={2018},
  organization={IEEE}
}

@inproceedings{zhang2022code,
  title={Code smells for machine learning applications},
  author={Zhang, Haiyin and Cruz, Lu{\'\i}s and Van Deursen, Arie},
  booktitle={Proceedings of the 1st international conference on AI engineering: software engineering for AI},
  pages={217--228},
  year={2022}
}

@inproceedings{liu2020using,
  title={Is using deep learning frameworks free? characterizing technical debt in deep learning frameworks},
  author={Liu, Jiakun and Huang, Qiao and Xia, Xin and Shihab, Emad and Lo, David and Li, Shanping},
  booktitle={Proceedings of the ACM/IEEE 42nd International Conference on Software Engineering: Software Engineering in Society},
  pages={1--10},
  year={2020}
}

@inproceedings{10628360,
  author={Costal, Dolors and Gómez, Cristina and del Rey, Santiago and Martínez-Fernández, Silverio},
  booktitle={2024 IEEE 21st International Conference on Software Architecture Companion (ICSA-C)}, 
  title={Using Metrics for Code Smells of ML Pipelines}, 
  year={2024},
  volume={},
  number={},
  pages={289-294},
  doi={10.1109/ICSA-C63560.2024.00055}}

@inproceedings{tang2021empirical,
  title={An empirical study of refactorings and technical debt in machine learning systems},
  author={Tang, Yiming and Khatchadourian, Raffi and Bagherzadeh, Mehdi and Singh, Rhia and Stewart, Ajani and Raja, Anita},
  booktitle={2021 IEEE/ACM 43rd international conference on software engineering (ICSE)},
  pages={238--250},
  year={2021},
  organization={IEEE}
}

@inproceedings{sklavenitis2024measuring,
  title={Measuring Technical Debt in AI-Based Competition Platforms},
  author={Sklavenitis, Dionysios and Kalles, Dimitris},
  booktitle={Proceedings of the 13th Hellenic Conference on Artificial Intelligence},
  pages={1--10},
  year={2024}
}

@article{bhatia2023empirical,
  title={An Empirical Study of Self-Admitted Technical Debt in Machine Learning Software},
  author={Bhatia, Aaditya and Khomh, Foutse and Adams, Bram and Hassan, Ahmed E},
  journal={arXiv preprint arXiv:2311.12019},
  year={2023}
}

@article{chen2023toward,
  title={Toward understanding deep learning framework bugs},
  author={Chen, Junjie and Liang, Yihua and Shen, Qingchao and Jiang, Jiajun and Li, Shuochuan},
  journal={ACM Transactions on Software Engineering and Methodology},
  volume={32},
  number={6},
  pages={1--31},
  year={2023},
  publisher={ACM New York, NY}
}

@inproceedings{washizaki2019studying,
  title={Studying software engineering patterns for designing machine learning systems},
  author={Washizaki, Hironori and Uchida, Hiromu and Khomh, Foutse and Gueheneuc, Yann-Gael},
  booktitle={2019 10th International Workshop on Empirical Software Engineering in Practice (IWESEP)},
  pages={49--495},
  year={2019},
  organization={IEEE}
}

@inproceedings{albuquerque2022comprehending,
  title={Comprehending the use of intelligent techniques to support technical debt management},
  author={Albuquerque, Danyllo and Guimaraes, Everton and Tonin, Graziela and Perkusich, Mirko and Almeida, Hyggo and Perkusich, Angelo},
  booktitle={Proceedings of the International Conference on Technical Debt},
  pages={21--30},
  year={2022}
}

@inproceedings{van2021prevalence,
  title={The prevalence of code smells in machine learning projects},
  author={Van Oort, Bart and Cruz, Lu{\'\i}s and Aniche, Maur{\'\i}cio and Van Deursen, Arie},
  booktitle={2021 IEEE/ACM 1st Workshop on AI Engineering-Software Engineering for AI (WAIN)},
  pages={1--8},
  year={2021},
  organization={IEEE}
}

@inproceedings{arpteg2018software,
  title={Software engineering challenges of deep learning},
  author={Arpteg, Anders and Brinne, Bj{\o}rn and Crnkovic-Friis, Luka and Bosch, Jan},
  booktitle={2018 44th euromicro conference on software engineering and advanced applications (SEAA)},
  pages={50--59},
  year={2018},
  organization={IEEE}
}

@inproceedings{li2023debtviz,
  title={DebtViz: A Tool for Identifying, Measuring, Visualizing, and Monitoring Self-Admitted Technical Debt},
  author={Li, Yikun and Soliman, Mohamed and Avgeriou, Paris and Van Ittersum, Maarten},
  booktitle={2023 IEEE International Conference on Software Maintenance and Evolution (ICSME)},
  pages={558--562},
  year={2023},
  organization={IEEE}
}

@article{li2023automatic,
  title={Automatic identification of self-admitted technical debt from four different sources},
  author={Li, Yikun and Soliman, Mohamed and Avgeriou, Paris},
  journal={Empirical Software Engineering},
  volume={28},
  number={3},
  pages={65},
  year={2023},
  publisher={Springer}
}

@article{polyzotis2018data,
  title={Data lifecycle challenges in production machine learning: a survey},
  author={Polyzotis, Neoklis and Roy, Sudip and Whang, Steven Euijong and Zinkevich, Martin},
  journal={ACM SIGMOD Record},
  volume={47},
  number={2},
  pages={17--28},
  year={2018},
  publisher={ACM New York, NY, USA}
}

@inproceedings{foidl2022data,
  title={Data smells: Categories, causes and consequences, and detection of suspicious data in ai-based systems},
  author={Foidl, Harald and Felderer, Michael and Ramler, Rudolf},
  booktitle={Proceedings of the 1st International Conference on AI Engineering: Software Engineering for AI},
  pages={229--239},
  year={2022}
}

@inproceedings{lenarduzzi2021software,
  title={Software quality for ai: Where we are now?},
  author={Lenarduzzi, Valentina and Lomio, Francesco and Moreschini, Sergio and Taibi, Davide and Tamburri, Damian Andrew},
  booktitle={Software Quality: Future Perspectives on Software Engineering Quality: 13th International Conference, SWQD 2021, Vienna, Austria, January 19--21, 2021, Proceedings 13},
  pages={43--53},
  year={2021},
  organization={Springer}
}

@inproceedings{foidl2019technical,
  title={Technical debt in data-intensive software systems},
  author={Foidl, Harald and Felderer, Michael and Biffl, Stefan},
  booktitle={2019 45th Euromicro conference on software engineering and advanced applications (SEAA)},
  pages={338--341},
  year={2019},
  organization={IEEE}
}

@inproceedings{breck2017ml,
  title={The ML test score: A rubric for ML production readiness and technical debt reduction},
  author={Breck, Eric and Cai, Shanqing and Nielsen, Eric and Salib, Michael and Sculley, D},
  booktitle={2017 IEEE international conference on big data (big data)},
  pages={1123--1132},
  year={2017},
  organization={IEEE}
}

@inproceedings{hutchinson2021towards,
  title={Towards accountability for machine learning datasets: Practices from software engineering and infrastructure},
  author={Hutchinson, Ben and Smart, Andrew and Hanna, Alex and Denton, Emily and Greer, Christina and Kjartansson, Oddur and Barnes, Parker and Mitchell, Margaret},
  booktitle={Proceedings of the 2021 ACM Conference on Fairness, Accountability, and Transparency},
  pages={560--575},
  year={2021}
}

@inproceedings{obrien202223,
  title={23 shades of self-admitted technical debt: An empirical study on machine learning software},
  author={OBrien, David and Biswas, Sumon and Imtiaz, Sayem and Abdalkareem, Rabe and Shihab, Emad and Rajan, Hridesh},
  booktitle={Proceedings of the 30th ACM Joint European Software Engineering Conference and Symposium on the Foundations of Software Engineering},
  pages={734--746},
  year={2022}
}

@article{perez2021technical,
  title={Technical debt payment and prevention through the lenses of software architects},
  author={P{\'e}rez, Boris and Castellanos, Camilo and Correal, Dar{\'\i}o and Rios, Nicolli and Freire, S{\'a}vio and Sp{\'\i}nola, Rodrigo and Seaman, Carolyn and Izurieta, Clemente},
  journal={Information and Software Technology},
  volume={140},
  pages={106692},
  year={2021},
  publisher={Elsevier}
}

@article{li2015systematic,
  title={A systematic mapping study on technical debt and its management},
  author={Li, Zengyang and Avgeriou, Paris and Liang, Peng},
  journal={Journal of Systems and Software},
  volume={101},
  pages={193--220},
  year={2015},
  publisher={Elsevier}
}

@inproceedings{menshawy2024navigating,
  title={Navigating Challenges and Technical Debt in Large Language Models Deployment},
  author={Menshawy, Ahmed and Nawaz, Zeeshan and Fahmy, Mahmoud},
  booktitle={Proceedings of the 4th Workshop on Machine Learning and Systems},
  pages={192--199},
  year={2024}
}

@inproceedings{moreschini2024towards,
  title={Towards a Technical Debt for AI-based Recommender System},
  author={Moreschini, Sergio and Lenarduzzi, Valentina and Coba, Ludovik},
  booktitle={Proceedings of the 7th ACM/IEEE International Conference on Technical Debt},
  pages={36--39},
  year={2024}
}

@inproceedings{nahar2022collaboration,
  title={Collaboration challenges in building ml-enabled systems: Communication, documentation, engineering, and process},
  author={Nahar, Nadia and Zhou, Shurui and Lewis, Grace and K{\"a}stner, Christian},
  booktitle={Proceedings of the 44th international conference on software engineering},
  pages={413--425},
  year={2022}
}

@inproceedings{shivashankar2022maintainability,
  title={Maintainability challenges in ML: A systematic literature review},
  author={Shivashankar, Karthik and Martini, Antonio},
  booktitle={2022 48th Euromicro Conference on Software Engineering and Advanced Applications (SEAA)},
  pages={60--67},
  year={2022},
  organization={IEEE}
}

@inproceedings{chang2022understanding,
  title={Understanding implementation challenges in machine learning documentation},
  author={Chang, Jiyoo and Custis, Christine},
  booktitle={Proceedings of the 2nd ACM Conference on Equity and Access in Algorithms, Mechanisms, and Optimization},
  pages={1--8},
  year={2022}
}

@inproceedings{roselli2019managing,
  title={Managing bias in AI},
  author={Roselli, Drew and Matthews, Jeanna and Talagala, Nisha},
  booktitle={Companion proceedings of the 2019 world wide web conference},
  pages={539--544},
  year={2019}
}

@article{petrozzino2021pays,
  title={Who pays for ethical debt in AI?},
  author={Petrozzino, Catherine},
  journal={AI and Ethics},
  volume={1},
  number={3},
  pages={205--208},
  year={2021},
  publisher={Springer}
}

@article{jebnoun2022clones,
  title={Clones in deep learning code: what, where, and why?},
  author={Jebnoun, Hadhemi and Rahman, Md Saidur and Khomh, Foutse and Muse, Biruk Asmare},
  journal={Empirical Software Engineering},
  volume={27},
  number={4},
  pages={84},
  year={2022},
  publisher={Springer}
}

@article{cote2024quality,
  title={Quality issues in machine learning software systems},
  author={C{\^o}t{\'e}, Pierre-Olivier and Nikanjam, Amin and Bouchoucha, Rached and Basta, Ilan and Abidi, Mouna and Khomh, Foutse},
  journal={Empirical Software Engineering},
  volume={29},
  number={6},
  pages={1--47},
  year={2024},
  publisher={Springer}
}

@inproceedings{belani2019requirements,
  title={Requirements engineering challenges in building AI-based complex systems},
  author={Belani, Hrvoje and Vukovic, Marin and Car, {\v{Z}}eljka},
  booktitle={2019 IEEE 27th International Requirements Engineering Conference Workshops (REW)},
  pages={252--255},
  year={2019},
  organization={IEEE}
}

@inproceedings{bavota2016large,
  title={A large-scale empirical study on self-admitted technical debt},
  author={Bavota, Gabriele and Russo, Barbara},
  booktitle={Proceedings of the 13th international conference on mining software repositories},
  pages={315--326},
  year={2016}
}

@article{yan2018automating,
  title={Automating change-level self-admitted technical debt determination},
  author={Yan, Meng and Xia, Xin and Shihab, Emad and Lo, David and Yin, Jianwei and Yang, Xiaohu},
  journal={IEEE Transactions on Software Engineering},
  volume={45},
  number={12},
  pages={1211--1229},
  year={2018},
  publisher={IEEE}
}

@article{liu2021exploratory,
  title={An exploratory study on the introduction and removal of different types of technical debt in deep learning frameworks},
  author={Liu, Jiakun and Huang, Qiao and Xia, Xin and Shihab, Emad and Lo, David and Li, Shanping},
  journal={Empirical Software Engineering},
  volume={26},
  pages={1--36},
  year={2021},
  publisher={Springer}
}

@inproceedings{khritankov2021hidden,
  title={Hidden feedback loops in machine learning systems: A simulation model and preliminary results},
  author={Khritankov, Anton},
  booktitle={Software Quality: Future Perspectives on Software Engineering Quality: 13th International Conference, SWQD 2021, Vienna, Austria, January 19--21, 2021, Proceedings 13},
  pages={54--65},
  year={2021},
  organization={Springer}
}

@article{sas2023architectural,
  title={An architectural technical debt index based on machine learning and architectural smells},
  author={Sas, Darius and Avgeriou, Paris},
  journal={IEEE Transactions on Software Engineering},
  volume={49},
  number={8},
  pages={4169--4195},
  year={2023},
  publisher={IEEE}
}

@inproceedings{simon2023algorithm,
  title={Algorithm Debt: Challenges and Future Paths},
  author={Simon, Emmanuel Iko-Ojo and Vidoni, Melina and Fard, Fatemeh H},
  booktitle={2023 IEEE/ACM 2nd International Conference on AI Engineering--Software Engineering for AI (CAIN)},
  pages={90--91},
  year={2023},
  organization={IEEE}
}

@inproceedings{wang2023technical,
  title={Technical Debt Management in Industrial ML-State of Practice and Management Model Proposal},
  author={Wang, Xiaofei and Schuster, Herbert and Borrison, Reuben and Kl{\o}pper, Benjamin},
  booktitle={2023 IEEE 21st International Conference on Industrial Informatics (INDIN)},
  pages={1--9},
  year={2023},
  organization={IEEE}
}

@article{alves2016identification,
  title={Identification and management of technical debt: A systematic mapping study},
  author={Alves, Nicolli SR and Mendes, Thiago S and De Mendon{\c{c}}a, Manoel G and Sp{\'\i}nola, Rodrigo O and Shull, Forrest and Seaman, Carolyn},
  journal={Information and Software Technology},
  volume={70},
  pages={100--121},
  year={2016},
  publisher={Elsevier}
}

@article{avgeriou2016managing,
  title={Managing technical debt in software engineering (dagstuhl seminar 16162)},
  author={Avgeriou, Paris and Kruchten, Philippe and Ozkaya, Ipek and Seaman, Carolyn},
  journal={Dagstuhl reports},
  volume={6},
  number={4},
  pages={110--138},
  year={2016},
  publisher={Schloss Dagstuhl--Leibniz-Zentrum f{\"u}r Informatik}
}

@article{ampatzoglou2015financial,
  title={The financial aspect of managing technical debt: A systematic literature review},
  author={Ampatzoglou, Areti and Ampatzoglou, Apostolos and Chatzigeorgiou, Alexander and Avgeriou, Paris},
  journal={Information and Software Technology},
  volume={64},
  pages={52--73},
  year={2015},
  publisher={Elsevier}
}

@article{kleinwaks2023technical,
  title={Technical debt in systems engineering—A systematic literature review},
  author={Kleinwaks, Howard and Batchelor, Ann and Bradley, Thomas H},
  journal={Systems Engineering},
  volume={26},
  number={5},
  pages={675--687},
  year={2023},
  publisher={Wiley Online Library}
}

@article{kleinwaks2023ontology,
  title={An Ontology for Technical Debt in Systems Engineering},
  author={Kleinwaks, Howard and Batchelor, Ann and Bradley, Thomas H},
  journal={IEEE Open Journal of Systems Engineering},
  year={2023},
  publisher={IEEE}
}

@inproceedings{alfayez2020systematic,
  title={A systematic literature review of technical debt prioritization},
  author={Alfayez, Reem and Alwehaibi, Wesam and Winn, Robert and Venson, Elaine and Boehm, Barry},
  booktitle={Proceedings of the 3rd international conference on technical debt},
  pages={1--10},
  year={2020}
}

@article{dave2023artificial,
  title={Artificial intelligence in healthcare and education},
  author={Dave, Manas and Patel, Neil},
  journal={British dental journal},
  volume={234},
  number={10},
  pages={761--764},
  year={2023},
  publisher={Nature Publishing Group UK London}
}

@article{giudici2023safe,
  title={SAFE Artificial Intelligence in finance},
  author={Giudici, Paolo and Raffinetti, Emanuela},
  journal={Finance Research Letters},
  volume={56},
  pages={104088},
  year={2023},
  publisher={Elsevier}
}

@article{bendiab2023autonomous,
  title={Autonomous vehicles security: Challenges and solutions using blockchain and artificial intelligence},
  author={Bendiab, Gueltoum and Hameurlaine, Amina and Germanos, Georgios and Kolokotronis, Nicholas and Shiaeles, Stavros},
  journal={IEEE Transactions on Intelligent Transportation Systems},
  volume={24},
  number={4},
  pages={3614--3637},
  year={2023},
  publisher={IEEE}
}

@incollection{thota2018centralized,
  title={Centralized fog computing security platform for IoT and cloud in healthcare system},
  author={Thota, Chandu and Sundarasekar, Revathi and Manogaran, Gunasekaran and Varatharajan, R and Priyan, MK},
  booktitle={Fog computing: Breakthroughs in research and practice},
  pages={365--378},
  year={2018},
  publisher={IGI global}
}

@inproceedings{katzenbeisser2019security,
  title={Security in autonomous systems},
  author={Katzenbeisser, Stefan and Polian, Ilia and Regazzoni, Francesco and St{\"o}ttinger, Marc},
  booktitle={2019 IEEE European Test Symposium (ETS)},
  pages={1--8},
  year={2019},
  organization={IEEE}
}

@inproceedings{foidl2019risk,
  title={Risk-based data validation in machine learning-based software systems},
  author={Foidl, Harald and Felderer, Michael},
  booktitle={proceedings of the 3rd ACM SIGSOFT international workshop on machine learning techniques for software quality evaluation},
  pages={13--18},
  year={2019}
}

@inproceedings{wang2021patchrnn,
  title={Patchrnn: A deep learning-based system for security patch identification},
  author={Wang, Xinda and Wang, Shu and Feng, Pengbin and Sun, Kun and Jajodia, Sushil and Benchaaboun, Sanae and Geck, Frank},
  booktitle={MILCOM 2021-2021 IEEE Military Communications Conference (MILCOM)},
  pages={595--600},
  year={2021},
  organization={IEEE}
}

@inproceedings{al2021quality,
  title={Quality assurance challenges for machine learning software applications during software development life cycle phases},
  author={Al Alamin, Md Abdullah and Uddin, Gias},
  booktitle={2021 IEEE International Conference on Autonomous Systems (ICAS)},
  pages={1--5},
  year={2021},
  organization={IEEE}
}

@article{nankya2024security,
  title={Security and Privacy in E-Health Systems: A Review of AI and Machine Learning Techniques},
  author={Nankya, Mary and Mugisa, Allan and Usman, Yusuf and Upadhyay, Aadesh and Chataut, Robin},
  journal={IEEE Access},
  year={2024},
  publisher={IEEE}
}

@inproceedings{dai2021artificial,
  title={Artificial intelligence technologies in building resilient machine learning},
  author={Dai, Guangye and Sthapit, Saurav and Epiphaniou, Gregory and Maple, Carsten},
  booktitle={Competitive Advantage in the Digital Economy (CADE 2021)},
  volume={2021},
  pages={50--55},
  year={2021},
  organization={IET}
}

@article{girdhar2023cybersecurity,
  title={Cybersecurity of autonomous vehicles: A systematic literature review of adversarial attacks and defense models},
  author={Girdhar, Mansi and Hong, Junho and Moore, John},
  journal={IEEE Open Journal of Vehicular Technology},
  volume={4},
  pages={417--437},
  year={2023},
  publisher={IEEE}
}

@article{li2023trustworthy,
  title={Trustworthy AI: From principles to practices},
  author={Li, Bo and Qi, Peng and Liu, Bo and Di, Shuai and Liu, Jingen and Pei, Jiquan and Yi, Jinfeng and Zhou, Bowen},
  journal={ACM Computing Surveys},
  volume={55},
  number={9},
  pages={1--46},
  year={2023},
  publisher={ACM New York, NY}
}

@inproceedings{farrell2021evolution,
  title={Evolution of the IEEE P7009 standard: Towards fail-safe design of autonomous systems},
  author={Farrell, Marie and Luckcuck, Matt and Pullum, Laura and Fisher, Michael and Hessami, Ali and Gal, Danit and Murahwi, Zvikomborero and Wallace, Ken},
  booktitle={2021 IEEE International Symposium on Software Reliability Engineering Workshops (ISSREW)},
  pages={401--406},
  year={2021},
  organization={IEEE}
}

@article{hoangexplainable,
  title={Explainable AI in Finance: an Overview},
  author={Hoang, Anh and Phan, Hien},
  journal={},
  year = {2024}
}

@article{chirra2020ai,
  title={AI-Based Real-Time Security Monitoring for Cloud-Native Applications in Hybrid Cloud Environments},
  author={Chirra, Dinesh Reddy},
  journal={Revista de Inteligencia Artificial en Medicina},
  volume={11},
  number={1},
  pages={382--402},
  year={2020}
}

@article{nallakaruppan2024explainable,
  title={An Explainable AI framework for credit evaluation and analysis},
  author={Nallakaruppan, MK and Balusamy, Balamurugan and Shri, M Lawanya and Malathi, V and Bhattacharyya, Siddhartha},
  journal={Applied Soft Computing},
  volume={153},
  pages={111307},
  year={2024},
  publisher={Elsevier}
}

@article{mohammed2024developing,
  title={Developing Transparent AI Models to Enhance Interpretability and Trust in Medical Diagnostics: Implementing explainable AI techniques to provide transparent explanations for medical diagnoses, enhancing trust and acceptance among healthcare professionals},
  author={Mohammed, Faisal},
  journal={Journal of Machine Learning for Healthcare Decision Support},
  volume={4},
  number={2},
  pages={36--43},
  year={2024}
}

@article{kumarml,
  title={ML and AI Based Healthcare Model to more Interpretable and Transparent in Medical Diagnosis},
  author={Kumar, P Vishnu and Ganguly, Tanaya and Gupta, Rolly and Pokkuluri, Kiran Sree and Mishra, Avinash Kumar V and Selvi, V}, 
journal = {African Journal of Biological Sciences},
 year={2024}
}

@article{tibebuframework,
  title={Framework for Data Protection, Security, and Privacy in AI Applications},
  author={Tibebu, Haileleol},
  journal = {The Broadcast Centre Here East, London},
  year = {2024}
}

@article{de2023guide,
  title={A guide to sharing open healthcare data under the General Data Protection Regulation},
  author={de Kok, Jip WTM and de la Hoz, Miguel {\'A} Armengol and de Jong, Ymke and Brokke, V{\'e}ronique and Elbers, Paul WG and Thoral, Patrick and Castillejo, Alejandro and Trenor, Tom{\'a}s and Castellano, Jose M and Bronchalo, Alberto E and others},
  journal={Scientific data},
  volume={10},
  number={1},
  pages={404},
  year={2023},
  publisher={Nature Publishing Group UK London}
}

@inproceedings{hernandez2019data,
  title={Data protection on FinTech platforms},
  author={Hern{\'a}ndez, Elena and {\"O}zt{\"u}rk, Mehmet and Sitt{\'o}n, In{\'e}s and Rodr{\'\i}guez, Sara},
  booktitle={Highlights of Practical Applications of Survivable Agents and Multi-Agent Systems. The PAAMS Collection: International Workshops of PAAMS 2019, {\'A}vila, Spain, June 26--28, 2019, Proceedings 17},
  pages={223--233},
  year={2019},
  organization={Springer}
}

@inproceedings{ruland2018access,
  title={Access control in safety critical environments},
  author={Ruland, Christoph and Sassmannshausen, Jochen},
  booktitle={2018 12th International Conference on Reliability, Maintainability, and Safety (ICRMS)},
  pages={223--229},
  year={2018},
  organization={IEEE}
}

@incollection{sujan2023looking,
  title={Looking at the Safety of AI from a Systems Perspective: Two Healthcare Examples},
  author={Sujan, Mark A},
  booktitle={Safety in the Digital Age: Sociotechnical Perspectives on Algorithms and Machine Learning},
  pages={79--90},
  year={2023},
  publisher={Springer Nature Switzerland Cham}
}

@article{shavit2023practices,
  title={Practices for governing agentic AI systems},
  author={Shavit, Yonadav and Agarwal, Sandhini and Brundage, Miles and Adler, Steven and O’Keefe, Cullen and Campbell, Rosie and Lee, Teddy and Mishkin, Pamela and Eloundou, Tyna and Hickey, Alan and others},
  journal={Research Paper, OpenAI, December},
  year={2023}
}

@article{chawla2024agentic,
  title={Agentic AI: The building blocks of sophisticated AI business applications},
  author={Chawla, Chhavi and Chatterjee, Siddharth and Gadadinni, Sanketh Siddanna and Verma, Pulkit and Banerjee, Sourav},
  journal={Journal of AI, Robotics \& Workplace Automation},
  volume={3},
  number={3},
  pages={1--15},
  year={2024},
  publisher={Henry Stewart Publications}
}

@article{chang2024survey,
  title={A survey on evaluation of large language models},
  author={Chang, Yupeng and Wang, Xu and Wang, Jindong and Wu, Yuan and Yang, Linyi and Zhu, Kaijie and Chen, Hao and Yi, Xiaoyuan and Wang, Cunxiang and Wang, Yidong and others},
  journal={ACM Transactions on Intelligent Systems and Technology},
  volume={15},
  number={3},
  pages={1--45},
  year={2024},
  publisher={ACM New York, NY}
}

@inproceedings{wohlin2014guidelines,
  title={Guidelines for snowballing in systematic literature studies and a replication in software engineering},
  author={Wohlin, Claes},
  booktitle={Proceedings of the 18th international conference on evaluation and assessment in software engineering},
  pages={1--10},
  year={2014}
}

@inproceedings{petersen2008systematic,
  title={Systematic mapping studies in software engineering},
  author={Petersen, Kai and Feldt, Robert and Mujtaba, Shahid and Mattsson, Michael},
  booktitle={12th international conference on evaluation and assessment in software engineering (EASE)},
  year={2008},
  organization={BCS Learning \& Development}
}

@article{salhab2024systematic,
  title={A systematic literature review on ai safety: Identifying trends, challenges and future directions},
  author={Salhab, Wissam and Ameyed, Darine and Jaafar, Fehmi and Mcheick, Hamid},
  journal={IEEE Access},
  year={2024},
  publisher={IEEE}
}

@article{wang2023distribution,
  title={Distribution-restrained Softmax Loss for the Model Robustness},
  author={Wang, Hao and Li, Chen and Jiang, Jinzhe and Zhang, Xin and Zhao, Yaqian and Gong, Weifeng},
  journal={arXiv preprint arXiv:2303.12363},
  year={2023}
}

@inproceedings{girard2022caisar,
  title={CAISAR: A platform for Characterizing Artificial Intelligence Safety and Robustness},
  author={Girard-Satabin, Julien and Alberti, Michele and Bobot, Fran{\c{c}}ois and Chihani, Zakaria and Lemesle, Augustin},
  booktitle={AISafety},
  year={2022}
}

@inproceedings{zhao2021detecting,
  title={Detecting operational adversarial examples for reliable deep learning},
  author={Zhao, Xingyu and Huang, Wei and Schewe, Sven and Dong, Yi and Huang, Xiaowei},
  booktitle={2021 51st Annual IEEE/IFIP International Conference on Dependable Systems and Networks-Supplemental Volume (DSN-S)},
  pages={5--6},
  year={2021},
  organization={IEEE}
}

@inproceedings{akram2022stadre,
  title={StaDRe and StaDRo: reliability and robustness estimation of ML-based forecasting using statistical distance measures},
  author={Akram, Mohammed Naveed and Ambekar, Akshatha and Sorokos, Ioannis and Aslansefat, Koorosh and Schneider, Daniel},
  booktitle={International Conference on Computer Safety, Reliability, and Security},
  pages={289--301},
  year={2022},
  organization={Springer}
}

@inproceedings{tarchoun2022investigating,
  title={Investigating the robustness of multi-view detection to current adversarial patch threats},
  author={Tarchoun, Bilel and Khalifa, Anouar Ben and Mahjoub, Mohamed Ali},
  booktitle={2022 6th International Conference on Advanced Technologies for Signal and Image Processing (ATSIP)},
  pages={1--6},
  year={2022},
  organization={IEEE}
}

@inproceedings{arnez2021improving,
  title={Improving robustness of deep neural networks for aerial navigation by incorporating input uncertainty},
  author={Arnez, Fabio and Espinoza, Huascar and Radermacher, Ansgar and Terrier, Fran{\c{c}}ois},
  booktitle={International Conference on Computer Safety, Reliability, and Security},
  pages={219--225},
  year={2021},
  organization={Springer}
}

@inproceedings{mziou2019safety,
  title={Safety and robustness of deep neural networks object recognition under generic attacks},
  author={Mziou Sallami, Mallek and Ibn Khedher, Mohamed and Trabelsi, Asma and Kerboua-Benlarbi, Samy and Bettebghor, Dimitri},
  booktitle={International conference on neural information processing},
  pages={274--286},
  year={2019},
  organization={Springer}
}

@article{choi2022argan,
  title={ARGAN: Adversarially robust generative adversarial networks for deep neural networks against adversarial examples},
  author={Choi, Seok-Hwan and Shin, Jin-Myeong and Liu, Peng and Choi, Yoon-Ho},
  journal={IEEE Access},
  volume={10},
  pages={33602--33615},
  year={2022},
  publisher={IEEE}
}

@inproceedings{schwartz2020regularization,
  title={Regularization and sparsity for adversarial robustness and stable attribution},
  author={Schwartz, Daniel and Alparslan, Yigit and Kim, Edward},
  booktitle={Advances in Visual Computing: 15th International Symposium, ISVC 2020, San Diego, CA, USA, October 5--7, 2020, Proceedings, Part I 15},
  pages={3--14},
  year={2020},
  organization={Springer}
}

@article{maabreh2022robustness,
  title={The robustness of popular multiclass machine learning models against poisoning attacks: Lessons and insights},
  author={Maabreh, Majdi and Maabreh, Arwa and Qolomany, Basheer and Al-Fuqaha, Ala},
  journal={International Journal of Distributed Sensor Networks},
  volume={18},
  number={7},
  pages={15501329221105159},
  year={2022},
  publisher={SAGE Publications Sage UK: London, England}
}

@inproceedings{kshetry2019safety,
  title={Safety in the face of unknown unknowns: Algorithm fusion in data-driven engineering systems},
  author={Kshetry, Nina and Varshney, Lav R},
  booktitle={ICASSP 2019-2019 IEEE International Conference on Acoustics, Speech and Signal Processing (ICASSP)},
  pages={8162--8166},
  year={2019},
  organization={IEEE}
}

@inproceedings{zhao2020safety,
  title={A safety framework for critical systems utilising deep neural networks},
  author={Zhao, Xingyu and Banks, Alec and Sharp, James and Robu, Valentin and Flynn, David and Fisher, Michael and Huang, Xiaowei},
  booktitle={Computer Safety, Reliability, and Security: 39th International Conference, SAFECOMP 2020, Lisbon, Portugal, September 16--18, 2020, Proceedings 39},
  pages={244--259},
  year={2020},
  organization={Springer}
}

@article{fisher2021towards,
  title={Towards a framework for certification of reliable autonomous systems},
  author={Fisher, Michael and Mascardi, Viviana and Rozier, Kristin Yvonne and Schlingloff, Bernd-Holger and Winikoff, Michael and Yorke-Smith, Neil},
  journal={Autonomous Agents and Multi-Agent Systems},
  volume={35},
  pages={1--65},
  year={2021},
  publisher={Springer}
}

@inproceedings{werner2023assurance,
  title={An Assurance Case for the DoD Ethical Principles of Artificial Intelligence},
  author={Werner, Benjamin D and Schumeg, Benjamin J and Mills, Tiffany M and Velilla, Elizabeth V},
  booktitle={2023 Annual Reliability and Maintainability Symposium (RAMS)},
  pages={1--7},
  year={2023},
  organization={IEEE}
}

@inproceedings{kim2020fair,
  title={Fair representation for safe artificial intelligence via adversarial learning of unbiased information bottleneck.},
  author={Kim, Jin-Young and Cho, Sung-Bae},
  booktitle={SafeAI@ AAAI},
  pages={105--112},
  year={2020}
}

@inproceedings{srinivasan2019understanding,
  title={Understanding Bias in Datasets using Topological Data Analysis.},
  author={Srinivasan, Ramya and Chander, Ajay},
  booktitle={AISafety@ IJCAI},
  year={2019}
}

@article{cooper2022believe,
  title={Believe the HiPe: Hierarchical perturbation for fast, robust, and model-agnostic saliency mapping},
  author={Cooper, Jessica and Arandjelovi{\'c}, Ognjen and Harrison, David J},
  journal={Pattern Recognition},
  volume={129},
  pages={108743},
  year={2022},
  publisher={Elsevier}
}

@inproceedings{krajna2022explainable,
  title={Explainable artificial intelligence: An updated perspective},
  author={Krajna, Agneza and Kovac, Mihael and Brcic, Mario and {\v{S}}ar{\v{c}}evi{\'c}, Ana},
  booktitle={2022 45th Jubilee International Convention on Information, Communication and Electronic Technology (MIPRO)},
  pages={859--864},
  year={2022},
  organization={IEEE}
}

@inproceedings{sokol2019counterfactual,
  title={Counterfactual explanations of machine learning predictions: opportunities and challenges for AI safety},
  author={Sokol, Kacper and Flach, Peter},
  booktitle={2019 AAAI Workshop on Artificial Intelligence Safety, SafeAI 2019},
  year={2019},
  organization={CEUR Workshop Proceedings}
}

@inproceedings{sheh2021explainable,
  title={Explainable artificial intelligence requirements for safe, intelligent robots},
  author={Sheh, Raymond},
  booktitle={2021 IEEE international conference on intelligence and safety for robotics (ISR)},
  pages={382--387},
  year={2021},
  organization={IEEE}
}

@inproceedings{buczak2022explainable,
  title={Explainable forecasts of disruptive events using recurrent neural networks},
  author={Buczak, Anna L and Baugher, Benjamin D and Berlier, Adam J and Scharfstein, Kayla E and Martin, Christine S},
  booktitle={2022 IEEE international conference on assured autonomy (ICAA)},
  pages={64--73},
  year={2022},
  organization={IEEE}
}

@article{chamola2023review,
  title={A review of trustworthy and explainable artificial intelligence (xai)},
  author={Chamola, Vinay and Hassija, Vikas and Sulthana, A Razia and Ghosh, Debshishu and Dhingra, Divyansh and Sikdar, Biplab},
  journal={IEEe Access},
  volume={11},
  pages={78994--79015},
  year={2023},
  publisher={IEEE}
}

@inproceedings{samadi2023safe,
  title={SAFE: Saliency-aware counterfactual explanations for DNN-based automated driving systems},
  author={Samadi, Amir and Shirian, Amir and Koufos, Konstantinos and Debattista, Kurt and Dianati, Mehrdad},
  booktitle={2023 IEEE 26th International Conference on Intelligent Transportation Systems (ITSC)},
  pages={5655--5662},
  year={2023},
  organization={IEEE}
}

@incollection{Dağlarli20,
author = {Evren Dağlarli},
title = {Explainable Artificial Intelligence (xAI) Approaches and Deep Meta-Learning Models},
booktitle = {Advances and Applications in Deep Learning},
publisher = {IntechOpen},
address = {Rijeka},
year = {2020},
editor = {Marco Antonio Aceves-Fernandez},
chapter = {5},
doi = {10.5772/intechopen.92172},
url = {https://doi.org/10.5772/intechopen.92172}
}

@inproceedings{jaipuria2022deeppic,
  title={deepPIC: Deep perceptual image clustering for identifying bias in vision datasets},
  author={Jaipuria, Nikita and Stevo, Katherine and Zhang, Xianling and Gaopande, Meghana L and Calle, Ian and Jain, Jinesh and Murali, Vidya N},
  booktitle={Proceedings of the IEEE/CVF Conference on Computer Vision and Pattern Recognition},
  pages={4793--4802},
  year={2022}
}

@inproceedings{bravo2022human,
  title={Human-in-the-loop online multi-agent approach to increase trustworthiness in ML models through trust scores and data augmentation},
  author={Bravo-Rocca, Gusseppe and Liu, Peini and Guitart, Jordi and Dholakia, Ajay and Ellison, David and Hodak, Miroslav},
  booktitle={2022 IEEE 46th Annual Computers, Software, and Applications Conference (COMPSAC)},
  pages={32--37},
  year={2022},
  organization={IEEE}
}

@article{he2021challenges,
  title={The challenges and opportunities of human-centered AI for trustworthy robots and autonomous systems},
  author={He, Hongmei and Gray, John and Cangelosi, Angelo and Meng, Qinggang and McGinnity, T Martin and Mehnen, J{\"o}rn},
  journal={IEEE Transactions on Cognitive and Developmental Systems},
  volume={14},
  number={4},
  pages={1398--1412},
  year={2021},
  publisher={IEEE}
}

@inproceedings{steimers2021sources,
  title={Sources of risk and design principles of trustworthy artificial intelligence},
  author={Steimers, Andr{\'e} and B{\"o}mer, Thomas},
  booktitle={International Conference on Human-Computer Interaction},
  pages={239--251},
  year={2021},
  organization={Springer}
}

@article{hagendorff2021linking,
  title={Linking human and machine behavior: A new approach to evaluate training data quality for beneficial machine learning},
  author={Hagendorff, Thilo},
  journal={Minds and Machines},
  volume={31},
  number={4},
  pages={563--593},
  year={2021},
  publisher={Springer}
}

@inproceedings{carlini2017towards,
  title={Towards evaluating the robustness of neural networks},
  author={Carlini, Nicholas and Wagner, David},
  booktitle={2017 ieee symposium on security and privacy (sp)},
  pages={39--57},
  year={2017},
  organization={Ieee}
}

@inproceedings{maabreh2022developing,
  title={On developing deep learning models with particle swarm optimization in the presence of poisoning attacks},
  author={Maabreh, Majdi and Darwish, Omar and Karajeh, Ola and Tashtoush, Yahya},
  booktitle={2022 International Arab Conference on Information Technology (ACIT)},
  pages={1--5},
  year={2022},
  organization={IEEE}
}

@INPROCEEDINGS{9256597,
  author={Ali, Muhammad and Hu, Yim-Fun and Luong, Doanh Kim and Oguntala, George and Li, Jian-Ping and Abdo, Kanaan},
  booktitle={2020 AIAA/IEEE 39th Digital Avionics Systems Conference (DASC)}, 
  title={Adversarial Attacks on AI based Intrusion Detection System for Heterogeneous Wireless Communications Networks}, 
  year={2020},
  volume={},
  number={},
  pages={1-6},
  doi={10.1109/DASC50938.2020.9256597}}

@inproceedings{10.1007/978-3-030-83906-2_20,
author = {Rajendran, Prajit T. and Espinoza, Huascar and Delaborde, Agnes and Mraidha, Chokri},
title = {Human-in-the-Loop Learning Methods Toward Safe DL-Based Autonomous Systems: A Review},
year = {2021},
isbn = {978-3-030-83905-5},
publisher = {Springer-Verlag},
address = {Berlin, Heidelberg},
url = {https://doi.org/10.1007/978-3-030-83906-2_20},
doi = {10.1007/978-3-030-83906-2_20},
booktitle = {Computer Safety, Reliability, and Security. SAFECOMP 2021 Workshops: DECSoS, MAPSOD, DepDevOps, USDAI, and WAISE, York, UK, September 7, 2021, Proceedings},
pages = {251–264},
numpages = {14},
location = {York, United Kingdom}
}

@inproceedings{Kamoi2020OutofDistributionDW,
  title={Out-of-Distribution Detection with Likelihoods Assigned by Deep Generative Models Using Multimodal Prior Distributions},
  author={Ryo Kamoi and Kei Kobayashi},
  booktitle={SafeAI@AAAI},
  year={2020},
  url={https://api.semanticscholar.org/CorpusID:212419400}
}

@InProceedings{10.1007/978-3-030-83906-2_17,
author="Arnez, Fabio
and Espinoza, Huascar
and Radermacher, Ansgar
and Terrier, Fran{\c{c}}ois",
editor="Habli, Ibrahim
and Sujan, Mark
and Gerasimou, Simos
and Schoitsch, Erwin
and Bitsch, Friedemann",
title="Improving Robustness of Deep Neural Networks for Aerial Navigation by Incorporating Input Uncertainty",
booktitle="Computer Safety, Reliability, and Security. SAFECOMP 2021 Workshops",
year="2021",
publisher="Springer International Publishing",
address="Cham",
pages="219--225",
isbn="978-3-030-83906-2"
}

@misc{rossolini2022increasingconfidencedeepneural,
      title={Increasing the Confidence of Deep Neural Networks by Coverage Analysis}, 
      author={Giulio Rossolini and Alessandro Biondi and Giorgio Buttazzo},
      
    journal={ACM Transactions on Software Engineering},
      volume={49},
      number={2},
      pages={802 -- 815},
      year={2022},
      publisher={ACM New York, NY}
      
}

@ARTICLE{Xu2019563,
	author = {Xu, Feiyu and Uszkoreit, Hans and Du, Yangzhou and Fan, Wei and Zhao, Dongyan and Zhu, Jun},
	title = {Explainable AI: A Brief Survey on History, Research Areas, Approaches and Challenges},
	year = {2019},
	journal = {Lecture Notes in Computer Science (including subseries Lecture Notes in Artificial Intelligence and Lecture Notes in Bioinformatics)},
	volume = {11839 LNAI},
	pages = {563 – 574},
	doi = {10.1007/978-3-030-32236-6_51},
	url = {https://www.scopus.com/inward/record.uri?eid=2-s2.0-85075832575&doi=10.1007%2f978-3-030-32236-6_51&partnerID=40&md5=6c3165fea2c868bb368323d50b1d32a6},
	type = {Conference paper},
	publication_stage = {Final},
	source = {Scopus},
	note = {Cited by: 486}
}

@CONFERENCE{Haider2023851,
	author = {Haider, Tom and Roscher, Karsten and da Roza, Felippe Schmoeller and Günnemann, Stephan},
	title = {Out-of-Distribution Detection for Reinforcement Learning Agents with Probabilistic Dynamics Models},
	year = {2023},
	journal = {Proceedings of the International Joint Conference on Autonomous Agents and Multiagent Systems, AAMAS},
	volume = {2023-May},
	pages = {851 – 859},
	url = {https://www.scopus.com/inward/record.uri?eid=2-s2.0-85171291933&partnerID=40&md5=eca3fa8f2d003fd46567c10148750abc},
	type = {Conference paper},
	publication_stage = {Final},
	source = {Scopus},
	note = {Cited by: 15}
}

@ARTICLE{Albahri2023156,
	author = {Albahri, A.S. and Duhaim, Ali M. and Fadhel, Mohammed A. and Alnoor, Alhamzah and Baqer, Noor S. and Alzubaidi, Laith and Albahri, O.S. and Alamoodi, A.H. and Bai, Jinshuai and Salhi, Asma and Santamaría, Jose and Ouyang, Chun and Gupta, Ashish and Gu, Yuantong and Deveci, Muhammet},
	title = {A systematic review of trustworthy and explainable artificial intelligence in healthcare: Assessment of quality, bias risk, and data fusion},
	year = {2023},
	journal = {Information Fusion},
	volume = {96},
	pages = {156 – 191},
	doi = {10.1016/j.inffus.2023.03.008},
	url = {https://www.scopus.com/inward/record.uri?eid=2-s2.0-85151265419&doi=10.1016%2fj.inffus.2023.03.008&partnerID=40&md5=ed8929df202fd77c56bb836cbd97bccd},
	type = {Article},
	publication_stage = {Final},
	source = {Scopus},
	note = {Cited by: 359}
}

@ARTICLE{8466590,
  author={Adadi, Amina and Berrada, Mohammed},
  journal={IEEE Access}, 
  title={Peeking Inside the Black-Box: A Survey on Explainable Artificial Intelligence (XAI)}, 
  year={2018},
  volume={6},
  number={},
  pages={52138-52160},
  doi={10.1109/ACCESS.2018.2870052}}

@article{ferrara2023fairness,
  title={Fairness and bias in artificial intelligence: A brief survey of sources, impacts, and mitigation strategies},
  author={Ferrara, Emilio},
  journal={Sci},
  volume={6},
  number={1},
  pages={3},
  year={2023},
  publisher={MDPI}
}

@article{shi2023towards,
  title={Towards fairness-aware federated learning},
  author={Shi, Yuxin and Yu, Han and Leung, Cyril},
  journal={IEEE Transactions on Neural Networks and Learning Systems},
  year={2023},
  publisher={IEEE}
}

@inproceedings{parmar2023review,
  title={A review on data balancing techniques and machine learning methods},
  author={Parmar, Gaurav and Gupta, Rimi and Bhatt, Tejas and Sahani, GJ and Panchal, Brijeshkumar Y and Patel, Hiren},
  booktitle={2023 5th International Conference on Smart Systems and Inventive Technology (ICSSIT)},
  pages={1004--1008},
  year={2023},
  organization={IEEE}
}

@inproceedings{10.5555/3157382.3157469,
author = {Hardt, Moritz and Price, Eric and Srebro, Nathan},
title = {Equality of opportunity in supervised learning},
year = {2016},
isbn = {9781510838819},
publisher = {Curran Associates Inc.},
address = {Red Hook, NY, USA},
booktitle = {Proceedings of the 30th International Conference on Neural Information Processing Systems},
pages = {3323–3331},
numpages = {9},
location = {Barcelona, Spain},
series = {NIPS'16}
}

@inproceedings{10.5555/3600270.3600944,
author = {Ziegler, Daniel M. and Nix, Seraphina and Chan, Lawrence and Bauman, Tim and Schmidt-Nielsen, Peter and Lin, Tao and Scherlis, Adam and Nabeshima, Noa and Weinstein-Raun, Ben and de Haas, Daniel and Shlegeris, Buck and Thomas, Nate},
title = {Adversarial training for high-stakes reliability},
year = {2022},
isbn = {9781713871088},
publisher = {Curran Associates Inc.},
address = {Red Hook, NY, USA},
booktitle = {Proceedings of the 36th International Conference on Neural Information Processing Systems},
articleno = {674},
numpages = {13},
location = {New Orleans, LA, USA},
series = {NIPS '22}
}

@Article{a15080283,
AUTHOR = {Zhao, Weimin and Alwidian, Sanaa and Mahmoud, Qusay H.},
TITLE = {Adversarial Training Methods for Deep Learning: A Systematic Review},
JOURNAL = {Algorithms},
VOLUME = {15},
YEAR = {2022},
NUMBER = {8},
ARTICLE-NUMBER = {283},
URL = {https://www.mdpi.com/1999-4893/15/8/283},
ISSN = {1999-4893},
DOI = {10.3390/a15080283}
}

@article{Meng2022AdversarialRO,
  title={Adversarial Robustness of Deep Neural Networks: A Survey from a Formal Verification Perspective},
  author={Mark Huasong Meng and Guangdong Bai and Sin Gee Teo and Zhe Hou and Yan Xiao and Yun Lin and Jin Song Dong},
  journal={ArXiv},
  year={2022},
  volume={abs/2206.12227},
  url={https://api.semanticscholar.org/CorpusID:249223202}
}

@InProceedings{pmlr-v161-corsi21a,
  title = 	 {Formal verification of neural networks for safety-critical tasks in deep reinforcement learning},
  author =       {Corsi, Davide and Marchesini, Enrico and Farinelli, Alessandro},
  booktitle = 	 {Proceedings of the Thirty-Seventh Conference on Uncertainty in Artificial Intelligence},
  pages = 	 {333--343},
  year = 	 {2021},
  editor = 	 {de Campos, Cassio and Maathuis, Marloes H.},
  volume = 	 {161},
  series = 	 {Proceedings of Machine Learning Research},
  month = 	 {27--30 Jul},
  publisher =    {PMLR},
  url = 	 {https://proceedings.mlr.press/v161/corsi21a.html}
}

@ARTICLE{9134370,
  author={Zhuang, Fuzhen and Qi, Zhiyuan and Duan, Keyu and Xi, Dongbo and Zhu, Yongchun and Zhu, Hengshu and Xiong, Hui and He, Qing},
  journal={Proceedings of the IEEE}, 
  title={A Comprehensive Survey on Transfer Learning}, 
  year={2021},
  volume={109},
  number={1},
  pages={43-76},
  doi={10.1109/JPROC.2020.3004555}}

@Inbook{Kamath2019,
author="Kamath, Uday
and Liu, John
and Whitaker, James",
title="Transfer Learning: Domain Adaptation",
bookTitle="Deep Learning for NLP and Speech Recognition ",
year="2019",
publisher="Springer International Publishing",
address="Cham",
pages="495--535",
isbn="978-3-030-14596-5",
doi="10.1007/978-3-030-14596-5_11",
url="https://doi.org/10.1007/978-3-030-14596-5_11"
}

@Inbook{Chennam2023,
author="Chennam, Krishna Keerthi
and Mudrakola, Swapna
and Maheswari, V. Uma
and Aluvalu, Rajanikanth
and Rao, K. Gangadhara",
editor="Mehta, Mayuri
and Palade	, Vasile
and Chatterjee, Indranath",
title="Black Box Models for eXplainable Artificial Intelligence",
bookTitle="Explainable AI: Foundations, Methodologies and Applications",
year="2023",
publisher="Springer International Publishing",
address="Cham",
pages="1--24",
isbn="978-3-031-12807-3",
doi="10.1007/978-3-031-12807-3_1",
url="https://doi.org/10.1007/978-3-031-12807-3_1"
}

@CONFERENCE{Shejwalkar20221354,
	author = {Shejwalkar, Virat and Houmansadr, Amir and Kairouz, Peter and Ramage, Daniel},
	title = {Back to the Drawing Board: A Critical Evaluation of Poisoning Attacks on Production Federated Learning},
	year = {2022},
	journal = {Proceedings - IEEE Symposium on Security and Privacy},
	volume = {2022-May},
	pages = {1354 – 1371},
	doi = {10.1109/SP46214.2022.9833647},
	url = {https://www.scopus.com/inward/record.uri?eid=2-s2.0-85125030551&doi=10.1109%2fSP46214.2022.9833647&partnerID=40&md5=ff822db90775a18b0525d815dc232b97},
	type = {Conference paper},
	publication_stage = {Final},
	source = {Scopus},
	note = {Cited by: 192; All Open Access, Green Open Access}
}

@ARTICLE{Zhang2022,
	author = {Zhang, Kaiyue and Song, Xuan and Zhang, Chenhan and Yu, Shui},
	title = {Challenges and future directions of secure federated learning: a survey},
	year = {2022},
	journal = {Frontiers of Computer Science},
	volume = {16},
	number = {5},
	doi = {10.1007/s11704-021-0598-z},
	url = {https://www.scopus.com/inward/record.uri?eid=2-s2.0-85120989981&doi=10.1007%2fs11704-021-0598-z&partnerID=40&md5=c5c5083147483aac4a1ec3cf616177d9},
	type = {Review},
	publication_stage = {Final},
	source = {Scopus},
	note = {Cited by: 94; All Open Access, Bronze Open Access, Green Open Access}
}

@CONFERENCE{Dang2022192,
	author = {Dang, Ning and Shao, Keyong and Chen, Long and Yang, Min},
	title = {Multi-model decision-making seizure types classification based on transfer learning},
	year = {2022},
	journal = {Proceedings - 2022 International Symposium on Control Engineering and Robotics, ISCER 2022},
	pages = {192 – 201},
	doi = {10.1109/ISCER55570.2022.00040},
	url = {https://www.scopus.com/inward/record.uri?eid=2-s2.0-85141409898&doi=10.1109%2fISCER55570.2022.00040&partnerID=40&md5=9ab11d856ba68f22b565106b113636ba},
	type = {Conference paper},
	publication_stage = {Final},
	source = {Scopus},
	note = {Cited by: 6}
}

@ARTICLE{Gyevnár2025531,
	author = {Gyevnár, Bálint and Kasirzadeh, Atoosa},
	title = {AI safety for everyone},
	year = {2025},
	journal = {Nature Machine Intelligence},
	volume = {7},
	number = {4},
	pages = {531 – 542},
	doi = {10.1038/s42256-025-01020-y},
	url = {https://www.scopus.com/inward/record.uri?eid=2-s2.0-105003409079&doi=10.1038%2fs42256-025-01020-y&partnerID=40&md5=afc7ca66d118f9c99ad814aed8cbf458},
	type = {Review},
	publication_stage = {Final},
	source = {Scopus},
	note = {Cited by: 0}
}

@ARTICLE{Bélisle-Pipon20231507,
	author = {Bélisle-Pipon, Jean-Christophe and Monteferrante, Erica and Roy, Marie-Christine and Couture, Vincent},
	title = {Artificial intelligence ethics has a black box problem},
	year = {2023},
	journal = {AI and Society},
	volume = {38},
	number = {4},
	pages = {1507 – 1522},
	doi = {10.1007/s00146-021-01380-0},
	url = {https://www.scopus.com/inward/record.uri?eid=2-s2.0-85122322223&doi=10.1007%2fs00146-021-01380-0&partnerID=40&md5=5cc19cca9544dd814a2f966a6032338a},
	type = {Article},
	publication_stage = {Final},
	source = {Scopus},
	note = {Cited by: 32}
}

@ARTICLE{Feng2022,
	author = {Feng, Jean and Phillips, Rachael V. and Malenica, Ivana and Bishara, Andrew and Hubbard, Alan E. and Celi, Leo A. and Pirracchio, Romain},
	title = {Clinical artificial intelligence quality improvement: towards continual monitoring and updating of AI algorithms in healthcare},
	year = {2022},
	journal = {npj Digital Medicine},
	volume = {5},
	number = {1},
	doi = {10.1038/s41746-022-00611-y},
	url = {https://www.scopus.com/inward/record.uri?eid=2-s2.0-85131510401&doi=10.1038%2fs41746-022-00611-y&partnerID=40&md5=09f7394d60d732fda84bd5dc1c8a133e},
	type = {Article},
	publication_stage = {Final},
	source = {Scopus},
	note = {Cited by: 175; All Open Access, Gold Open Access, Green Open Access}
}

@CONFERENCE{Xia20232630,
	author = {Xia, Boming and Bi, Tingting and Xing, Zhenchang and Lu, Qinghua and Zhu, Liming},
	title = {An Empirical Study on Software Bill of Materials: Where We Stand and the Road Ahead},
	year = {2023},
	journal = {Proceedings - International Conference on Software Engineering},
	pages = {2630 – 2642},
	doi = {10.1109/ICSE48619.2023.00219},
	url = {https://www.scopus.com/inward/record.uri?eid=2-s2.0-85164615895&doi=10.1109%2fICSE48619.2023.00219&partnerID=40&md5=270e4c09670c816f1a428846cd36721a},
	type = {Conference paper},
	publication_stage = {Final},
	source = {Scopus},
	note = {Cited by: 37; All Open Access, Green Open Access}
}

@ARTICLE{Martínez-Fernández2022,
	author = {Martínez-Fernández, Silverio and Bogner, Justus and Franch, Xavier and Oriol, Marc and Siebert, Julien and Trendowicz, Adam and Vollmer, Anna Maria and Wagner, Stefan},
	title = {Software Engineering for AI-Based Systems: A Survey},
	year = {2022},
	journal = {ACM Transactions on Software Engineering and Methodology},
	volume = {31},
	number = {2},
	doi = {10.1145/3487043},
	url = {https://www.scopus.com/inward/record.uri?eid=2-s2.0-85130727000&doi=10.1145%2f3487043&partnerID=40&md5=fbf7ac61842908f81cd945ae6a1009e4},
	type = {Article},
	publication_stage = {Final},
	source = {Scopus},
	note = {Cited by: 138; All Open Access, Green Open Access}
}

@ARTICLE{Guo2025197,
	author = {Guo, Jiajia and Ma, Shaodan and Wen, Chao-Kai and Jin, Shi},
	title = {Performance Monitoring-Enabled Reliable AI-Based CSI Feedback},
	year = {2025},
	journal = {IEEE Transactions on Wireless Communications},
	volume = {24},
	number = {1},
	pages = {197 – 212},
	doi = {10.1109/TWC.2024.3490600},
	url = {https://www.scopus.com/inward/record.uri?eid=2-s2.0-85209903858&doi=10.1109%2fTWC.2024.3490600&partnerID=40&md5=e9a5266423f22cedd5569dfeeb1bc8f1},
	type = {Article},
	publication_stage = {Final},
	source = {Scopus},
	note = {Cited by: 1}
}

@CONFERENCE{Sujatha2023,
	author = {Sujatha, B. and Faraz, K. Anas and Pranathi, N. and Saranya, Ch. B. R. and Chaitanya, B.S.V.},
	title = {Securing data with blockchain and AI},
	year = {2023},
	journal = {AIP Conference Proceedings},
	volume = {2492},
	doi = {10.1063/5.0115385},
	url = {https://www.scopus.com/inward/record.uri?eid=2-s2.0-85161443034&doi=10.1063%2f5.0115385&partnerID=40&md5=cfa4e2264eead8b407277cb3306b9a5c},
	type = {Conference paper},
	publication_stage = {Final},
	source = {Scopus},
	note = {Cited by: 1}
}

@CONFERENCE{Hjerppe2019265,
	author = {Hjerppe, Kalle and Ruohonen, Jukka and Leppänen, Ville},
	title = {The general data protection regulation: Requirements, architectures, and constraints},
	year = {2019},
	journal = {Proceedings of the IEEE International Conference on Requirements Engineering},
	volume = {2019-September},
	pages = {265 – 275},
	doi = {10.1109/RE.2019.00036},
	url = {https://www.scopus.com/inward/record.uri?eid=2-s2.0-85076928916&doi=10.1109%2fRE.2019.00036&partnerID=40&md5=d85865859a886ae0c67148ea6bef1f53},
	type = {Conference paper},
	publication_stage = {Final},
	source = {Scopus},
	note = {Cited by: 39; All Open Access, Green Open Access}
}

@Inbook{Srinivasa2022,
author="Srinivasa, K. G.
and Kurni, Muralidhar
and Saritha, Kuppala",
title="Harnessing the Power of AI to Education",
bookTitle="Learning, Teaching, and Assessment Methods for Contemporary Learners: Pedagogy for the Digital Generation",
year="2022",
publisher="Springer Nature Singapore",
address="Singapore",
pages="311--342",
isbn="978-981-19-6734-4",
doi="10.1007/978-981-19-6734-4_13",
url="https://doi.org/10.1007/978-981-19-6734-4_13"
}

@article{habbal2024artificial,
  title={Artificial Intelligence Trust, risk and security management (AI trism): Frameworks, applications, challenges and future research directions},
  author={Habbal, Adib and Ali, Mohamed Khalif and Abuzaraida, Mustafa Ali},
  journal={Expert Systems with Applications},
  volume={240},
  pages={122442},
  year={2024},
  publisher={Elsevier}
}

@article{groombridge2022gartner,
  title={Gartner top 10 strategic technology trends for 2023},
  author={Groombridge, David and others},
  journal={https://www. gartner. com/en/articles/gartner-top-10-strategic-technology-trends-for-2023},
  year={2022},
  publisher={Gartner}
}

@misc{ibm2025aitrism,
  author       = {{IBM}},
  title        = {What is AI TRiSM?},
  year         = {2025},
  url          = {https://www.ibm.com/think/topics/ai-trism},
  note         = {Accessed: 16 January 2026}
}

@article{avivah2024aitrism,
  title={Tackling Trust, Risk and Security in AI Models},
  author={Avivah, Litan},
  journal={https://www.gartner.com/en/articles/ai-trust-and-ai-risk},
  year={2024},
  publisher={Gartner}
}

@article{nist2024aitrism,
  title={AI Risks and Trustworthiness},
  author={NIST},
  journal={https://airc.nist.gov/airmf-resources/airmf/3-sec-characteristics/},
  year={2024},
  publisher={National Institute of Standards and Technology}
}

@article{aleksandra2025evaluating,
  title={Evaluating Trustworthiness in AI: Risks, Metrics, and Applications Across Industries},
  author={Aleksandra, Nastoska and Bojana, Jancheska and Maryan, Rizinski and Dimitar, Trajanov},
  journal={Electronics},
  volume={14},
  number={13},
  pages={2717},
  year={2025},
  publisher={MDPI AG}
}

@inproceedings{khanvilkar2025automated,
  title={Automated Identification of Machine Learning Technical Debt Code Comments},
  author={Khanvilkar, Omkar and Mkaouer, Mohamed Wiem and AlOmar, Eman Abdullah and ElSaid, Abdelrahman and Chaaben, Amal and Touati, Mohamed},
  booktitle={2025 International Conference on Emerging Technologies and Computing (IC\_ETC)},
  pages={1--6},
  year={2025},
  organization={IEEE}
}

@inproceedings{mailach2023socio,
  title={Socio-technical anti-patterns in building ML-enabled software: insights from leaders on the forefront},
  author={Mailach, Alina and Siegmund, Norbert},
  booktitle={2023 IEEE/ACM 45th International Conference on Software Engineering (ICSE)},
  pages={690--702},
  year={2023},
  organization={IEEE}
}

@inproceedings{nikanjam2021design,
  title={Design smells in Deep Learning programs: an empirical study},
  author={Nikanjam, Amin and Khomh, Foutse},
  booktitle={2021 IEEE International conference on software maintenance and evolution (ICSME)},
  pages={332--342},
  year={2021},
  organization={IEEE}
}

@article{li2022identifying,
  title={Identifying self-admitted technical debt in issue tracking systems using machine learning},
  author={Li, Yikun and Soliman, Mohamed and Avgeriou, Paris},
  journal={Empirical Software Engineering},
  volume={27},
  number={6},
  pages={131},
  year={2022},
  publisher={Springer}
}

@inproceedings{perez2019proposed,
  title={A proposed model-driven approach to manage architectural technical debt life cycle},
  author={P{\'e}rez, Boris and Correal, Dar{\'\i}o and Astudillo, Hern{\'a}n},
  booktitle={2019 IEEE/ACM International Conference on Technical Debt (TechDebt)},
  pages={73--77},
  year={2019},
  organization={IEEE}
}

@article{annunziata2025uncovering,
  title={Uncovering community smells in machine learning-enabled systems: Causes, effects, and mitigation strategies},
  author={Annunziata, Giusy and Lambiase, Stefano and Tamburri, Damian A and Van Den Heuvel, Willem-Jan and Palomba, Fabio and Catolino, Gemma and Ferrucci, Filomena and De Lucia, Andrea},
  journal={ACM Transactions on Software Engineering and Methodology},
  volume={34},
  number={6},
  pages={1--48},
  year={2025},
  publisher={ACM New York, NY}
}

@inproceedings{recupito2024unmasking,
  title={Unmasking data secrets: An empirical investigation into data smells and their impact on data quality},
  author={Recupito, Gilberto and Rapacciuolo, Raimondo and Di Nucci, Dario and Palomba, Fabio},
  booktitle={Proceedings of the IEEE/ACM 3rd International Conference on AI Engineering-Software Engineering for AI},
  pages={53--63},
  year={2024}
}

@inproceedings{cunha2020investigating,
  title={Investigating non-usually employed features in the identification of architectural smells: A machine learning-based approach},
  author={Cunha, Warteruzannan Soyer and Armijo, Guisella Angulo and de Camargo, Valter Vieira},
  booktitle={Proceedings of the 14th Brazilian Symposium on Software Components, Architectures, and Reuse},
  pages={21--30},
  year={2020}
}

@article{de2025software,
  title={Software fairness debt: Building a research agenda for addressing bias in AI systems},
  author={de Souza Santos, Ronnie and Fronchetti, Felipe and Freire, S{\'a}vio and Spinola, Rodrigo},
  journal={ACM Transactions on Software Engineering and Methodology},
  volume={34},
  number={5},
  pages={1--21},
  year={2025},
  publisher={ACM New York, NY}
}

@inproceedings{sutoyo2024satdaug,
  title={SATDAUG-A Balanced and Augmented Dataset for Detecting Self-Admitted Technical Debt},
  author={Sutoyo, Edi and Capiluppi, Andrea},
  booktitle={Proceedings of the 21st International Conference on Mining Software Repositories},
  pages={289--293},
  year={2024}
}

@inproceedings{akgul2025aligning,
  title={Aligning Data Debt with AI-Integrated Software Project Lifecycle Processes: A Standard-Based Mapping Approach},
  author={Akg{\"u}l, Nilay Yorganc{\i}lar and Temizel, Tu{\u{g}}ba Ta{\c{s}}kaya and Top, {\"O}zden {\"O}zcan and Akman, Pelin Dayan},
  booktitle={2025 IEEE/ACM International Conference on Technical Debt (TechDebt)},
  pages={1--11},
  year={2025},
  organization={IEEE}
}

@inproceedings{ximenes2025investigating,
  title={Investigating Issues that Lead to Code Technical Debt in Machine Learning Systems},
  author={Ximenes, Rodrigo and Alves, Antonio Pedro Santos and Escovedo, Tatiana and Spinola, Rodrigo and Kalinowski, Marcos},
  booktitle={2025 IEEE/ACM 4th International Conference on AI Engineering--Software Engineering for AI (CAIN)},
  pages={173--183},
  year={2025},
  organization={IEEE}
}

@article{akman2025people,
  title={People and Management Debt in ML-Integrated Software Projects: Structuring Industry Insights},
  author={Akman, Pelin Dayan and Top, {\"O}zden {\"O}zcan and Temizel, Tugba Taskaya},
  journal={IEEE Access},
  year={2025},
  publisher={IEEE}
}

@inproceedings{shukla2022challenges,
  title={Challenges faced by industries and their potential solutions in deploying machine learning applications},
  author={Shukla, Raj Mani and Cartlidge, John},
  booktitle={2022 IEEE 12th Annual Computing and Communication Workshop and Conference (CCWC)},
  pages={0119--0124},
  year={2022},
  organization={IEEE}
}

@inproceedings{shome2022data,
  title={Data smells in public datasets},
  author={Shome, Arumoy and Cruz, Luis and Van Deursen, Arie},
  booktitle={Proceedings of the 1st International Conference on AI Engineering: Software Engineering for AI},
  pages={205--216},
  year={2022}
}

@inproceedings{moldovan2024python,
  title={The python software quality dataset},
  author={Moldovan, Vasilica-Andreea and Berciu, Liviu-Marian and Patcas, Rares-Danut},
  booktitle={2024 50th Euromicro Conference on Software Engineering and Advanced Applications (SEAA)},
  pages={395--398},
  year={2024},
  organization={IEEE}
}

@article{hamon2024three,
  title={Three challenges to secure AI systems in the context of AI regulations},
  author={Hamon, Ronan and Junklewitz, Henrik and Garrido, Josep Soler and Sanchez, Ignacio},
  journal={Ieee Access},
  volume={12},
  pages={61022--61035},
  year={2024},
  publisher={IEEE}
}

@article{gnitko2024systematic,
  title={Systematic overview of AI security standards},
  author={Gnitko, Kseniia},
  journal={Available at SSRN 4922592},
  year={2024}
}

@article{spelda2025security,
  title={Security practices in AI development},
  author={Spelda, Petr and Stritecky, Vit},
  journal={AI \& SOCIETY},
  pages={1--11},
  year={2025},
  publisher={Springer}
}

@article{ouyang2022training,
  title={Training language models to follow instructions with human feedback},
  author={Ouyang, Long and Wu, Jeffrey and Jiang, Xu and Almeida, Diogo and Wainwright, Carroll and Mishkin, Pamela and Zhang, Chong and Agarwal, Sandhini and Slama, Katarina and Ray, Alex and others},
  journal={Advances in neural information processing systems},
  volume={35},
  pages={27730--27744},
  year={2022}
}

@article{bai2022training,
  title={Training a helpful and harmless assistant with reinforcement learning from human feedback},
  author={Bai, Yuntao and Jones, Andy and Ndousse, Kamal and Askell, Amanda and Chen, Anna and DasSarma, Nova and Drain, Dawn and Fort, Stanislav and Ganguli, Deep and Henighan, Tom and others},
  journal={arXiv preprint arXiv:2204.05862},
  year={2022}
}

@online{ISOIEC27002,
  author       = {{ISO/IEC 27002}},
  title        = {List of {ISO} 27002:2022 Controls — What Changed in 2022?},
  year         = {2022},
  howpublished = {\url{https://sprinto.com/blog/iso-27002-controls/}},
  note         = {Accessed: Dec 02, 2025}
}

@article{simonetta2024iso,
  title={ISO/IEC Standards and Design of an Artificial Intelligence System},
  author={Simonetta, Alessandro and Paoletti, Maria Cristina},
  year={2024}
}

@article{rajbahadur2025building,
  title={Building an Open AIBOM Standard in the Wild},
  author={Rajbahadur, Gopi Krishnan and Gallaba, Keheliya and Rashno, Elyas and Suriyawongkul, Arthit and Bennet, Karen and Stewart, Kate and Hassan, Ahmed E},
  journal={arXiv preprint arXiv:2510.07070},
  year={2025}
}

@article{soundararajansecure,
  title={Secure Configuration Management for Microservices Architecture},
  author={Soundararajan, Balaji}
}

@inproceedings{sheeba2025decentralized,
  title={Decentralized Data Validation for Ethical AI Training},
  author={Sheeba, R and Mahto, Jay Prakash and Ansari, Syed Sabith and Surani, Zian Rajeshkumar and Chinnasamy, P and Alagarsamy, Manjunathan},
  booktitle={2025 International Conference on Computational Robotics, Testing and Engineering Evaluation (ICCRTEE)},
  pages={1--6},
  year={2025},
  organization={IEEE}
}

@article{ranjitsingh2025establish,
  title={Establish legal and regulatory standards for the testing and validation of AI systems to ensure their reliability and safety in operational environments},
  author={Ranjitsingh, Legha Mamta and Rao, TV Subba},
  journal={International Journal of System Assurance Engineering and Management},
  volume={16},
  number={10},
  pages={3338--3353},
  year={2025},
  publisher={Springer}
}

@article{loncar2024secure,
  title={SECURE CODING GUIDELINES AND STANDARDS.},
  author={Loncar, Kristina and Redzepagic, Jasmin and Dakic, Vedran},
  journal={Annals of DAAAM \& Proceedings},
  volume={35},
  year={2024}
}

@inproceedings{jawhar2024ai,
  title={AI-based cybersecurity policies and procedures},
  author={Jawhar, Shadi and Miller, Jeremy and Bitar, Zeina},
  booktitle={2024 IEEE 3rd International Conference on AI in Cybersecurity (ICAIC)},
  pages={1--5},
  year={2024},
  organization={IEEE}
}

@inproceedings{matsuda2019cyber,
  title={Cyber security risk assessment on industry 4.0 using ics testbed with ai and cloud},
  author={Matsuda, Wataru and Fujimoto, Mariko and Aoyama, Tomomi and Mitsunaga, Takuho},
  booktitle={2019 IEEE conference on application, information and network security (AINS)},
  pages={54--59},
  year={2019},
  organization={IEEE}
}

@article{jones2025analysing,
  title={Analysing the role of LLMs in cybersecurity incident management},
  author={Jones, Gavin and Kasimatis, Dimitrios and Pitropakis, Nikolaos and Macfarlane, Richard and Buchanan, William J},
  journal={International Journal of Information Security},
  volume={24},
  number={6},
  pages={1--14},
  year={2025},
  publisher={Springer}
}

@incollection{sarker2024introduction,
  title={Introduction to AI-driven cybersecurity and threat intelligence},
  author={Sarker, Iqbal H},
  booktitle={AI-driven cybersecurity and threat intelligence: Cyber automation, intelligent decision-making and explainability},
  pages={3--19},
  year={2024},
  publisher={Springer}
}

@article{camilo2024ai,
  title={AI-DRIVEN THREAT INTELLIGENCE: ENHANCING CYBERSECURITY IN MODERN SOFTWARE SYSTEMS},
  author={Camilo, Rojas and Yuki, Sato and Eleanor, Bennett},
  journal={Journal of Adaptive Learning Technologies},
  volume={1},
  number={8},
  pages={53--68},
  year={2024},
  publisher={Scientific Bulletin}
}

@inproceedings{feffer2024red,
  title={Red-teaming for generative AI: Silver bullet or security theater?},
  author={Feffer, Michael and Sinha, Anusha and Deng, Wesley H and Lipton, Zachary C and Heidari, Hoda},
  booktitle={Proceedings of the AAAI/ACM Conference on AI, Ethics, and Society},
  volume={7},
  pages={421--437},
  year={2024}
}

@article{walter2024red,
  title={A red teaming framework for securing AI in maritime autonomous systems},
  author={Walter, Mathew J and Barrett, Aaron and Tam, Kimberly},
  journal={Applied Artificial Intelligence},
  volume={38},
  number={1},
  pages={2395750},
  year={2024},
  publisher={Taylor \& Francis}
}

@inproceedings{jaeyalakshmi2023self,
  title={A Self-learning Ai-Based Information Leak Protection System},
  author={Jaeyalakshmi, M and Gangadhar, P Rohit and Srivatsan, M and Bhavani, M},
  booktitle={International Conference on Advances in Artificial Intelligence and Machine Learning in Big Data Processinging},
  pages={68--78},
  year={2023},
  organization={Springer}
}

@article{alneyadi2016survey,
  title={A survey on data leakage prevention systems},
  author={Alneyadi, Sultan and Sithirasenan, Elankayer and Muthukkumarasamy, Vallipuram},
  journal={Journal of Network and Computer Applications},
  volume={62},
  pages={137--152},
  year={2016},
  publisher={Elsevier}
}

@article{gaddam2024ai,
  title={AI-POWERED DATA MASKING FOR PRIVACY-PRESERVING CLOUD DATA SHARING},
  author={Gaddam, Narayana},
  journal={International Journal of Advanced Research in Cloud Computing},
  volume={5},
  number={2},
  pages={12--22},
  year={2024}
}

@article{bilakantisecure,
  title={Secure Data Masking for Healthcare Data Protection},
  author={Bilakanti, Goutham},
journal = {}
}

@article{martinez2022software,
  title={Software engineering for AI-based systems: a survey},
  author={Mart{\'\i}nez-Fern{\'a}ndez, Silverio and Bogner, Justus and Franch, Xavier and Oriol, Marc and Siebert, Julien and Trendowicz, Adam and Vollmer, Anna Maria and Wagner, Stefan},
  journal={ACM Transactions on Software Engineering and Methodology (TOSEM)},
  volume={31},
  number={2},
  pages={1--59},
  year={2022},
  publisher={ACM New York, NY}
}

@inproceedings{banerjee2025securing,
  title={Securing the future of AI: a holistic approach to trust and robustness},
  author={Banerjee, Sourav and Thomas, Mathews and Chavan, Vinod and Mangla, Utpal and Tummalapenta, Srinivas},
  booktitle={Assurance and Security for AI-enabled Systems 2025},
  volume={13476},
  pages={121--142},
  year={2025},
  organization={SPIE}
}

@article{mekhfioui2025optimized,
  title={Optimized digital watermarking for robust information security in embedded systems},
  author={Mekhfioui, Mohcin and El Bazi, Nabil and Laayati, Oussama and Satif, Amal and Bouchouirbat, Marouan and Kissi, Cha{\"\i}ma{\^a} and Boujiha, Tarik and Chebak, Ahmed},
  journal={Information},
  volume={16},
  number={4},
  pages={322},
  year={2025},
  publisher={MDPI}
}

@article{narula2025exploring,
  title={Exploring AI Security: A Systematic Mapping Study},
  author={Narula, Sidhant and Ghasemigol, Mohammad and Carnerero-Cano, Javier and Minnich, Amanda and Lupu, Emil and Takabi, Daniel},
  journal={IEEE Access},
  year={2025},
  publisher={IEEE}
}

@article{pooyandeh2022cybersecurity,
  title={Cybersecurity in the AI-Based metaverse: A survey},
  author={Pooyandeh, Mitra and Han, Ki-Jin and Sohn, Insoo},
  journal={Applied Sciences},
  volume={12},
  number={24},
  pages={12993},
  year={2022},
  publisher={MDPI}
}

@phdthesis{xia2025operationalising,
  title={Operationalising Safe and Responsible AI: A System Level Perspective},
  author={Xia, Boming},
  year={2025},
  school={UNSW Sydney}
}

@article{sood2025malicious,
  title={Malicious AI Models Undermine Software Supply-Chain Security},
  author={Sood, Aditya K and Zeadally, Sherali},
  journal={Communications of the ACM},
  volume={68},
  number={6},
  pages={62--71},
  year={2025},
  publisher={ACM New York, NY, USA}
}

@inproceedings{riggio2021ai,
  title={Ai@ edge: A secure and reusable artificial intelligence platform for edge computing},
  author={Riggio, Roberto and Coronado, Estefan{\'\i}a and Linder, Neiva and Jovanka, Adzic and Mastinu, Gianpiero and Goratti, Leonardo and Rosa, Miguel and Schotten, Hans and Pistore, Marco},
  booktitle={2021 Joint European Conference on Networks and Communications \& 6G Summit (EuCNC/6G Summit)},
  pages={610--615},
  year={2021},
  organization={IEEE}
}

@article{celepija2025towards,
  title={Towards a structured AI development lifecycle for reusable AI products in the public sector},
  author={Celepija, Albana and Lepri, Bruno and Kazhamiakin, Raman},
  year={2025}
}

@incollection{gujar2025data,
  title={Data standardization and interoperability},
  author={Gujar, Praveen},
  booktitle={Data usability in the enterprise: how usability leads to optimal digital experiences},
  pages={89--110},
  year={2025},
  publisher={Springer}
}

@inproceedings{zhi2024algorithm,
  title={Algorithm for Data Format Conversion and Compatibility Guarantee in Technical Platforms for Cross Platform Application Integration},
  author={Zhi, Lanyue and Liu, Shaoguang and Dai, Haoqi and Liu, Mingwei and Wang, Jinhe},
  booktitle={2024 IEEE 4th International Conference on Data Science and Computer Application (ICDSCA)},
  pages={893--897},
  year={2024},
  organization={IEEE}
}

@inproceedings{naeem2021trends,
  title={Trends and future perspective challenges in big data},
  author={Naeem, Muhammad and Jamal, Tauseef and Diaz-Martinez, Jorge and Butt, Shariq Aziz and Montesano, Nicolo and Tariq, Muhammad Imran and De-la-Hoz-Franco, Emiro and De-La-Hoz-Valdiris, Ethel},
  booktitle={Advances in intelligent data analysis and applications: Proceeding of the sixth euro-China conference on intelligent data analysis and applications, 15--18 October 2019, Arad, Romania},
  pages={309--325},
  year={2021},
  organization={Springer}
}

@book{siriwardena2019advanced,
  title={Advanced API security: OAuth 2.0 and beyond},
  author={Siriwardena, Prabath},
  year={2019},
  publisher={Apress}
}

@article{paidy2024securing,
  title={Securing AI-driven APIs: Authentication and abuse prevention},
  author={Paidy, Pavan and Chaganti, Krishna},
  journal={International Journal of Emerging Research in Engineering and Technology},
  volume={5},
  number={1},
  pages={27--37},
  year={2024}
}

@article{shneiderman2020bridging,
  title={Bridging the gap between ethics and practice: guidelines for reliable, safe, and trustworthy human-centered AI systems},
  author={Shneiderman, Ben},
  journal={ACM Transactions on Interactive Intelligent Systems (TiiS)},
  volume={10},
  number={4},
  pages={1--31},
  year={2020},
  publisher={ACM New York, NY, USA}
}

@article{aagaard2024discrepancies,
  title={Discrepancies between Promised and Actual AI Capabilities in the Continuous Vital Sign Monitoring of In-Hospital Patients: A Review of the Current Evidence},
  author={Aagaard, Nikolaj and Aasvang, Eske K and Meyhoff, Christian S},
  journal={Sensors (Basel, Switzerland)},
  volume={24},
  number={19},
  pages={6497},
  year={2024}
}

@article{renard2024understanding,
  title={Understanding prediction discrepancies in classification},
  author={Renard, Xavier and Laugel, Thibault and Detyniecki, Marcin},
  journal={Machine Learning},
  volume={113},
  number={10},
  pages={7997--8026},
  year={2024},
  publisher={Springer}
}

@article{shahin2025automated,
  title={Automated multi-model framework for malaria detection using deep learning and feature fusion},
  author={Shahin, Osama R and Alshammari, Hamoud H and Alabdali, Raed N and Salaheldin, Ahmed M and Saleh, Neven},
  journal={Scientific Reports},
  volume={15},
  number={1},
  pages={25672},
  year={2025},
  publisher={Nature Publishing Group UK London}
}

@article{gupta2025optimizing,
  title={Optimizing Cryptocurrency Trading Strategies through Artificial Intelligence and Blockchain Integration: A Multi-Model Framework for Predictive Analytics.},
  author={Gupta, Pooja and Singh, Ravinder and Kaur, Harshdeep and Subramanian, N and Jain, Neha and others},
  journal={Advances in Consumer Research},
  volume={2},
  number={3},
  year={2025}
}

@article{musunuru2025optimizing,
  title={Optimizing Hot Standby Redundancy Using AI for Network Traffic Balancing and Failover Management},
  author={Musunuru, Mohan Vamsi and Devi, Chiranjeevi and Sethuraman, Swaminathan},
  journal={Journal of Knowledge Learning and Science Technology ISSN: 2959-6386 (online)},
  volume={4},
  number={3},
  pages={14--26},
  year={2025}
}

@article{gupta2025gpu,
  title={GPU Reliability in AI Clusters: A Study of Failure Modes and Effects},
  author={Gupta, Sameeksha},
  journal={Journal Of Engineering And Computer Sciences},
  volume={4},
  number={6},
  pages={298--306},
  year={2025}
}

@article{tutuncuoglu2024zero,
  title={Zero-Downtime AI: Predictive and Autonomous Server Restoration Without Human Input},
  author={Tutuncuoglu, Bekir Tolga},
  journal={Available at SSRN 5249062},
  year={2024}
}

@article{olutimehin2025adversarial,
  title={Adversarial threats to AI-driven systems: Exploring the attack surface of machine learning models and countermeasures},
  author={Olutimehin, Abayomi Titilola and Ajayi, Adekunbi Justina and Metibemu, Olufunke Cynthia and Balogun, Adebayo Yusuf and Oladoyinbo, Tunboson Oyewale and Olaniyi, Oluwaseun Oladeji},
  journal={Available at SSRN 5137026},
  year={2025}
}

@article{samuel2025adversarial,
  title={Adversarial AI the New Frontier in Cybersecurity Threats and Defenses},
  author={Samuel, Olusoji John},
  journal={Journal of Science, Technology and Engineering Research},
  volume={3},
  number={1},
  pages={1--13},
  year={2025}
}

@inproceedings{barlas2022exploiting,
  title={Exploiting input sanitization for regex denial of service},
  author={Barlas, Efe and Du, Xin and Davis, James C},
  booktitle={Proceedings of the 44th International Conference on Software Engineering},
  pages={883--895},
  year={2022}
}

@article{kiranbabu2025challenge,
  title={The Challenge of Adversarial Attacks on AI-Driven Cybersecurity Systems.},
  author={Kiranbabu, MNV and Viji, A Jeraldine and Chandanan, Amit Kumar and Birchha, Vijay and Pandey, Tushar Kumar and Sar, Sumit Kumar},
  journal={Journal of Cybersecurity \& Information Management},
  volume={15},
  number={1},
  year={2025}
}

@inproceedings{franzoni2023black,
  title={From black box to glass box: advancing transparency in artificial intelligence systems for ethical and trustworthy AI},
  author={Franzoni, Valentina},
  booktitle={International Conference on Computational Science and Its Applications},
  pages={118--130},
  year={2023},
  organization={Springer}
}

@inproceedings{pedreschi2019meaningful,
  title={Meaningful explanations of black box AI decision systems},
  author={Pedreschi, Dino and Giannotti, Fosca and Guidotti, Riccardo and Monreale, Anna and Ruggieri, Salvatore and Turini, Franco},
  booktitle={Proceedings of the AAAI conference on artificial intelligence},
  volume={33},
  number={01},
  pages={9780--9784},
  year={2019}
}

@article{ge2024mode,
  title={A mode-switched control architecture for human-in-the-loop teleoperation of multislave robots via data-training-based observer},
  author={Ge, Ming-Feng and Xu, Jing-Zhe and Liu, Zhi-Wei and Huang, Jian},
  journal={IEEE Transactions on Systems, Man, and Cybernetics: Systems},
  volume={54},
  number={4},
  pages={2471--2483},
  year={2024},
  publisher={IEEE}
}

@article{lu2023responsible,
  title={Responsible-AI-by-design: A pattern collection for designing responsible artificial intelligence systems},
  author={Lu, Qinghua and Zhu, Liming and Xu, Xiwei and Whittle, Jon},
  journal={Ieee Software},
  volume={40},
  number={3},
  pages={63--71},
  year={2023},
  publisher={IEEE}
}

@inproceedings{lu2022software,
  title={Software engineering for responsible AI: An empirical study and operationalised patterns},
  author={Lu, Qinghua and Zhu, Liming and Xu, Xiwei and Whittle, Jon and Douglas, David and Sanderson, Conrad},
  booktitle={Proceedings of the 44th International Conference on Software Engineering: Software Engineering in Practice},
  pages={241--242},
  year={2022}
}

@article{lu2023developing,
  title={Developing responsible chatbots for financial services: a pattern-oriented responsible artificial intelligence engineering approach},
  author={Lu, Qinghua and Luo, Yuxiu and Zhu, Liming and Tang, Mingjian and Xu, Xiwei and Whittle, Jon},
  journal={IEEE Intelligent Systems},
  volume={38},
  number={6},
  pages={42--51},
  year={2023},
  publisher={IEEE}
}

@article{yang2024multi,
  title={Multi-language software development: Issues, challenges, and solutions},
  author={Yang, Haoran and Nong, Yu and Wang, Shaowei and Cai, Haipeng},
  journal={IEEE Transactions on Software Engineering},
  volume={50},
  number={3},
  pages={512--533},
  year={2024},
  publisher={IEEE}
}

@article{ebad2022exploring,
  title={Exploring how to apply secure software design principles},
  author={Ebad, Shouki A},
  journal={IEEE Access},
  volume={10},
  pages={128983--128993},
  year={2022},
  publisher={IEEE}
}

@article{patel2021creating,
  title={Creating Safe and Secure AI-From Computer Design to Cloud Technology},
  author={Patel, Jay and Shah, Harshal},
  journal={INTERNATIONAL RESEARCH JOURNAL OF ENGINEERING \& APPLIED SCIENCES},
  volume={9},
  number={4},
  pages={10--55083},
  year={2021}
}

@book{ernst2021technical,
  title={Technical Debt in Practice: How to Find It and Fix It},
  author={Ernst, Neil and Kazman, Rick and Delange, Julien},
  year={2021},
  publisher={MIT Press}
}

@article{bucaioni2025checklist,
  title={A checklist of quality concerns for architecting ML-intensive systems},
  author={Bucaioni, Alessio and Kazman, Rick and Pelliccione, Patrizio},
  journal={Journal of Systems and Software},
  pages={112612},
  year={2025},
  publisher={Elsevier}
}

\end{document}